\documentclass[nofootinbib,preprintnumbers]{revtex4}
\usepackage{graphicx}% Include figure files
\usepackage{dcolumn}% Align table columns on decimal point
\usepackage{amsmath,amssymb,epsfig}
\usepackage{paralist}
\usepackage{comment}
\usepackage{graphicx}
\usepackage{multirow}
\usepackage{wrapfig}

\renewcommand{\vec}[1]{\boldsymbol{\mathrm{#1}}}

\allowdisplaybreaks

\begin{document}

\title{General relativistic observables of the GRAIL mission}

\author{Slava
%Vyacheslav
G. Turyshev$^{1,3}$, Viktor T. Toth$^2$, and Mikhail V. Sazhin$^3$
}

\affiliation{\vskip 3pt
$^1$Jet Propulsion Laboratory, California Institute of Technology, 4800 Oak Grove Drive, Pasadena, CA 91109-0899, USA
}%

\affiliation{\vskip 3pt
$^2$Ottawa, ON  K1N 9H5, Canada
}%

\affiliation{\vskip 3pt
$^3$Sternberg Astronomical Institute, Lomonosov Moscow State University,  Moscow, Russia
}%

\date{\today}% It is always \today, today,
             %  but any date may be explicitly specified

\begin{abstract}

We present a realization of astronomical relativistic reference frames in the Solar System and its application to the GRAIL mission. We model the necessary spacetime coordinate transformations for light-trip time computations and address some practical aspects of the implementation of the resulting model.  We develop all the relevant relativistic coordinate transformations that are needed to describe the motion of the GRAIL spacecraft and to compute all observable quantities. We take into account major relativistic effects contributing to the dual one-way range observable, which is derived from one-way signal travel times between the two GRAIL spacecraft. We develop a general relativistic model for this fundamental observable of GRAIL, accurate to 1~$\mu$m. We develop and present a relativistic model for another key observable of this experiment, the dual one-way range-rate, accurate to 1~$\mu$m/s.  The presented formulation justifies the basic assumptions behind the design of the GRAIL mission. It may also be used to further improve the already impressive results of this lunar gravity recovery experiment after the mission is complete. Finally, we present transformation rules for frequencies and gravitational potentials and their application to GRAIL.

\end{abstract}

% insert suggested PACS numbers in braces on next line
\pacs{03.30.+p, 04.25.Nx, 04.80.-y, 06.30.Gv, 95.10.Eg, 95.10.Jk, 95.55.Pe}
% insert suggested keywords - APS authors don't need to do this
%\keywords{}

\maketitle

\section{Introduction}

Several past, present and planned space missions utilize a pair of spacecraft orbiting a celestial body in a tight formation. Continuous high-precision range and range-rate measurements between the spacecraft yield detailed information about the gravity field of the target body. Missions of this type include the Gravity Recovery and Climate Experiment (GRACE) mission \cite{Kim:2000} in orbit around the Earth; the Gravity Recovery and Interior Laboratory (GRAIL) mission, which comprises two spacecraft in orbit around the Moon \cite{Roncoli-Fujii:2010,Park-etal:2012,Zuber2012,Asmar2012}; and planned missions such as a GRACE Follow-on mission or a proposal for a GRAIL-like mission in orbit around Mars.

Of these, the mission of particular current interest is GRAIL, as the two GRAIL spacecraft are presently (2012) orbiting the Moon. In this paper, we therefore focus on the GRAIL mission and its science observables. However, the lessons learned are also applicable to other, similar experiments.

To reach its science objectives, the GRAIL mission relies on precision navigation of both spacecraft and accurate range measurements between the two lunar orbiters performed with their on-board Ka-band ranging (KBR) system. The  instantaneous one-way range measurements performed at each spacecraft are time-tagged and processed on the ground to form dual one-way range (DOWR) measurements \cite{TimingMemo:2010}.  The mission relies on precision timing of all critical events (using the on-board ultra-stable oscillator, or USO) related to the transmission and reception of various microwave signals used on GRAIL for formation tracking and navigation. The resulting time series of highly accurate radio-metric data will allow for a major increase in accuracy when studying the gravity field of the Moon. The differential nature of the science measurements allows for the removal of a number of measurement errors introduced in the process. In particular, the approach compensates for errors due to long-term instabilities of the on-board USOs. This allows for an improvement in accuracy by about two orders of magnitude when compared to other techniques. In fact, the anticipated accuracies are of the order of 1~$\mu$m in range and 1~$\mu$m/s in range rate.

It was recognized early on during the mission development that due to the expected high accuracy of ranging data on GRAIL,  models of its observables must be formulated within the framework of Einstein's general theory of relativity. In fact, a naive application of the observable models developed for the GRACE mission \cite{Kim:2000} may have led to significant model discrepancy (as emphasized in Ref.~\cite{TimingMemo:2010}), as these models do not take into account relativistic contributions that are critical for GRAIL. The ultimate observable model for GRAIL must correctly describe all the timing events occurring during the science operations of the mission, including both the navigation observables (S- and X-band, $\sim2$~GHz and $\sim8$~GHz correspondingly) and inter-spacecraft tracking (Ka-band, $\sim32$~GHz) data.

The model must represent the different times at which the events are computed, involving the time of transmission of the Ka-band signal at one of the spacecraft, say GRAIL-A, at $t_{\rm A0}$, and the reception of this signal by its twin, GRAIL-B, at time $t_{\rm B}$. In addition, the model must include a description of the process of transmitting S-band and X-band navigation signals from either spacecraft and reception of this signal at a Deep Space Network (DSN) tracking station at time $t_{\rm C}$.

\begin{figure}[t]
\includegraphics[width=0.45\linewidth,angle=0]{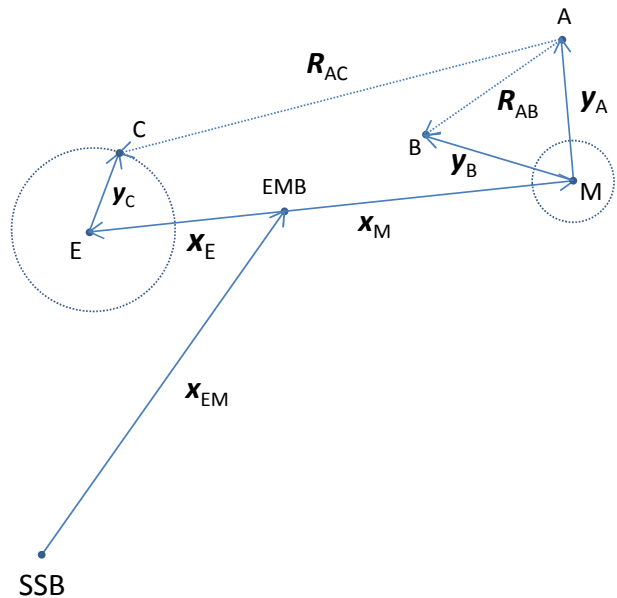}
\vskip -6pt
\caption{\label{fig:grail} Representative geometry (not to scale) of the vectors involved in the computation of the GRAIL observables. ``SSB'' is the Solar System barycenter, ``E'' is the center of the Earth, ``M'' is the center of Moon, ``EMB'' is the Earth-Moon Barycenter. ``A'' and ``B'' are the positions of the GRAIL-A and GRAIL-B spacecraft, respectively, and ``C'' is the position of the DSN tracking antenna on the surface of the Earth.}
\end{figure}

We model the range $R_{\rm AB} = |\vec{R}_{\rm AB}|$ between the two spacecraft $A$ and $B$ (see Fig.~\ref{fig:grail} for geometry and notations) as:
\begin{equation}
R_{\rm AB} = |\vec{R}_{\rm AB}|=|\vec{x}_{\rm B}-\vec{x}_{\rm A}| = |(\vec{x}_{\rm EM}+\vec{x}_{\rm M}+\vec{y}_{\rm B}) - (\vec{x}_{\rm EM} + \vec{x}_{\rm M} + \vec{y}_{\rm A})|,
\label{eq(1)}
\end{equation}
where $\vec{x}_{\rm EM}$ is the vector connecting the Solar System barycenter (SSB) with the Earth-Moon barycenter (EMB), $\vec{x}_{\rm M}$ is the vector from the EMB to the Moon's (M) center of mass, $\vec{x}_{\rm A}$ and $\vec{x}_{\rm B}$ are vectors connecting the SSB with the positions of the two GRAIL orbiters and vectors $\vec{y}_{\rm A}$ and $\vec{y}_{\rm B}$ connect the Moon's center of mass with the orbiters.

For navigation purposes, both orbiters maintain communication links with a ground-based DSN antenna. The range $R_{\rm AC}=|\vec{R}_{\rm AC}|$ between a GRAIL spacecraft (GRAIL-A, for instance) and a ground-based antenna can be modeled as:
{}
\begin{equation}
R_{\rm AC} = |\vec{R}_{\rm AC}|=|\vec{x}_{\rm C}-\vec{x}_{\rm A}| = |(\vec{x}_{\rm EM}+\vec{x}_{\rm E}+\vec{y}_{\rm C}) - (\vec{x}_{\rm EM} + \vec{x}_{\rm M} + \vec{y}_{\rm A})|,
\label{eq(1a)}
\end{equation}
where $\vec{x}_{\rm E}$ is the vector from the EMB to the geocenter (E), $\vec{x}_{\rm C}$ is the vector connecting the SSB with the ground antenna whereas the vector $\vec{y}_{\rm C}$ determines the geocentric position of the ground antenna's reference point.

For actual computations, we use several different reference systems\footnote{Following Refs.~\cite{Soffel-etal:2003,Tommei-etal:2010}, we use the term ``reference system'' to describe a purely mathematical construction, while a ``reference frame'' is a physical realization of such.}. The Solar System Barycentric Coordinate Reference System (BCRS) has its origin at the SSB. The origin of the Geocentric Coordinate Reference System (GCRS) is the Earth's center of mass. Positions of DSN ground stations are given with respect to another terrestrial coordinate system, the Topocentric Coordinate Reference System (TCRS; see also Ref.~\cite{IERS2010}). We also consider the Lunicentric Coordinate Reference System (LCRS; for additional discussion, see Ref.~\cite{Kopeikin2010}), the origin of which is fixed at the Moon's center of mass. Finally, we attach to each spacecraft its Satellite Coordinate Reference System (SCRS; for a similar approach aimed to construct a reference frame for the GAIA project, see Ref.~\cite{Klioner2004}). (We discuss these reference frames and their relationships in depth in Sec.~\ref{sec:celest-RF}.)

%%%%%%%%%%%%%%%%%%%%%%%%%%%
%%%%%%%%%%%%%%%%%%%%%%%%%%%

Equations (\ref{eq(1)}) and (\ref{eq(1a)}) offer a good starting point to develop an appropriate relativistic formulation for the experiment. The six vectors involved in Eqs.~(\ref{eq(1)})--(\ref{eq(1a)}) can be expressed in terms of their respective points of origin: e.g., $\vec{x}_{\rm EM}$ would be expressed in the BCRS, $\vec{y}_{\rm A}$ and $\vec{y}_{\rm B}$ in the LCRS, $\vec{R}_{\rm AB}$ and $\vec{R}_{\rm AC}$ in the SCRS of GRAIL-A, etc. Each of these coordinate systems has a corresponding time coordinate. To compute the vector sums and differences, all vectors involved must be converted to a common relativistic space-time reference system. Although in general relativity one can introduce any reference frame to describe the experiment, the best practical choice is offered by  some realization of the BCRS. We will use a realization of the BCRS that is called the SSB reference frame. The coordinate time associated with the BCRS is TCB (Barycentric Coordinate Time). For practical applications, it is often preferable to use another time scale, the TDB (Barycentric Dynamical Time). Currently published planetary ephemerides are provided using TDB. TDB and TCB differ only by a linear scaling. The advantage of using TDB is that the difference between it and terrestrial timescales (e.g., TT, defined in Sec.~\ref{sec:timekeeping}) is as small as possible and periodic. The choice of the TDB as the SSB time coordinate is realized by the appropriate linear scaling of space coordinates and planetary masses (see \cite{Klioner:2008,Klioner-etal:2010,Kopeikin-book} for review).

The vectors $\vec{x}_{\rm E}, \vec{x}_{\rm M},$ and $\vec{x}_{\rm EM}$ are readily available in the SSB reference frame, obtained by numerical integration and from Solar System ephemerides \cite{FolknerWilliamsBoggs:2009}. The vectors $\vec{y}_{\rm A}$, $\vec{y}_{\rm B}$ and $\vec{y}_{\rm C}$ have to be transformed to the SSB frame from geocentric and lunicentric reference systems, respectively. Clearly, the required conversion between reference systems also involves conversion of the relativistic time coordinate. The equations of motion of the Moon and Earth, including all the relativistic effects at an accuracy even exceeding that of the GRAIL experiment, have already been discussed elsewhere \cite{Turyshev:2012}; here we concentrate on the computation of observables.

In this paper, we focus on the formulation of a relativistic model for computing the observables of the GRAIL mission, with results that are applicable to other past and planned missions with similar observables. We address some practical aspects of the implementation of these computations. In Sec.~\ref{sec:celest-RF} we discuss all relevant relativistic four-dimensional reference systems and the transformations that are required to make the vector sums in Eqs.~(\ref{eq(1)}) and (\ref{eq(1a)}) computable. In Sec.~\ref{sec:KBRR} we discuss the process of forming the inter-satellite Ka-band range (KBR) observables of GRAIL and derive a model for the dual one-way range (DOWR) observable. We also develop a relativistic model for another fundamental observable on GRAIL: the dual one-way range-rate (DOWRR). We conclude with a set of recommendations and an outlook in Sec.~\ref{sec:sonc}.

In order to keep the main body of the paper focused, we chose to present some calculational details in the form of appendices. In Appendix~\ref{sec:phase-delay} we present some important derivations: In Appendix \ref{sec:post-Mink} we derive the solution for the post-Minkowskian space-time in general relativity, in Appendix \ref{sec:em-phase} we derive analytic expressions to describe the phase of an electromagnetic signal in gravitational field, and in Appendix \ref{sec:gr-delay} we discuss the coordinate gravitational time delay. In Appendix \ref{sec:int-error} contains a discussion on the evaluation of the integral that is needed to assess the full accuracy of the DOWR observable. Finally, in Appendix~\ref{sec:frqxfr} we briefly address the transfer of a precision frequency reference between the spacecraft and a ground station.

The notational conventions used in this paper are as follows. Latin indices from the beginning of the alphabet, $a,b,c,...$, are used to denote Solar System bodies. Latin indices from the second half of the alphabet ($m,n,...$) are space-time indices that run from 0 to 3. Greek indices $\alpha,\beta,...$ are spatial indices that run from 1 to 3. In case of repeated indices in products, the Einstein summation rule applies: e.g., $a_mb^m=\sum_{m=0}^3a_mb^m$. Bold letters denote spatial (three-dimensional) vectors: e.g., ${\vec a} = (a_1, a_2, a_3), {\vec b} = (b_1, b_2, b_3)$. The dot is used to indicate the Euclidean inner product of spatial vectors: e.g., $(\vec a \cdot \vec b) = a_1b_1 + a_2b_2 + a_3b_3$. Latin indices are raised and lowered using the metric $g_{mn}$. The Minkowski (flat) space-time metric is given by $\gamma_{mn} = {\rm diag} (1, -1, -1, -1)$, so that $\gamma_{\mu\nu}a^\mu b^\nu=-({\vec a}\cdot{\vec b})$. We use powers of the inverse of the speed of light, $c^{-1}$, and the gravitational constant, $G$ as bookkeeping devices for order terms: in the low-velocity ($v\ll c$), weak-field ($GM/r\ll c^2$) approximation, a quantity of ${\cal O}(c^{-2})\simeq{\cal O}(G)$, for instance, has a magnitude comparable to $v^2/c^2$ or $GM/c^2r$. The notation ${\cal O}(a^k,b^\ell)$ is used to indicate that the preceding expression is free of terms containing powers of $a$ greater than or equal to $k$, and powers of $b$ greater than or equal to $\ell$.

\section{Space-time reference frames and transformations}
\label{sec:celest-RF}

The theory of general relativity is generally covariant. In the Riemannian geometry that underlies the theory, coordinate charts are merely labels. One may choose an arbitrary coordinate system to describe the results of a particular experiment. Space-time coordinates have no direct physical meaning and it is essential to construct physical observables as coordinate-independent quantities.

On the other hand, some of the available coordinate systems have important practical advantages. These systems are usually associated with a particular celestial body, ground-based facility or spacecraft, thereby yielding a material realization of a reference system to be used to describe the results of precision experiments. In order to interpret the results of observations or experiments, one picks a specific coordinate system that is chosen for the sake of convenience and calculational expediency, formulates a coordinate picture of the measurement procedure, and then derives the observable. It is also known that an ill-defined reference frame may lead to the appearance of non-physical terms that may significantly complicate the interpretation of the data. Therefore, in practical problems involving relativistic reference frames, choosing the right coordinate system with clearly understood properties is of paramount importance, even as we recognize that in principle, all (non-degenerate) coordinate systems are created equal \cite{Soffel-etal:2003}.

In a recent study \cite{Turyshev:2012}, we presented a new approach to investigate the dynamics of an isolated, gravitationally bound astronomical $N$-body system in the weak field, slow-motion approximation of the general theory of relativity. Celestial bodies are described using an arbitrary energy-momentum tensor and assumed to possess any number of internal multipole moments. Using the harmonic gauge conditions together with a requirement for preserving conservation laws, we were able to construct the relativistic proper reference frame associated with a particular body. We also were able to determine explicitly all the terms of the resulting coordinate transformations and their inverses. In this paper we rely on the results obtained in Refs.~\cite{Turyshev:2012,Turyshev-etal:2012} and develop a set of coordinate reference frames for GRAIL.

To reach its scientific objectives, in addition to the BCRS, GRAIL will have to utilize a set of several fundamental coordinate reference frames. These include terrestrial reference systems, namely the GCRS and the TCRS, and lunar reference systems, the LCRS and SCRS. In Ref.~\cite{Turyshev:2012}, we presented the detailed structure of the representations of the metric tensor corresponding to the various reference frames involved, the rules for transforming relativistic gravitational potentials, the coordinate transformations between the frames and the resulting relativistic equations of motion. The accuracy that is achievable by these calculations is sufficient to accommodate modern-day experiments in the Solar System and exceeds that needed for GRAIL. Here, we present the essential part of these transformations between various coordinate systems involved and dealing with transformations of relativistic time scales and position vectors, at the level of accuracy required by GRAIL.

\subsection{Barycentric Coordinate Reference Frame (BCRS)}
\label{sec:BCRS}

The Barycentric Celestial Reference System  (BCRS) is defined with coordinates $\{x^m\}\equiv(ct, \vec{x}=x^\alpha)$, where $t$ is TCB. The BCRS is a particular implementation of a barycentric reference system in the Solar System. The metric tensor $g_{mn}(x)$ of the BCRS satisfies the harmonic gauge condition. It can be written \cite{Turyshev:2012} as
\begin{eqnarray}
g_{00}&=& 1-\frac{2}{c^2}{w}+
\frac{2}{c^4}{w}^2+{\cal O}(c^{-6}),\qquad
g_{0\alpha}= -\frac{4}{c^3}\gamma_{\alpha\lambda}
{w}^\lambda+{\cal O}(c^{-5}),\qquad
g_{\alpha\beta}= \gamma_{\alpha\beta}+\gamma_{\alpha\beta}\frac{2}{c^2} {w}+{\cal O}(c^{-4}),
\label{eq:gab-BCRS}
\end{eqnarray}
where ${w}$ and ${w}^\lambda$ are harmonic gauge potentials that can be presented, at the level of accuracy suitable for the purposes of the GRAIL mission (i.e., neglecting higher order mass- and current-multipole moments), in the form \cite{Soffel-etal:2003,Turyshev:2012,Turyshev-etal:2012}:
{}
\begin{eqnarray}
{w}&=&\sum_b
\frac{GM_b}{r^{}_{b}}\Big(1+\frac{1}{c^2}\Big\{2{v}_{b}^2-\sum_{c\not=b}\frac{GM_c}{r_{cb}}-{\textstyle\frac{1}{2}}({\vec n}_{b}\cdot{\vec v}_{b})^2-{\textstyle\frac{1}{2}}({\vec r}_{b}\cdot{\vec a}_{b})\Big\}\Big)+{\cal O}(c^{-3}),~~~~~
\label{eq:w_0-BCRS}\\
{\vec w}&=&\sum_b\frac{GM_b}{r_{b}}{\vec v}_{b}+{\cal O}(c^{-2}),
\label{eq:w_a-BCRS}
\end{eqnarray}
where ${\vec r}_{b}={\vec x}-{\vec z}_{b}$, $r_{b}=|{\vec r}_{b}|$, and ${\vec n}_{b}={\vec r}_{b}/r_{b}$, with  ${\vec z}_{b}$ being the barycentric position of body $b$, and we use ${\vec r}_{ab}={\vec r}_{b}-{\vec r}_{a}$ to denote the vector separating two bodies $a$ and $b$. Also, the overdot denotes ordinary differentiation with respect to $t$, ${\vec v}_{b}=\dot{\vec z}_{b}$ ($v_b=|\vec{v}_b|$) and ${\vec a}_{b}=\ddot{\vec z}_{b}$ ($a_b=|\vec{a}_b|$) are the barycentric velocity and  acceleration of body $b$, and $M_b$ is its rest mass. Lastly, the summation in (\ref{eq:w_0-BCRS})--(\ref{eq:w_a-BCRS}) is being performed over all the bodies $b=1,2..., N$ in the Solar System. The metric tensor (\ref{eq:gab-BCRS}) and the gravitational potentials (\ref{eq:w_0-BCRS})--(\ref{eq:w_a-BCRS}) have sufficient accuracy for modern precision experiments in the Solar System \cite{Soffel-etal:2003,Turyshev:2012}.

From Fig.~\ref{fig:grail}, we can read off the barycentric positions of the Earth, ${\vec z}_{\rm E}$, and the Moon, ${\vec z}_{\rm M}$: ${\vec z}_{\rm E}=\vec{x}_{\rm EM}+\vec{x}_{\rm E}$ and ${\vec z}_{\rm M}=\vec{x}_{\rm EM}+\vec{x}_{\rm M}$, respectively. Both of these vectors, and corresponding velocities ${\vec v}_{\rm E}=\dot{\vec z}_{\rm E}$ and ${\vec z}_{\rm M}=\dot{\vec z}_{\rm M}$ can be computed in the first post-Newtonian approximation using the Einstein-Infeld-Hoffmann (EIH) equations in the coordinates of the BCRS \cite{EIH1938,Moyer:2003,MG2005,Turyshev:2008,Turyshev:2012}:
{}
\begin{eqnarray}
\ddot{\bf z}_a&=&\sum_{b\not=a}\frac{GM_b{\bf r}_{ab}}{r_{ab}^3}\Big\{1+\frac{1}{c^2}\Big(-4\sum_{c\not=a}\frac{GM_c}{r_{ac}}-
\sum_{c\not=b}\frac{GM_c}{r_{bc}}+
{\dot r}_a^2+2{\dot r}_b^2-4(\dot{\bf r}_a \cdot\dot{\bf r}_b)-{\textstyle\frac{3}{2}}({{\bf n}_{ab}\cdot{\dot{\bf r}}_b})^2+{\textstyle\frac{1}{2}}({\bf r}_{ab}\cdot{\ddot{\bf r}}_b)\Big)\Big\}+\nonumber\\
&&\hskip 67pt +\,
\frac{1}{c^2}\Big(\sum_{b\not=a}\frac{GM_b}{r_{ab}^3}
\big({\bf r}_{ab}\cdot(4{\dot {\bf r}}_a-3{\dot {\bf r}}_b)\big){\dot{\bf r}}_{ab}+{\textstyle\frac{7}{2}}\sum_{b\not=a}\frac{GM_b{\ddot {\bf r}}_b}{r_{ab}}\Big)+{\cal O}(c^{-4}),
\label{eq:EIH}
\end{eqnarray}
where $r_{ab}=|\vec{r}_{ab}|$ and $\vec{n}_{ab}=\vec{r}_{ab}/r_{ab}$.
When describing the motion of spacecraft in the Solar System, the models also include forces of attraction between the zonal harmonics of the bodies of interest and forces from asteroids and planetary satellites (see details in Ref.~\cite{Standish-Williams:2012}).

To determine the orbits of planets and the spacecraft, one must also describe the propagation of electromagnetic signals between any two points in space.  The light-time equation corresponding to the metric tensor (\ref{eq:gab-BCRS}) and written to the accuracy sufficient for GRAIL has the form (see also Ref.~\cite{Teyssandier2008,Ashby2010}):
\begin{equation}
t_2-t_1= \frac{|{\vec r}_2-{\vec r}_1|}{c}+(1+\gamma)\sum_b \frac{G M_b}{c^3}
\ln\left[\frac{r_1^b+r_2^b+r_{12}^b}{r_1^b+r_2^b-r_{12}^b}\right]+{\cal O}(c^{-5}),
\label{eq:light-time}
\end{equation}
\noindent  where $t_1$ refers to the signal transmission time and $t_2$ refers to the reception time, while  ${\vec r}_{1,2}$ are the barycentric positions of the transmitter and receiver. Also, $r^b_{1,2}$ are the distances of the transmitter and receiver from the body $b$ and $r^b_{12}$ is their spatial separation \cite{Moyer:2003,Turyshev:2008}. The logarithmic contribution in (\ref{eq:light-time}) is the Shapiro gravitational time delay that, in the case of GRAIL, is mostly due to the Moon, the Earth, and the Sun. (Note that the ${\cal O}(c^{-5})$ terms are beyond GRAIL's sensitivity; see the analysis in Sec.~\ref{sec:dowr}.)

The general relativistic equations of motion (\ref{eq:EIH}) and light-time equation (\ref{eq:light-time}) are used to produce numerical codes for the purposes of constructing Solar System ephemerides, spacecraft navigation \cite{Moyer:2003,Standish-Williams:2012} and analysis of gravitational experiments in the Solar System \cite{Turyshev:2008,Will}. GRAIL also relies on these equations to compute its range and range-rate observables between the two spacecraft in lunar orbit. The numerical algorithm developed for this purpose \cite{Park-etal:2012} iteratively solves the light-time equation (\ref{eq:light-time}) in the SSB frame in terms of the instantaneous distance between the two spacecraft.  Our objective is to develop an explicit analytical model for all the quantities involved in these high-precision computations.  For this purpose, we need a clearly defined set of  astronomical reference frames, which we discuss next.

\subsection{Relativistic coordinate transformations between various reference frames}

To describe the dynamics of an $N$-body system (such as the Solar System) in general relativity, one may choose to introduce $N+1$ reference frames, each with its own coordinate chart. We need one {\it global} coordinate chart defined for the inertial reference frame that covers the entire system under consideration (e.g., BCRS).  In the immediate vicinity of each of the $N$ bodies in the system we can also introduce a set of {\it local} coordinates defined in the frame associated with this body (body-centric system). In the remainder of this paper, we use $\{x^m\}$ to represent the coordinates of the global inertial frame and $\{y^m_a\}$ to be the local coordinates of the accelerated proper reference frame of body $a$.

In Ref.~\cite{Turyshev:2012}, we showed that the transformations between the harmonic coordinates of the BCRS $\{x^m\}$ and non-rotating body-centric reference systems $\{y^m_a\}$ (such as the GCRS or LCRS) may be written in the following form:
{}
\begin{eqnarray}
x^0&=& y^0_a+
c^{-2}\Big\{c({\vec v}_{a}\cdot {\vec y}_a)+\int_{y^0_{a_0}}^{y^0_a}\!\!
\Big[{\textstyle\frac{1}{2}}{\vec v}^2_{a}
+U^a_{\rm ext}
\Big]dy'^0_a\Big\}+{\cal O}(c^{-4}),
\label{eq:trans++x0}\\[3pt]
{\vec x}&=& {\vec y}^{}_a+{\vec z}^{}_{a}+
c^{-2}\Big\{{\textstyle\frac{1}{2}}{\vec v}^{}_{a} ({\vec v}^{}_{a}\cdot{\vec y}^{}_a)-{\vec y}^{}_a
U_{\rm ext}^a
+[{\boldsymbol\omega}^{}_{a}\times{\vec y}_a] + {\textstyle\frac{1}{2}}{\vec a}^{}_{a}{y}_a^2-{\vec y}^{}_a({\vec y}^{}_a\cdot {\vec a}^{}_{a})
\Big\}+{\cal O}(c^{-4}),
\label{eq:trans++xa}
\end{eqnarray}
where ${\vec z}_{a}$ is the vector that connects the origin of the $\{x^m\}$ reference system with the origin of the $\{y^m_a\}$ reference system. Note that the accuracy of timing for GRAIL is limited by the performance of the on-board USO, which have an error of ${\cal O}(10^{-13})$ for $10^3$~s of integration time \cite{TimingMemo:2010}. Therefore, the $c^{-4}$ terms in Eq.~(\ref{eq:trans++x0}), which are at most of order $\sim v^4/c^4\simeq 10^{-16}$, are negligible for GRAIL, even in the absolute sense. The differential nature of the observables on GRAIL further reduces the sensitivity of the mission to such small terms in the transformations. For a complete post-Newtonian form of these transformations, including the terms $c^{-4}$ and their explicit derivation, consult Ref.~\cite{Turyshev:2012}.

The inverses of the transformations (\ref{eq:trans++x0})--(\ref{eq:trans++xa}) can be written as
{}
\begin{eqnarray}
y^0_a&=& x^0-c^{-2}\Big\{c\,({\vec v}^{}_{a}\cdot {\vec r}^{}_a)+\!\int_{x^0_{0}}^{x^0}\!\!\! \Big[
{\textstyle\frac{1}{2}}{\vec v}^2_{a}+
U_{\rm ext}^a
\Big]dx'^0\Big\}+{\cal O}({c^{-4}}),
\label{eq:trans++y0}\\[3pt]
{\vec y}_a&=& {\vec r}_a+
c^{-2}\Big\{{\textstyle\frac{1}{2}}{\vec v}_{a} ({\vec v}_{a}\cdot{\vec r}_a)+{\vec r}_a
U_{\rm ext}^a
+[{\boldsymbol\omega}^{}_{a}\times {\vec r}^{}_{a}] + {\vec r}^{}_a ({\vec r}{}_a\cdot{\vec a}_{a})-{\textstyle\frac{1}{2}}{\vec a}^{}_{a}{r}^2_a
\Big\}+{\cal O}(c^{-4}),
\label{eq:trans++ya}
\end{eqnarray}
where ${\vec r}_{a}={\vec x}-{\vec z}_{a}$. The quantity $U^a_{\rm ext}$ in Eqs.~(\ref{eq:trans++x0})--(\ref{eq:trans++xa}) and (\ref{eq:trans++y0})--(\ref{eq:trans++ya}) is the Newtonian gravitational potential (including, if necessary, multipole corrections) due to all bodies in the Solar System other than body $a$, at the location of body $a$. Furthermore, ${\vec a}_{a}$ is the Newtonian acceleration of body $a$ due to the combined gravity of all other bodies. Later in this section, we will present the expressions for ${U}^a_{\rm ext}$ and ${\vec a}_{a}$ for each of the chosen reference frames.

Finally, ${\boldsymbol\omega}_{a}$ in Eqs.~(\ref{eq:trans++xa}) and (\ref{eq:trans++ya}) is the vector associated with the relativistic precession given as ${\omega}^\alpha_{a}={\textstyle\frac{1}{2}}\epsilon^\alpha_{\mu\nu}{\omega}_{a}^{\mu\nu}$, with $\epsilon^\alpha_{\mu\nu}$ being the fully antisymmetric Levi-Civita symbol, normalized as $\epsilon^1_{23}=1$, and the matrix ${\omega}_{a}^{\alpha\beta}$ having the form \cite{Turyshev:2012}:
{}
\begin{equation}
\dot{\omega}_{a}^{\alpha\beta}=-\sum_{b\not=a}\frac{GM_b}{r^{2}_{ba}}\Big\{n^\alpha_{ba}({\textstyle\frac{3}{2}}v^\beta_{a}-2v^\beta_{b})-n^\beta_{ba}({\textstyle\frac{3}{2}}v^\alpha_{a}-2v^\alpha_{b})\Big\}+{\cal O}(c^{-2}). \label{eq:omega-pot}
\end{equation}
The expression for the relativistic precession matrix is given here only for the sake of completeness. Because of their small magnitude ($\sim 10^{-15}$~m), these terms will not affect the GRAIL measurements (see discussion in Sec.~\ref{sec:pos-vectors}).

In the rest of this section, we discuss four fundamental body-centric reference frames that are useful for collection and interpretation of GRAIL data.

\subsection{Coordinate systems used in the vicinity of the Earth}

In the vicinity of the Earth, two standard coordinate systems are utilized: the Geocentric Coordinate Reference System (GCRS), centered at the Earth's center of mass is used to track orbits in the vicinity of the Earth. The positions of objects on the surface of the Earth, such as DSN ground stations, are usually given in the Topocentric Coordinate Reference System (TCRS).

\subsubsection{Geocentric Coordinate Reference System (GCRS)}
\label{sec:GCRS}

When constructing a body-centric coordinate reference frame for a body $a$ at the level of accuracy anticipated for GRAIL, it is sufficient to consider only monopole contributions to the external potential ${U}^a_{\rm ext}$ of all the bodies in the Solar System (for the Earth it is mostly the Sun and the Moon) excluding body $a$ itself \cite{Turyshev:2012}. Thus, for the GCRS, the Newtonian potential of the external bodies (i.e., excluding the Earth) ${U}^a_{\rm ext}$ and the corresponding acceleration ${\vec a}_{a}$ that are present in the coordinate transformations Eqs.~(\ref{eq:trans++x0})--(\ref{eq:trans++xa}), have the form \cite{Turyshev:2012}:
{}
\begin{equation}
{U}_{\rm ext}^{\rm E}=\sum_{b\not={\rm E}} \frac{GM_b}{r^{}_{b{\rm E}}}+{\cal O}(c^{-2}), \qquad
{\vec a}_{\rm E}=-\nabla U_{\rm ext}^{\rm E}=-\sum_{b\not={\rm E}} GM_b\frac{{\vec r}_{b{\rm E}}}{r^3_{b{\rm E}}}+{\cal O}(c^{-2}),
\label{eq:U-est-E}
\end{equation}
where summation is performed over all the bodies excluding the Earth (symbolically, $b\not= {\rm E}$), the vector that connects body $b$ with the Earth's center of mass is represented by ${\vec r}_{b{\rm E}}={\vec x}_{\rm E}-{\vec x}_{b}$ and the contributions of the higher multipole moments of mass distribution within the bodies are neglected due to their smallness.

The transformations given by Eqs.~(\ref{eq:trans++x0})--(\ref{eq:trans++xa}), together with the potential ${U}^a_{\rm ext}$ and the acceleration  ${\vec a}_{\rm E}$ given by Eq.~(\ref{eq:U-est-E}), determine the metric tensor $g^{\rm E}_{mn}$ of the non-rotating GCRS \cite{Turyshev:2012}. We denote the coordinates of this reference frame as $\{y^m_{\rm E}\}\equiv(y^0_{\rm E}, {\vec y}_{\rm E})$ and present the metric tensor $g^{\rm E}_{mn}$  in the following form:
\begin{eqnarray}
g^{\rm E}_{00}&=& 1-\frac{2}{c^2}w_{\rm [E]}+\frac{2}{c^4}w^2_{\rm [E]}+O(c^{-6}), \quad
g^{\rm E}_{0\alpha}= -\gamma_{\alpha\lambda}\frac{4}{c^3}w^\lambda_{\rm E}+O(c^{-5}),\quad
g^{\rm E}_{\alpha\beta}= \gamma_{\alpha\beta}+\gamma_{\alpha\beta}\frac{2}{c^2} w_{\rm E}+O(c^{-4}),~~~
\label{eq:gab-E}
\end{eqnarray}
where $w_{\rm E}$ and  $w^\lambda_{\rm E}$ are the scalar and vector harmonic potentials that are given by
{}
\begin{eqnarray}
w_{\rm E}&=&U_{\rm E}+ u^{\tt tidal}_{\rm E}+{\cal O}(c^{-4}),~~~
\label{eq:pot_loc-w_0}\\[3pt]
{\vec w}_{\rm E}&=&-\frac{G}{2y^3_{\rm E}}[{\vec y}_{\rm E}\times{\vec S}_{\rm E}]+{\cal O}(c^{-2}),
\label{eq:pot_loc-w_a+}
\end{eqnarray}
where ${\vec S}_{\rm E}$ in Eq.~(\ref{eq:pot_loc-w_a+}) is the Earth's angular momentum. The scalar potential $w_{\rm E}$ is formed as a linear superposition of the gravitational potential $U_{\rm E}$ of the isolated Earth and the tidal potential $u^{\tt tidal}_{\rm E}$ produced by all the Solar System bodies (excluding the Earth itself, $b\not={\rm E}$) evaluated at the origin of the GCRF. The Earth's gravitational potential $U_{\rm E}$ at a location defined by spherical coordinates $(y_{\rm E},\phi,\theta)$ is given by
{}
\begin{eqnarray}
U_{\rm E}&=&\frac{GM_{\rm E}}{y_{\rm E}}\Big\{1+\sum_{\ell=2}^\infty\sum_{k=0}^{+\ell}\Big(\frac{R_{0\rm E}}{y_{\rm E}}\Big)^\ell P_{\ell k}(\cos\theta)(C^{{\rm E}}_{\ell k}\cos k\phi+S^{{\rm E}}_{\ell k}\sin k\phi)\Big\}+
{\cal O}(c^{-4}),
\label{eq:pot_w_0}
\end{eqnarray}
where $R_{0\rm E}$ being the Earth's radius, $P_{\ell k}$ are the Legendre polynomials, while $C^{{\rm E}}_{\ell k}$ and $S^{{\rm E}}_{\ell k}$ are spherical harmonic coefficients that characterize the Earth. At the level of sensitivity of GRAIL, only the lowest order spherical harmonic coefficients need to be accounted for, and time-dependent contributions due to the elasticity of the Earth can be ignored.  Insofar as the tidal potential $u^{\tt tidal}_{\rm E}$ is concerned, for GRAIL it is sufficient to keep only its Newtonian contribution (primarily due to the Moon and the Sun) which can be given as usual:
{}
\begin{eqnarray}
u^{\tt tidal}_{\rm E}&=&\sum_{b\not={\rm E}}\Big(U_b({\vec r}^{}_{b{\rm E}}+\vec{y}^{}_{\rm E})-U_b({\vec r}^{}_{b{\rm E}}) - \vec{y}_{\rm E}\cdot {\vec \nabla} U_b ({\vec r}^{}_{b{\rm E}})\Big)\simeq\sum_{b\not={\rm E}}\frac{GM_b}{2r^3_{b{\rm E}}}\Big(3(\vec{n}^{}_{b{\rm E}}\cdot\vec{y}_{\rm E})^2-\vec{y}_{\rm E}^2\Big)+
{\cal O}(y_{\rm E}^3,c^{-2}),
\label{eq:u-tidal-E}
\end{eqnarray}
where $U_b$ is the Newtonian gravitational potential of body $b$, $\vec{r}_{b{\rm E}}$ is the vector connecting the center of mass of body $b$ with that of the Earth, and $\nabla$ denotes the divergence with respect to $\vec{y}_{\rm E}$.
Note that in Eq.~(\ref{eq:u-tidal-E}) we omitted relativistic tidal contributions of ${\cal O}(c^{-2})$ that are produced by the external gravitational potentials. These are of the order of $10^{-16}$ compared to $U_{\rm E}$ and, thus, completely negligible for GRAIL. In addition, we present only the largest term in the tidal potential of the order of $\sim y_{\rm E}^2$; however, using the explicit form of the tidal potential Eq.~(\ref{eq:u-tidal-E}), one can easily evaluate this expression to any order needed for a particular problem.

The proper time at the origin of the GCRS is called the Geocentric Coordinate Time (TCG), denoted here as $t_{\rm TCG}$. It relates to the barycentric time TCB $t$ as
{}
\begin{eqnarray}
\frac{dt_{\rm TCG}}{dt}&=& 1-\frac{1}{c^2}\Big[\frac{{\vec v}_{\rm E}^2}{2}+\sum_{b\not={\rm E}}\frac{GM_b}{r_{b{\rm E}}}
\Big]+{\cal O}({c^{-4}})\approx 1-1.48\times 10^{-8}.
\label{eq:proper-t-C}
\end{eqnarray}
The Earth's barycentric velocity $\vec{v}_{\rm E}$ and position ${\vec z}_{\rm E}$ can be computed from Eq.~(\ref{eq:EIH}).

\subsubsection{Topocentric Coordinate Reference System (TCRS): proper and coordinate times}
\label{sec:TCRS}

To obtain the metric of the topocentric coordinate reference system, the TCRS, one can transform the metric $g_{mn}^{\rm E}$ of the GCRS using coordinate transformations given by Eqs.~(\ref{eq:trans++x0})--(\ref{eq:trans++xa}), where the ``external'' potential ${U}_{\rm ext}^{\rm C}$ is the gravitational potential $w^{}_{\rm E}$ given by Eq.~(\ref{eq:pot_loc-w_0}) and evaluated at the surface of the Earth:
{}
\begin{eqnarray}
{U}_{\rm ext}^{\rm C}&=&w_{\rm E}({\vec y}_{\rm C})=U_{\rm E}({\vec y}_{\rm C})+
\sum_{b\not={\rm E}}\Big(U_b({\vec r}^{}_{b{\rm E}}+\vec{y}^{}_{\rm C})-U_b({\vec r}^{}_{b{\rm E}}) - \vec{y}_{\rm C}\cdot {\vec \nabla} U_b ({\vec r}^{}_{b{\rm E}})\Big) +{\cal O}(c^{-2}),~~~
\label{eq:TCRS-w_0}\\
\vec{a}_{\rm C}&=&-\nabla U^{\rm C}_{\rm ext}=-\nabla U_{\rm E}({\vec y}_{\rm C})-\sum_{b\not={\rm E}}\Big({\vec \nabla} U_b({\vec r}^{}_{b{\rm E}}+\vec{y}^{}_{\rm C})- {\vec \nabla} U_b ({\vec r}^{}_{b{\rm E}})\Big) +{\cal O}(c^{-2}),
\label{eq:TCRS-accel}
\end{eqnarray}
where ${\vec y}_{\rm C}$ is the position vector of the DSN station in the GCRS. Note that $U_{\rm E}({\vec y}_{\rm C})$ must be treated as the potential of an extended body and include a multipolar expansion with sufficient accuracy, taking into account time-dependent terms due to tidal effects on the elastic Earth.

The proper time $\tau_{\rm C}$, kept by a clock located at the GCRS coordinate position ${\vec y}_{\rm C}(t)$, and moving with the coordinate velocity ${\vec v}_{\rm C0} = d{\vec y}_{\rm C}/dt_{\rm TCG}=[{\boldsymbol\Omega}^{}_{\rm E}\times{\vec y}_{\rm C}]$, where ${\boldsymbol\Omega}_{\rm E}$ is the angular rotational velocity of the Earth at $C$, is determined by
{}
\begin{equation}
\frac{d\tau_{\rm C}}{dt_{\rm TCG}}= 1-\frac{1}{c^2}\Big[{\textstyle\frac{1}{2}}\vec{v}_{\rm C0}^2+U_{\rm E}({\vec y}_{\rm C})+
\sum_{b\not={\rm E}}\frac{GM_b}{2r^3_{b{\rm E}}}\Big(3(\vec{n}_{b{\rm E}}\cdot\vec{y}_{\rm C})^2-\vec{y}_{\rm C}^2\Big)+({\vec a}^{}_{\rm E}\cdot {\vec y}^{}_{\rm C})\Big]+{\cal O}({y^3_{\rm C},c^{-4}}),
\label{eq:proper-t-C-comp}
\end{equation}
where ${\vec a}_{\rm E}$ is the Earth's acceleration in the BCRS, Eq.~(\ref{eq:U-est-E}), and $\vec{n}_{b{\rm E}}$ is a unit spatial vector in the body-Earth direction, i.e., $\vec{n}_{b{\rm E}}=\vec{r}_{b{\rm E}}/|\vec{r}_{b{\rm E}}|$, where $\vec{r}_{b{\rm E}}$ is the vector connecting body $b$ with the Earth. The term within the square brackets in Eq.~(\ref{eq:proper-t-C-comp}) is the sum of Newtonian tides due to the Sun, the Moon, and other bodies at the clock location ${\vec y}_{\rm C}$. These terms are small for Earth stations (of order $2 \times 10^{-17}$) and are negligible for GRAIL. The last term is due to non-inertiality of the GCRS and accounts for the Earth's finite size. This term is evaluated to be of the order  of $4.2\times 10^{-13}$, which is about $10^{-3}$ smaller compared to the gravity potential on the surface of the Earth and, thus, it can be omitted.

Therefore, at the accuracy required for GRAIL, it is sufficient to keep only the first two terms in Eq.~(\ref{eq:proper-t-C-comp}) when defining the relationship between the proper time $\tau_{\rm C}$ and the coordinate time $t_{\rm TCG}$:
\begin{equation}
\frac{d\tau_{\rm C}}{dt_{\rm TCG}} = 1 -\frac{1}{c^2}\Big[{\textstyle\frac{1}{2}}{\vec v}^2_{\rm C0}
+ U_{\rm E}({\vec y}_{\rm C})\Big]+{\cal O}(c^{-4}).
\label{eq:proper-coord-t+}
\end{equation}
At the level of accuracy required for GRAIL, it is important to account in Eq.~(\ref{eq:proper-coord-t+}) for the oblateness (non-sphericity) of the Earth's Newtonian potential, which is given in the form of Eq.~(\ref{eq:pot_w_0}). In fact, when we model the Earth's gravity potential, we need to take into account quadrupole and higher moments, time-dependent terms due to tides as well as the tidal displacement of the DSN station. For example, for a clock situated on the surface of the Earth, the relativistic correction term appearing in Eq.~(\ref{eq:proper-coord-t+}) is given at the needed precision by
{}
\begin{equation}
\frac{{\vec v}^2_{\rm C0}}{2} + U_{\rm E}({\vec y}_{\rm C})= W_0-\int_0^{h_{\rm C}}gdh,
\label{eq:proper-coord-t++}
\end{equation}
where $W_0 = 6.2636856\times 10^7$ m$^2$/s$^2$ is the Earth's potential at the reference geoid while $g$ denotes the Earth's acceleration (gravitational plus centrifugal), and where $h_{\rm C}$ is the clock's altitude above the reference geoid.

Finally, we present the relation of the proper time read by the clock on the surface of the Earth at point $C$ with  respect to the TCB. Expressing ${d\tau_{\rm C}}/{dt}=({d\tau_{\rm C}}/{dt_{\rm TCG}})({dt_{\rm TCG}}/{dt})$, with the help of Eq.~(\ref{eq:trans++x0}) together with Eqs.~(\ref{eq:proper-t-C}) and (\ref{eq:proper-coord-t+}), at the level of accuracy sufficient for GRAIL, we have:
{}
\begin{equation}
\frac{d\tau_{\rm C}}{dt}=
1 -\frac{1}{c^2}\Big[{\textstyle\frac{1}{2}}({\vec v}_{\rm E}+[{\boldsymbol\Omega}^{}_{\rm E}\times{\vec y}_{\rm C}])^2+ U_{\rm E}({\vec y}_{\rm C})+
\sum_{b\not={\rm E}}U_b({\vec r}^{}_{b{\rm E}}+\vec{y}^{}_{\rm C})
\Big]+{\cal O}(10^{-17}),
\label{eq:proper-t-C+}
\end{equation}
where the first term in the brackets, ${\vec v}_{\rm E}+[{\boldsymbol\Omega}^{}_{\rm E}\times{\vec y}_{\rm C}]\equiv{\vec v}_{\rm E}+{\vec v}_{\rm C0}={\vec v}_{\rm C}$, is the barycentric velocity of the DSN station.
For details on the recommended relativistic formulation of GCRS consult Refs.~\cite{Moyer:2003,Klioner:2008,Kopeikin-book}. Coordinate transformations (in particular, transformations involving topocentric coordinates) are discussed extensively in the IERS Conventions\footnote{
All software, technical specification and other relevant materials associated with the IERS Conventions (2010) can be found at {\tt http://www.iers.org/}}.

\subsection{Relativistic timekeeping in the Solar System}
\label{sec:timekeeping}

Spacecraft radio science observations are clock and frequency measurements made at Earth stations \cite{Moyer:2003}. For this purpose, the time coordinate called Terrestrial Time (TT) is defined. TT is related to TCG linearly by definition:
\begin{equation}
\frac{dt_{\rm TT}}{dt_{\rm TCG}}=1-L_{\rm G},
\end{equation}
where $L_{\rm G}=6.969290134\times 10^{-10}$ by definition. This definition accounts for the secular term due to the Earth's potential when converting between TCG and the time measured by an idealized clock on the Earth geoid \cite{Moyer:2003,Klioner:2008,Klioner-etal:2010,Kopeikin-book}. Using Eq.~(\ref{eq:proper-coord-t+}), we also have
\begin{equation}
\frac{d\tau_{\rm C}}{dt_{\rm TT}}=\frac{d\tau_{\rm C}}{dt_{\rm TCG}}\frac{dt_{\rm TCG}}{dt_{\rm TT}}=1+L_{\rm G}-\frac{1}{c^2}\Big[{\textstyle\frac{1}{2}}{\vec v}^2_{\rm C0}
+ U_{\rm E}({\vec y}_{\rm C})\Big]+{\cal O}(c^{-4}).
\end{equation}

On the other hand, equations of motion in the Solar System are often evaluated using another defined time scale, TDB. TDB time ($t_{\rm TDB}$) is also related to TCB time $t$ linearly:
\begin{equation}
\frac{dt_{\rm TDB}}{dt}=1-L_{\rm B},
\end{equation}
where $L_{\rm B}=1.550519768\times 10^{-8}$ by definition, accounting for all secular terms due to the solar gravitational field, the Earth's orbital velocity, and the Earth potential on the geoid.

The relationship between TCB and TCG is nonlinear; these are the coordinate times of two coordinate systems related to one another by the space-time transformations (\ref{eq:trans++x0})--(\ref{eq:trans++xa}) and their inverses (\ref{eq:trans++y0})--(\ref{eq:trans++ya}).

The relationship between TT and TDB, therefore, is also nonlinear. The difference is dominated by an annual periodic term with an amplitude of $\sim 1.6 \times 10^{-3}$~s. The definition of TT and TDB ensures the absence of a significant linear term.

For accurate computations in the SSB reference frame, observed times of transmission and reception need to be converted from TT to TDB.

\subsection{Coordinate reference frames in the vicinity of the Moon}

In the vicinity of the Moon, once again we consider two coordinate systems. The lunicentric LCRS is a coordinate system used, for instance, to represent lunar orbits. To describe experiments carried out on board the GRAIL spacecraft, we use the SCRS.

\subsubsection{Lunar Coordinate Reference System (LCRS)}
\label{sec:LCRS}

In complete analogy to the formulation of the GCRS (discussed in Sec.~\ref{sec:GCRS}) and similarly to the approach advocated in Ref.~\cite{Kopeikin2010}, the metric tensor of the LCRS may be obtained by transforming the metric (\ref{eq:gab-BCRS}) and the potentials (\ref{eq:w_0-BCRS})--(\ref{eq:w_a-BCRS}) of the BCRS using the coordinate transformations given by Eqs.~(\ref{eq:trans++x0})--(\ref{eq:trans++xa}) where it is sufficient to consider only the monopole contribution to ${U}^a_{\rm ext}$ from all the bodies of the Solar System excluding the Moon:
{}
\begin{equation}
{U}_{\rm ext}^{\rm M}=\sum_{b\not={\rm M}} \frac{GM_b}{r^{}_{b{\rm M}}}+{\cal O}(c^{-2}), \qquad
{\vec a}_{\rm M}=-\nabla U^{\rm M}_{\rm ext}=-\sum_{b\not=a} GM_b\frac{{\vec r}_{b{\rm M}}}{r^3_{b{\rm M}}}+{\cal O}(c^{-2}),
\label{eq:U-est-M}
\end{equation}
where summation is performed over all the bodies excluding the Moon ($b\not={\rm M}$) and  the contributions due to the higher multipole moments of the mass distributions within the bodies are neglected.

Applying the coordinate transformations given by Eqs.~(\ref{eq:trans++x0})--(\ref{eq:trans++xa}) together with the external potential and acceleration (\ref{eq:U-est-M}), one can derive the  metric tensor $g^{\rm M}_{mn}$ of the  non-rotating lunar coordinate reference system (LCRS). Denoting the coordinates of the LCRS as $\{y^m_{\rm M}\}\equiv(y^0_{\rm M}, {\vec y}_{\rm M})$, this tensor may be presented in the following form (which is identical to the expressions in Sec.~\ref{sec:GCRS} after making the substitution ${\rm E}\rightarrow {\rm M}$):
\begin{eqnarray}
g^{\rm M}_{00}&=& 1-\frac{2}{c^2}w_{\rm M}+\frac{2}{c^4}{w}^2_{\rm M}+O(c^{-6}), \quad
g^{\rm M}_{0\alpha}= -\gamma_{\alpha\lambda}\frac{4}{c^3}w^\lambda_{\rm M}+O(c^{-5}),\quad
g^{\rm M}_{\alpha\beta}= \gamma_{\alpha\beta}+\gamma_{\alpha\beta}\frac{2}{c^2} w^{}_{\rm M}+O(c^{-4}),~~~~
\label{eq:gab-M}
\end{eqnarray}
where the scalar and vector potentials $w_{\rm M}$ and $w^\lambda_{\rm M}$ are given as:
{}
\begin{eqnarray}
w_{\rm M}&=&U_{\rm M}+ u^{\tt tidal}_{\rm M}+
{\cal O}(c^{-4}),~~~
\label{eq:pot_M}\\[3pt]
{\vec w}_{\rm M}&=&-\frac{G}{2y^3_{\rm M}}[{\vec y}_{\rm M}\times{\vec S}_{{\rm M}}]+{\cal O}(c^{-2}),
\label{eq:pot_loc-w_a}
\end{eqnarray}
where ${\vec S}_{\rm M}$ in Eq.~(\ref{eq:pot_loc-w_a}) is the Moon's angular momentum. Similarly to the GCRS, the scalar potential $w_{\rm M}$ (\ref{eq:pot_M}) is a linear superposition of the proper gravitational potential of the Moon (with $R_{0\rm M}$ being the Moon's radius, $M_{\rm M}$ its mass, while $C^{{\rm M}}_{\ell k}$ and $S^{{\rm M}}_{\ell k}$ are the Moon's spherical harmonic coefficients):
{}
\begin{eqnarray}
U_{\rm M}&=&\frac{GM_{\rm M}}{y_{\rm M}}\Big\{1+\sum_{\ell=2}^\infty\sum_{k=0}^{+\ell}\Big(\frac{R_{0\rm M}}{y_{\rm M}}\Big)^\ell P_{\ell k}(\cos\theta)(C^{{\rm M}}_{\ell k}\cos k\phi+S^{{\rm M}}_{\ell k}\sin k\phi)\Big\}+
{\cal O}(c^{-4})
\label{eq:pot_w_0+}
\end{eqnarray}
plus tidal contributions $u^{\tt tidal}_{\rm M}$ produced by all the Solar System bodies (excluding the Moon itself, $b\not={\rm M}$) evaluated at the origin of the LCRF. For the GRAIL's accuracy, it is sufficient to keep only the Newtonian contribution to the tidal potential produced by the external bodies which can be presented as
{}
\begin{eqnarray}
u^{\tt tidal}_{\rm M}&=&\sum_{b\not={\rm M}}\Big(U_b({\vec r}^{}_{b{\rm M}}+\vec{y}^{}_{\rm M})-U_b({\vec r}^{}_{b{\rm M}}) - {\vec{y}}^{}_{\rm M}\cdot{\vec \nabla} U_b ({\vec r}^{}_{b{\rm M}})\Big)\simeq\sum_{b\not={\rm M}}\frac{GM_b}{2r^3_{b{\rm M}}}\Big(3(\vec{n}^{}_{b{\rm M}}\cdot\vec{y}^{}_{\rm M})^2-\vec{y}_{\rm M}^2\Big)+
{\cal O}(y_{\rm M}^3,c^{-2}),~~~
\label{eq:u-tidal-M}
\end{eqnarray}
where $\vec{n}_{b{\rm M}}$ is a unit spatial vector in the body-Moon direction, i.e., $\vec{n}_{b{\rm M}}=\vec{r}_{b{\rm M}}/|\vec{r}_{b{\rm M}}|$, where $\vec{r}_{b{\rm M}}$ is the vector connecting body $b$ with the Moon. Note that the relativistic tidal contributions of $1/c^2$ order that are due to external potentials have a magnitude of $10^{-16}$ when compared to $U_{\rm M}$ and, thus, they were omitted in Eq.~(\ref{eq:u-tidal-M}). In addition, we present only the largest term in the tidal potential of the order of $\sim y_{\rm M}^2$; however, using the explicit form of the tidal potential Eq.~(\ref{eq:u-tidal-M}), one easily evaluate this expression to any order needed for a particular problem.

At the same time, we must account for deformations of the elastic Moon, expressed in the form of corrections $\Delta C^{{\rm M}}_{\ell k}$ and $\Delta S^{{\rm M}}_{\ell k}$ to the lunar spherical coefficients, due to the tidal potential of body $b$, located at lunicentric spherical coordinates $(r_{b{\rm M}},\phi_{b{\rm M}},\theta_{b{\rm M}})$ \cite{Williams-etal:2001,MG2005}:
\begin{equation}
\left\{\begin{matrix}
\Delta C_{\ell k}\\
\Delta S_{\ell k}\end{matrix}\right\}=4k_\ell^{\rm M}\frac{M_b}{M_{\rm M}}\left(\frac{R_{\rm 0M}}{r_{b{\rm M}}}\right)^{\ell+1}\sqrt{\frac{(\ell+2)[(\ell-k)!]^3}{[(\ell+k)!]^3}}P_{\ell k}(\cos\theta_{b{\rm M}})\left\{\begin{matrix}\cos k\phi_{b{\rm M}}\\
\sin k\phi_{b{\rm M}}\end{matrix}\right\}
\label{eq:LOVE}.
\end{equation}
The lunar Love number $k_2^{\rm M}\simeq 0.025$ \cite{Goossens-Matsumoto:2008}
 thus introduces a significant time-dependent contribution to the spherical harmonic coefficients $C^{{\rm M}}_{\ell k}$ and $S^{{\rm M}}_{\ell k}$, which must be written as the sums
\begin{align}
C^{{\rm M}}_{\ell k}&=C^{{\rm M0}}_{\ell k} + \Delta C^{{\rm M}}_{\ell k}\qquad {\rm and} \qquad
S^{{\rm M}}_{\ell k}=S^{{\rm M0}}_{\ell k} + \Delta S^{{\rm M}}_{\ell k},\label{eq:DS}
\end{align}
where we used $C^{{\rm M0}}_{\ell k}$ and $S^{{\rm M0}}_{\ell k}$ to denote the constant part of the lunar spherical harmonic coefficients.

\subsubsection{Lunar Coordinate Time (TCL)}

There are several different time coordinates to be considered for GRAIL. In addition to the terrestrial time scales defined in Sec.~\ref{sec:timekeeping}, GRAIL also relies on the timing events reported at proper times measured by clocks on board the lunar orbiters. Thus, one would need to introduce a realization of lunar coordinate time (TCL) and a spacecraft proper time (ST).

The lunicentric orbits of the GRAIL spacecraft are coupled to the orbit of the Moon mostly through the difference between the acceleration of the probe and that of the Moon due to the gravitational pull of the Earth and the Sun (the Earth's and the Sun's tidal terms). This coupling is weak because the Earth and Sun tides are, respectively, just $2.5\times10^{-7}$ and $4.7\times10^{-8}$ times the monopole acceleration due to the Moon. Relativistic perturbations containing the mass of the Moon are small ($\sim4.6\times 10^{-11}$~m/s$^2$) to the point that they are not measurable, being easily absorbed into the much larger non-gravitational perturbations (for instance, solar radiation pressure). Should we conclude that general relativity does not matter in the computation of the the lunicentric orbit of the spacecraft? The answer is negative, but the main relativistic effect does not appear in the equation of motion.

According to Eq.~(\ref{eq:trans++x0}), the differential equation that gives the local  proper time $t_{\rm TCL}$ at the origin of the LCRS as it relates to the barycentric time TCB $t$ is
{}
\begin{eqnarray}
\frac{dt_{\rm TCL}}{dt}&=& 1-\frac{1}{c^2}\Big[\frac{{\vec v}_{\rm M}^2}{2}+\sum_{b\not={\rm M}}U^{}_{\rm b}({\vec r}^{}_{b{\rm M}})\Big]+{\cal O}({c^{-4}}),
\label{eq:proper-TCL}
\end{eqnarray}
where the Moon's barycentric velocity ${\vec v}_{\rm M}$ and position ${\vec z}_{\rm M}$ can be computed from the EIH equations (\ref{eq:EIH}).
Equation~(\ref{eq:proper-TCL}) establishes the relationship between the TCL ($t_{\rm TCL}$) and TCB $(t)$ time scales. Truncated to the first post-Newtonian (1PN) order (we put the clock at the origin of its proper reference system, ${\vec y}_{\rm M}=0$ and drop on the right hand side ${\cal O}(c^{-4})$ terms that are in principle known, but certainly not needed for our purposes), it is given by a differential equation
\begin{equation}
\frac{dt_{\rm TCL}}{dt}= 1 - \frac{1}{c^2}\Big[\frac{{\vec v}_{\rm M}^2}{2}+\sum_{b\not={\rm M}}\frac{GM_b}{r_{b{\rm M}}}
\Big]+{\cal O}(c^{-4})\approx 1-1.48\times 10^{-8},
\label{eq:TCL}
\end{equation}
which can be solved by a quadrature formula once the orbits of the Moon, the Sun and the other planets are known.

\subsubsection{Satellite Coordinate Reference System (SCRS)}
\label{sec:SCRS}

To determine the metric tensor for the satellite coordinate reference system, the SCRS,  we perform the coordinate transformation given by  Eqs.~(\ref{eq:trans++x0})--(\ref{eq:trans++xa}), where the ``external'' potential and acceleration determined by the potential $w^{}_{\rm M}$ given by Eq.~(\ref{eq:pot_M}), taken at the lunicentric position $\vec{y}_{\rm A}$ of the spacecraft GRAIL-A are (the equations are identical for GRAIL-B, except for the substitution ${\rm A}\rightarrow{\rm B}$):
{}
\begin{eqnarray}
{U}_{\rm ext}^{\rm A}&=&w_{\rm M}({\vec y}_{\rm A})=U_{\rm M}({\vec y}_{\rm A})+
\sum_{b\not={\rm M}}\Big(U_b({\vec r}^{}_{b{\rm M}}+\vec{y}^{}_{\rm A})-U_b({\vec r}^{}_{b{\rm M}}) - {\vec{y}}^{}_{\rm A}\cdot{\vec \nabla} U_b ({\vec r}^{}_{b{\rm M}})\Big)
+{\cal O}(c^{-2}),~~~
\label{eq:SCRS-w_0}
\end{eqnarray}
where ${\vec y}_{\rm A}$ is the solution of the equations of motion of the GRAIL-A spacecraft in the lunicentric frame. This equation can be obtained from equations of geodesics and the metric tensor of the LCRS (\ref{eq:gab-M}) with relativistic gravitational potentials given by (\ref{eq:pot_M})--(\ref{eq:pot_loc-w_a}), (\ref{eq:pot_w_0+}), and (\ref{eq:SCRS-w_0}). Including all the terms of the order of $\sim 10^{-12}$~m/s$^2$ and larger, the equation of spacecraft motion of the GRAIL spacecraft in the LCRS takes the form:
{}
\begin{eqnarray}
\vec{a}_{\rm A0}&=&-{\vec \nabla}U_{\rm M}(\vec{y}^{}_{\rm A})-
\sum_{b\not={\rm M}}\Big({\vec \nabla}U_b({\vec r}^{}_{b{\rm M}}+\vec{y}^{}_{\rm A})-{\vec \nabla} U_b ({\vec r}^{}_{b{\rm M}})\Big)+\nonumber\\
&&\hskip 55pt +\,
\frac{GM_{\rm M}}{c^2y^2_{\rm A}}\Big(\frac{4GM_{\rm M}}{y_{\rm A}}{\vec n}_{\rm A}-v^2_{\rm A0}{\vec n}_{\rm A} +4({\vec n}_{\rm A}\cdot{\vec v}_{\rm A0}){\vec v}_{\rm A0}\Big)+{\vec a}_{\rm NG}+{\cal O}(10^{-13}~{\rm m/s}^2),
\label{eq:eq-sc-mot-lcrs}
\end{eqnarray}
where $\vec{a}_{\rm A0}\equiv d^2\vec{y}_{\rm A}/dt^2_{\rm TCL}$ and ${\vec v}_{\rm A0}=d\vec{y}_{\rm A}/dt_{\rm TCL}$ are the lunicentric acceleration and velocity of the spacecraft in a non-rotating LCRS, also $\vec{n}^{}_{\rm A}=\vec{y}^{}_{\rm A}/{y}^{}_{\rm A}$ is the unit vector in the direction of the spacecraft, $y^{}_{\rm A}=|{\vec y}^{}_{\rm A}|$ and ${\vec a}_{\rm NG}$ is the contribution of non-gravitational forces affecting the motion of a spacecraft (e.g., solar radiation pressure, thermal imbalance, outgassing, etc).

The first term in Eq.~(\ref{eq:eq-sc-mot-lcrs}) is the contribution of the lunar gravity potential $U_{\rm M}({\vec y}_{\rm A})$, given by Eq.~(\ref{eq:pot_w_0+}). Note that the potential $U_{\rm M}({\vec y}_{\rm A})$ must be treated as the potential of an extended body and include a multipolar expansion with sufficient accuracy.
The second term in Eq.~(\ref{eq:eq-sc-mot-lcrs}) is due to the tidal potential at the location of the spacecraft produced by external bodies $u^{\tt tidal}_{\rm M}$ (mostly the Earth and the Sun) which is given by Eq.~(\ref{eq:u-tidal-M}). To reach GRAIL's accuracy requirement of $\sim 10^{-12}$~m/s$^2$, one would have to account for several terms in the expansion beyond the second order one $\sim y^2_{\rm M}$ given in Eq.~(\ref{eq:u-tidal-M}). In fact, terms up to $\sim y^5_{\rm M}$ in the tidal potential $u^{\tt tidal}_{\rm M}$ are needed.
The group of terms on the second line of Eq.~(\ref{eq:eq-sc-mot-lcrs}) is the relativistic Schwarzschild perturbation due to the spherically symmetrical component of the Moon's gravitational field. The first two terms in this group are of the order of $\sim 1.86\times 10^{-10}$~m/s$^2$ and $\sim 4.63\times 10^{-11}$~m/s$^2$, respectively. These are large enough to be in the equations of motion. Given the nearly-circular orbit of the GRAIL spacecraft, the magnitude of the last term in this group is reduced by the orbital eccentricity, which is $e_{\rm A}\sim 0.018$. This fact reduces the contribution of this term by nearly two orders of magnitude when compared to the first two terms, making it barely observable with GRAIL.
Note that the lunar angular momentum ${\vec S}_{{\rm M}}$ present in the relativistic vector gravity potential Eq.~(\ref{eq:pot_loc-w_a}) produces contribution to Eq.~(\ref{eq:eq-sc-mot-lcrs}) of the order of $\sim 10^{-14}$~m/s$^2$, which makes it negligible for GRAIL.

A a result, the metric tensor $g^{\rm A}_{mn}$ representing space-time in the coordinates $\{\hat y^m_{\rm A}\}\equiv(\hat y^0_{\rm A}, \hat{\vec y}_{\rm A})$ of the proper non-rotating spacecraft coordinate reference system (SCRS) may be given in the following form:
{}
\begin{eqnarray}
g^{\rm A}_{00}&=& 1-\frac{2}{c^2}w_{\rm A}
+O(c^{-6}), \qquad
g^{\rm A}_{0\alpha}= O(c^{-5}),\qquad
g^{\rm A}_{\alpha\beta}= \gamma_{\alpha\beta}+\gamma_{\alpha\beta}\frac{2}{c^2} w^{}_{\rm A}+O(c^{-4}),
\label{eq:gab-MA}
\end{eqnarray}
where $w_{\rm A}$ is the tidal contribution produced by the Moon on the world-line of the spacecraft:
{}
\begin{eqnarray}
w_{\rm A}&=&
\frac{GM_{\rm M}}{2\hat y^3_{\rm A}}\Big(
3(\vec{n}_{\rm A}\cdot\hat{\vec y}_{\rm A})^2-\hat{\vec y}_{\rm A}^2\Big)+{\cal O}(\hat y^3_{\rm A},c^{-4}).
\label{eq:pot_w_0*}
\end{eqnarray}

We can also determine the differential equation that relates the rate of the spacecraft proper $\tau_{\rm A}$ time, as measured by an on-board clock in lunar orbit, to the time in LCRS, $t_{\rm TCL}$:
{}
\begin{eqnarray}
\frac{d\tau_{\rm A}}{dt_{\rm TCL}}&=& 1-\frac{1}{c^2}\Big[\frac{{\vec v}^2_{\rm A0}}{2}+U_{\rm M}({\vec y}_{\rm A})+
\sum_{b\not={\rm M}}\Big(U_b({\vec r}^{}_{b{\rm M}}+\vec{y}^{}_{\rm A})-U_b({\vec r}^{}_{b{\rm M}}) - {\vec{y}}^{}_{\rm A}\cdot{\vec \nabla} U_b ({\vec r}^{}_{b{\rm M}})\Big)
+({\vec a}^{}_{\rm M}\cdot {\vec y}^{}_{\rm A})\Big]+{\cal O}(c^{-4}),~~~
\label{eq:proper-coord-t}
\end{eqnarray}
where ${\vec a}^{}_{\rm M}$ is the barycentric acceleration of the Moon, Eq.~(\ref{eq:U-est-M}), see Ref.~\cite{Turyshev:2012}.

As a result, we can establish the rate of the spacecraft proper time with respect to the time of the BCRS, $t=t_{\rm TDB}$:
{}
\begin{eqnarray}
\frac{d\tau_{\rm A}}{dt}
&=& 1-\frac{1}{c^2}\Big[{\textstyle\frac{1}{2}}({\vec v}_{\rm M}+{\vec v}_{\rm A0})^2+U_{\rm M}({\vec y}_{\rm A})+
\sum_{b\not={\rm M}}U_b({\vec r}^{}_{b{\rm M}}+\vec{y}^{}_{\rm A})\Big]+{\cal O}({c^{-4}}).
\label{eq:proper-A-t}
\end{eqnarray}

This result summarizes the relationship of the proper time of an on-board clock in lunar orbit $\tau_{\rm A}$ and the TDB.

\subsubsection{Transformation of gravitational potentials}
\label{sec:pot-trans}

To complete the description of the LCRS, we present the transformation rules for the relativistic gravitational potentials. In Ref.~\cite{Turyshev:2012}, we obtained the structure of the metric tensors corresponding to the local space-times in the reference frames relevant for GRAIL, expressed in terms of harmonic gauge potentials $w$ and $\vec w$. We also derived the rules for transforming relativistic gravitational potentials, the coordinate transformations between the frames and resulting relativistic equations of motion. Applying these results to GRAIL, we see that the scalar and vector gravitational potentials of the Moon
$w_{\rm M}({\vec y}^{}_{\rm M})$ and $\vec w^{}_{\rm M}({\vec y}^{}_{\rm M})$ (as measured at the LCRS) relate to those measured in the coordinates of the BCRS $w_{\rm M}({\vec r}^{}_{\rm BCRS})$ and $\vec w^{}_{\rm M}({\vec r}^{}_{\rm BCRS})$ as
\begin{eqnarray}
w_{\rm M}({\vec y}^{}_{\rm M})&=& \Big(1+\frac{2{v}^2_{\rm M}}{c^2}\Big)
w_{\rm M}({\vec r}^{}_{\rm BCRS})+\frac{4}{c^2}\big({\vec v}_{\rm M}\cdot{\vec w}^{}_{\rm M}({\vec r}^{}_{\rm BCRS})\big)+{\cal O}(c^{-4}),
\label{eq:pot_loc_grav-w_0-cov}\\[3pt]
{\vec w}^{}_{\rm M}({\vec y}^{}_{\rm M})&=& {\vec w}^{}_{\rm M}({\vec r}^{}_{\rm BCRS})-{\vec v}^{}_{\rm M} \,w_{\rm M}({\vec r}^{}_{\rm BCRS})+{\cal O}(c^{-2}).
\label{eq:pot_loc_grav-w_a-cov}
\end{eqnarray}

We estimate the magnitude of ${\vec w}^{}_{\rm M}$ given by Eq.~(\ref{eq:pot_loc-w_a}) as $\sim 2.4\times 10^{6}~{\rm m}^3/{\rm s}^{3}$. This results in a value of $\sim 3.2\times 10^{-6}~{\rm m}^2/{\rm s}^{2}$ for the third term in (\ref{eq:pot_loc_grav-w_0-cov}), which is four orders of magnitude too small compared to the second term in that expression, the scalar potential of the Moon multiplied by $2{v}^2_{\rm M}/c^2$ that was evaluated to be $\sim 5.5\times 10^{-2}~{\rm m}^2/{\rm s}^{2}$.  Thus, to determine the relationship between the LCRS-defined mass of the Moon and its barycentrically defined mass, we must multiply the latter by the factor $(1+2{v}^2_{\rm M}/c^2)\approx(1+2\times 10^{-8}$). Such a transformation results in a small, but observable effect. As far as the GRAIL's accuracy in concerned, contributions to other multipoles of the lunar gravity field are not sensitive to such a small correction factor.

As we know, GRAIL determines the lunar gravity field relying on the EIH equations of motion (\ref{eq:EIH}) for the bodies of the Solar System, including the Moon, the Earth, and the GRAIL spacecraft. For this, Eq.~(\ref{eq:EIH}) must also includes the gravitational potential of the extended Moon (\ref{eq:pot_w_0+}) and tidal potentials due to the Earth and the Sun (\ref{eq:u-tidal-M}). Therefore, one would have to transform the resulting barycentrically defined gravitational potential of the Moon from coordinates of the BCRS to those of the proper lunicentric frame LCRS. A concern was that such a procedure may lead to some unwanted biases in the determination of the lunar gravity field.

Our approach allows one to evaluate the general relativistic effects on the largest  coefficients to the lunar gravity potential corresponding to this transformation. Substituting Eq.~(\ref{eq:pot_loc_grav-w_0-cov}) into Eq.~(\ref{eq:eq-sc-mot-lcrs}), we essentially modify the equation of motion of the GRAIL spacecraft by accounting for the $(1+{2{v}^2_{\rm M}}/{c^2})$ factor.
Now, we can represent the barycentric velocity of the Moon as  ${\vec v}_{\rm M}={\vec v}_{\rm EM}+{\vec v}'_{\rm M}$, where ${\vec v}_{\rm EM}$ is the barycentric velocity of the Earth-Moon barycenter and ${\vec v}'_{\rm M}$ is the velocity of the Moon in the EMB frame. Therefore, $v^2_{\rm M}=v^2_{\rm EM}+v'^2_{\rm M}+2v_{\rm EM}v'_{\rm M}\cos D$, where $D=\theta-\theta'$ is the the difference between the longitudes of the mean Moon and the mean Sun with a period of 29.531 days. The constant part in the barycentric velocity of the Moon $v_{\rm M}$ may be easily absorbed in the determination of the lunar mass as a bias with magnitude of $2\times 10^{-8} M_{\rm M}$. The variability in $v_{\rm M}$, if not properly removed in accordance with Eq.~(\ref{eq:pot_loc_grav-w_0-cov}), may introduce an additional time-dependent bias with a magnitude of $2.2\times 10^{-9} M_{\rm M}\cos D$, which can be removed in the data analysis. Finally, given the nearly circular obit of the GRAIL spacecraft around the Moon, the $1/c^2$ terms in the second line of Eq.~(\ref{eq:eq-sc-mot-lcrs}) can be seen and a modification of the Newtonian point-mass acceleration of the Moon ${\vec a}^{\rm N}_{\rm A}=GM_{\rm M}{\vec n}_{\rm A}/{y^2_{\rm A}}$. These terms are nearly constant, have combined magnitude of $\sim 1\times 10^{-10}{\vec a}^{\rm N}_{\rm A}$ and would be easily absorbed in the determination of the lunar mass as a small bias of $1\times 10^{-10} M_{\rm M}$.

For spacecraft with lesser sensitivity, these corrections are irrelevant. However, at the micron-level sensitivity of the GRAIL mission, they become noticeable. It is, of course, possible to absorb small constant or periodic terms into constants such as $M_{\rm M}$ during data analysis, with no impact on mission objectives or the quality of the mission's results. Nonetheless, pursuing these small corrections is worthwhile, demonstrating that a spacecraft with GRAIL's sensitivity is already a practical instrument for relativistic geodesy in the lunar environment, and paving the way for future missions that will operate at even greater accuracy.

\subsection{Transformations of position vectors}
\label{sec:pos-vectors}

Equation~(\ref{eq:trans++xa}) establishes the relationship between the coordinates of the local body-centric coordinate reference frame and the coordinates of the global BCRS \cite{Turyshev:2012}:
{}
\begin{eqnarray}
{\vec r}_a&=& {\vec y}_a+
c^{-2}\Big\{-{\textstyle\frac{1}{2}}{\vec v}_{a} ({\vec v}_{a}\cdot{\vec y}_a)-{\vec y}_a{\bar U}_{\rm ext}^a +
[{\boldsymbol\omega}^{}_{a}\times{\vec y}_a] + {\textstyle\frac{1}{2}}{\vec a}^{}_{a}{y}_a^2-{\vec y}_a({\vec y}_a\cdot {\vec a}^{}_{a})
\Big\}+{\cal O}(c^{-4}).
\label{eq:trans+r}
\end{eqnarray}

Considering the anticipated accuracy of the GRAIL experiment, we can simplify Eq.~(\ref{eq:trans+r}) by noting that the last three terms in this expression are much smaller than needed for GRAIL. Indeed, we can evaluate the magnitude of the third term \cite{Turyshev:2012} as $[{\boldsymbol\omega}^{}_{a}\times{\vec y}_a]\simeq {\vec y}_a GM_\odot v_{\rm E}\Delta t/{\rm AU}^2$, where $\Delta t$ is the signal propagation time, $M_\odot$ is the mass of the Sun and ${\rm AU}\simeq 1.5\times 10^{11}$~m is the astronomical unit. The Moon-Earth radio-signal propagation time is $\Delta t\simeq 1.3$~s. However, even for $\Delta t=10^3$~s, we have $[{\boldsymbol\omega}^{}_{a}\times{\vec y}_a]\lesssim 2\times 10^{-4}{\vec y}_a {U}_{\rm ext}^a$; therefore, the third term within the curly braces in Eq.~(\ref{eq:trans+r}) is negligible for GRAIL.
The last two terms within the curly braces in Eq.~(\ref{eq:trans+r}) are dependent on the acceleration of the planet center, and are ignored for the reasons that we discuss at the end of this subsection.

The required space-time transformations relate the position of the ground-based antenna and that of the spacecraft. Using superscript indices to indicate explicitly the dependence on the various time scales TT, TDB, etc., the terrestrial (geocentric) coordinates $\vec{y}^{\rm TT}_{\rm C}$ of the antenna must be transformed into TDB-compatible (barycentric) coordinates $\vec{r}^{\rm TDB}_{\rm C}=\vec{z}_{\rm C}-\vec{z}_{\rm E}+{\cal O}(c^{-4})$, where $\vec{z}_{\rm C}$ being the barycentric position of the DSN station $\vec{z}_{\rm C}={\vec z}_{\rm E}+{\vec y}_{\rm C}$. Similarly to the approach developed in Ref.~\cite{Tommei-etal:2010} for the BepiColombo mission, this transformation is expressed by
\begin{equation}
\vec{r}^{\rm TDB}_{\rm C} = \vec{y}^{\rm TT}_{\rm C}\Big(1 -\frac{1}{c^2}\sum_{b\not={\rm E}}U_b({\vec r}_{b\rm E})\Big)-
\frac{1}{2c^2}\left({\vec{v}^{\rm TDB}_{\rm E}\cdot \vec{y}^{\rm TT}_{\rm C}}\right)
\vec{v}^{\rm TDB}_{\rm E},\label{eq:posxform}
\end{equation}
where $\sum_{b\not={\rm E}}U_b({\vec x}_{\rm E})$ is the Newtonian gravitational potential due to bodies other than the Earth at the geocenter, $\vec{x}_{\rm E}$ is the SSB position and $\vec{v}^{\rm TDB}_{\rm E}=d\vec{z}_{\rm E}/dt$ is the SSB velocity of the Earth (as defined in the paragraph before Eq.~(\ref{eq:EIH})).

The time coordinate must also be changed consistently together with the spatial coordinates. The effect of this change on velocities is given by:
\begin{equation}
\vec{v}^{\rm TDB}_{\rm C}-\vec{v}^{\rm TDB}_{\rm E} = \Big[\vec{v}^{\rm TT}_{\rm C0}\Big(1 - \frac{1}{c^2}\sum_{b\not={\rm E}}U_b({\vec r}_{b\rm E})
\Big)- \frac{1}{2c^2}\Big({\vec{v}^{\rm TDB}_{\rm E}\cdot \vec{v}^{\rm TT}_{\rm C0}}\Big) \vec{v}^{\rm TDB}_{\rm E}\Big]\frac{d\tau_{\rm C}}{dt},\label{eq:velxform}
\end{equation}
where $\vec{v}^{\rm TT}_{\rm C0}=d\vec{y}^{\rm TT}_{\rm C}/d\tau_{\rm C}$.
Note that Eq.~(\ref{eq:velxform}) contains the factor $d\tau_{\rm C}/dt$, identical to Eq.~(\ref{eq:proper-t-C}), that deals with time transformation: $\tau_{\rm C}$ is the local time for a ground-based antenna, that is, TT, and $t$ is the corresponding TDB time.

Similar to Eq.~(\ref{eq:posxform}), the lunicentric coordinates of the orbiter ${\vec y}^{\rm TCL}_{\rm A}$ are transformed into BCRS coordinates $\vec{r}^{\rm TDB}_{\rm A}=\vec{z}_{\rm A}-\vec{z}_{\rm M}+{\cal O}(c^{-4})$, where $\vec{z}_{\rm A}$ being the barycentric position of the spacecraft $\vec{z}_{\rm A}={\vec z}_{\rm M}+{\vec y}_{\rm A}$:
\begin{equation}
\vec{r}^{\rm TDB}_{\rm A} = \vec{y}^{\rm TCL}_{\rm A}\Big(1 - \frac{1}{c^2}\sum_{b\not={\rm M}}U_b({\vec r}_{b\rm M})
\Big)-\frac{1}{2c^2}\Big({\vec{v}^{\rm TDB}_{\rm M}\cdot \vec{y}^{\rm TCL}_{\rm A}}\Big) \vec{v}^{\rm TDB}_{\rm M},
\end{equation}
where $\sum_{b\not={\rm M}}U_b({\vec x}_{\rm M})$ is the gravitational potential due to bodies other than the Moon at the Moon's barycenter, $\vec{x}_{\rm M}$ is the Moon's position in SSB coordinates and $\vec{v}^{\rm TDB}_{\rm M}=d\vec{z}_{\rm M}/dt$ is the Moon's SSB velocity.

The corresponding velocity transformation is given by
\begin{equation}
\vec{v}^{\rm TDB}_{\rm A}-\vec{v}^{\rm TDB}_{\rm M} = \Big[\vec{v}^{\rm TCL}_{\rm A0}\Big(1 -\frac{1}{c^2}\sum_{b\not={\rm M}}U_b({\vec r}_{b\rm M})
\Big)-\frac{1}{2c^2}\Big({\vec{v}^{\rm TDB}_{\rm M}\cdot \vec{v}^{\rm TCL}_{\rm A0}}\Big) \vec{v}^{\rm TDB}_{\rm M}\Big]\frac{dt_{\rm TCL}}{dt},
\end{equation}
with $\vec{v}^{\rm TCL}_{\rm A0}=d\vec{y}^{\rm TCL}_{\rm A}/dt_{\rm TCL}$ and $dt_{\rm TCL}/dt_{\rm TDB}$ given by Eq.~(\ref{eq:TCL}). The relations for the other spacecraft is obtained by replacing ${\rm A}\rightarrow {\rm B}$.

Note that in all the coordinate transformations presented in this section, we neglected terms that contain the SSB acceleration of the planet center, as these have an additional small parameter $({R_b}/{z_{b}})$, where $R_b$ is distance from the planetary barycenter and $z_{b}$ is the distance of the planet from the SSB. Even for the Earth-Moon distance, these acceleration-dependent terms are at most of the order of $10^{-3}$ compared to the other $1/c^2$ terms, both velocity- and Newtonian potential-dependent ones, making the acceleration-dependent terms negligible for the results above.

\section{Forming K\lowercase{a}-band range (KBR) observables for GRAIL}
\label{sec:KBRR}

One can demonstrate that in the post-Minkowskian approximation, appropriate for most Solar System experiments including GRAIL, as seen from BCRS, the phase of an electromagnetic wave that is passing by a gravitating body with the mass $M$ can be presented as (see a detailed derivation in Appendix~\ref{sec:em-phase}, with the result given by Eq.~(\ref{eq:eq_eik-phi0})):
\begin{equation}
\varphi(t,{\vec x}) = k_0\Big(ct-R_{\rm A}-\frac{2GM}{c^2}\ln\Big[\frac{r_{\rm A}+r+R_{\rm A}}{r_{\rm A}+r-R_{\rm A}}\Big]\Big)+{\cal O}(G^2, c^{-3}),
\label{eq:eq-phase}
\end{equation}
where $k^m = k^0(1, {\vec k})$ is a constant null vector directed along the trajectory of propagation of the unperturbed electromagnetic wave such that $\gamma_{mn}k^mk^n=0$, also $k^0=\omega/c$ where $\omega$ is the constant frequency of the unperturbed wave. We also use the following notations:
\begin{equation}
{\vec k}=\frac{{\vec R}_{\rm A}}{R_{\rm A}}, \quad {\vec R}_{\rm A}={\vec x}-{\vec x}_{\rm A}, \quad R_{\rm A}=|{\vec R}_{\rm A}|, \qquad {\rm also} \qquad {\vec r}={\vec x}-{\vec z}, \quad r=|{\vec r}|\quad {\vec r}_{\rm A}={\vec x}_{\rm A}-{\vec z}, \quad r_{\rm A}=|{\vec r}_{\rm A}|,
\end{equation}
where $\vec z$ is the time-dependent spatial coordinate of the massive body.

The phase $\varphi$ of an electromagnetic wave that was emitted at the point $x^m_{\rm A0}= (ct_{\rm A0}, {\vec x}_{\rm A0})$ and received at the point $x^m_{\rm B}= (ct_{\rm B}, {\vec x}_{\rm B})$ remains constant along the path of this wave \cite{Landau-Lifshitz:1988,MTW:1973}. In particular, if $\lambda$ is an affine parameter along the wave's path, the derivative of the phase satisfies the equation
\begin{equation}
\frac{d\varphi}{d\lambda}= \frac{\partial \varphi}{\partial x^m}\frac{d x^m}{d\lambda}= K_mK^m = 0,
\label{eq:cosnt}
\end{equation}
which suggests that $\varphi[x^m(\lambda)] = {\rm const.}$ In other words, along the signal's world-line the phase stays constant and equal to its initial value $\varphi(t,{\vec x})=k_0 ct_{\rm A0}= \omega_{\rm A0}t_{\rm A0}$.

Equating the values of the phase given by Eq.~(\ref{eq:eq-phase}) at two points $A_0$ and $B$ as $\varphi(t_{\rm A0},{\vec x}_{\rm A0})=\varphi(t_{\rm B},{\vec x_{\rm B}})$, we can determine the gravitational delay of the signal moving through a particular space-time. Indeed, up to ${\cal O}(c^{-3})$ the coordinate time transfer, which is defined as  $t_{\rm B} - t_{\rm A0}=T_{\rm AB}$, is given by \cite{Turyshev:2012}:
\begin{equation}
T_{\rm AB}=t_{\rm B} - t_{\rm A0}= \frac{R_{\rm AB}}{c}+ \frac{2GM}{c^3}\ln\Big[\frac{
r_{\rm A0} + r_{\rm B} + R_{\rm AB}}{r_{\rm A0} + r_{\rm B} - R_{\rm AB}}\Big]+{\cal O}(c^{-4}),
\label{eq:t-AB=!}
\end{equation}
where the logarithmic term represents the Shapiro time delay. Also, ${\vec r}_{\rm B}={\vec x}_{\rm B}-{\vec z}$ and ${\vec R}_{\rm AB}={\vec x}_{\rm B}-{\vec x}_{\rm A0}$.

With these results, we can now formulate a relativistic model for the fundamental timing observables on GRAIL.

\subsection{The inter-spacecraft ranging observables}
\label{sec:kbr}

Consider a clock with proper frequency $f_{\rm A0}$, located at moving point $\rm A_0$, that emits a signal with frequency $f_{\rm A0}$ at an instant of proper time $\tau_{\rm A0}$ measured on the world-line of the clock. This signal is received by the moving point $\rm B$ at an instant of the proper time $\tau_{\rm B}$ taken at the world-line body $\rm B$ and the instantaneous phase of this signal is compared with the phase of the local oscillator with proper frequency $f_{\rm B0}$ of the clock located at point $\rm B$.

The measurable quantity is the difference between the instantaneous phases of the two signals compared at point $\rm B$. Instrumentally, at point $\rm B$ one measures the fractional difference $d n_{\rm AB}$ in the number of cycles $dn_{\rm A0}^{\rm B}$ received from the clock at point $\rm A_0$ and the number of the locally generated cycles $dn_{\rm B}$. Mathematically, this quantity may be expressed at the point $\rm B$ at an instance of the proper time $d\tau_{\rm B}$ as:
{}
\begin{equation}
d n_{\rm AB}=dn_{\rm B}-dn_{\rm A0}^{\rm B}= f_{\rm B0}d\tau_{\rm B}-f_{\rm A0}^{\rm B}d\tau_{\rm B},
\label{eq:dn-ab}
\end{equation}
where $f_{\rm A0}^{\rm B}$ is the frequency of the oscillator $\rm A$ as detected at $\rm B$.

Assuming that the number of pulses sent from spacecraft $\rm A$, $dn_{\rm A0}$, and received on spacecraft $\rm B$, $dn_{\rm A0}^{\rm B}$, are the same, or $dn_{\rm A0}^{\rm B}=dn_{\rm A0}$, we can express $f_{\rm A0}^{\rm B}$ via its value at the proper time of emission on spacecraft $\rm A$ (see also Eq.~(\ref{eq:fA-fB*+}) below):
{}
\begin{equation}
\frac{f_{\rm A0}^{\rm B}}{f_{\rm A0}}=\frac{dn_{\rm A0}^{\rm B}}{d\tau_{\rm B}}\frac{d\tau_{\rm A0}}{dn_{\rm A0}}=\frac{d\tau_{\rm A0}}{d\tau_{\rm B}}.
\label{eq:dn-ab+}
\end{equation}
Furthermore, the instantaneous difference of the number of cycles measured on spacecraft $\rm B$, as given by Eq.~(\ref{eq:dn-ab}), takes the form:
{}
\begin{equation}
d n_{\rm AB}=f_{\rm B0}d\tau_{\rm B}-f_{\rm A0}d\tau_{\rm A0}.
\label{eq:dn-ab++}
\end{equation}
Eq.~(\ref{eq:dn-ab++}) is the difference in the number of cycles generated by the two oscillators during the given proper time intervals along the world-lines of the two clocks.

\begin{figure}[t]
\includegraphics[width=0.55\linewidth,angle=0]{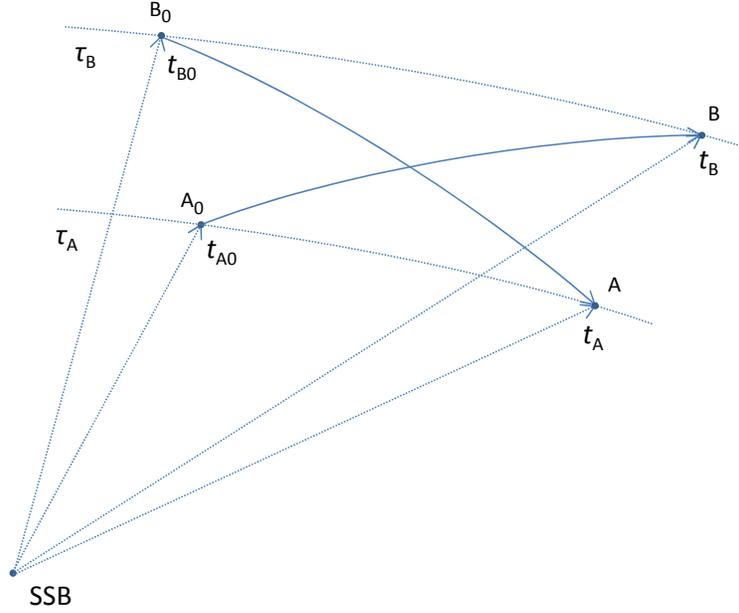}
\vskip -6pt
\caption{\label{fig:grail-timing} Timing events on GRAIL:
Depicted (not to scale) are the world-lines of the GRAIL-A and GRAIL-B spacecraft with corresponding proper times $\tau_{\rm A}$ and $\tau_{\rm B}$. Ka-band signals that are emitted from points $A_0$ and $B_0$ are received at $B$ and $A$, respectively. The times $t_{\rm A0}, t_{\rm B0}, t_{\rm A}$ and $t_{\rm B}$ are the coordinate times at these points, as measured in the BCRS.}
\end{figure}

We can now express Eq.~(\ref{eq:dn-ab++}) in terms of the coordinate time:
{}
\begin{equation}
d n_{\rm AB}=f_{\rm B0}\Big(\frac{d\tau_{\rm B}}{dt_{\rm B}}\Big)dt_{\rm B}-f_{\rm A0}\Big(\frac{d\tau_{\rm A0}}{dt_{\rm A0}}\Big)dt_{\rm A0}.
\label{eq:dn-ab+-+}
\end{equation}

Note that Eq.~(\ref{eq:dn-ab+-+}) cannot be integrated in general case if the two time variables $t_{\rm A0}$ and $t_{\rm B}$ are treated as independent. However, in our case the points $\rm A_0$ and $\rm B$ are connected by a time-like geodesic, and therefore, the coordinate times $t_{\rm A0}$ and $t_{\rm B}$ are connected by the light-time equation (\ref{eq:t-AB=!}) that reads:
{}
\begin{equation}
t_{\rm B}-t_{\rm A0}=T_{\rm AB}(t_{\rm A0},t_{\rm B})=
\frac{1}{c}|{\vec r}_{\rm B}(t_{\rm B}) - {\vec r}_{\rm A}(t_{\rm A0})|+ \frac{2GM}{c^3}\ln\Big[\frac{
r_{\rm A} + r_{\rm B} + R_{\rm AB}}{r_{\rm A} + r_{\rm B} - R_{\rm AB}}\Big].
\label{eq:dn-lt}
\end{equation}

We note that Eq.~(\ref{eq:dn-lt}) can be used to express either $t_{\rm A0}$ as a function of $t_{\rm B}$ or vice versa. Observables on GRAIL are time-stamped using the time of reception (that is, $t_{\rm B}$). We therefore have more direct access to ${\vec x}_{\rm A}(t_{\rm B})$ rather than ${\vec x}_{\rm A}(t_{\rm A0})$, and the first term on the right hand side of Eq.~(\ref{eq:dn-lt}) gets modified by Sagnac correction terms (as observed in Ref.~\cite{Blanchet-etal:2001}) consistently to the order $1/c^3$:
\begin{equation}
R_{\rm AB} = d_{\rm AB}+\frac{({\vec d}_{\rm AB}\cdot {\vec v}_{\rm A})}{c}
+\frac{d_{\rm AB}}{2c^2}\left({\vec v}_{\rm A}^2+({{\vec n}}_{\rm AB}\cdot {\vec v}_{\rm A})^2-({\vec d}_{\rm AB}\cdot {\vec a}_{\rm A})\right)+{\cal O}(c^{-3}),
\label{eq:nt-AB+**}
\end{equation}
where ${\vec d}_{\rm AB} = {\vec x}_{\rm B}(t_{\rm B})-{\vec x}_{\rm A}(t_{\rm B})$ is the coordinate distance between $A$ and $B$ at the moment of reception at $B$ (we have $d_{\rm AB} = |{\vec d}_{\rm AB}|$ and ${{\vec n}}_{\rm AB}={\vec d}_{\rm AB}/d_{\rm AB}$), where ${\vec v}_{\rm A}={\vec v}_{\rm A}(t_{\rm B})$ denotes the coordinate velocity of spacecraft $A$ at that instant, and where ${\vec a}_{\rm B}$ is the acceleration of $A$ (in all the order $1/c^3$ terms we can use quantities at $t_{\rm A0}$ or $t_{\rm B}$). In this case, Eq.~(\ref{eq:dn-lt}) becomes
{}
\begin{equation}
T_{\rm AB}(t_{\rm B}) = \frac{d_{\rm AB}}{c}+\frac{({\vec d}_{\rm AB}\cdot {\vec v}_{\rm A})}{c^2}
+\frac{d_{\rm AB}}{2c^3}\left({\vec v}_{\rm A}^2+({{\vec n}}_{\rm AB}\cdot {\vec v}_{\rm A})^2-({\vec d}_{\rm AB}\cdot {\vec a}_{\rm A})\right)+ \frac{2GM}{c^3}\ln\Big[\frac{
r_{\rm A} + r_{\rm B} + d_{\rm AB}}{r_{\rm A} + r_{\rm B} - d_{\rm AB}}\Big]+{\cal O}(c^{-4}),
\label{eq:nt-BA+**}
\end{equation}
where all quantities here are taken at the instant of reception $t_{\rm B}$. In the case of the GRAIL mission, when signal transmission between the two spacecraft is concerned the first term in Eq.~(\ref{eq:nt-AB+**}), which is of order $1/c$, is $\sim$6.67~$\mu$s. The second term in Eq.~(\ref{eq:nt-AB+**}) represents the Sagnac term of order $1/c^2$ and can amount to $\sim$3.67~ns at GRAIL's orbit around the Moon; the third Sagnac term, of order $1/c^3$, is $\sim$0.02~ps (comparable to the lunar Shapiro term, which is $\sim$0.04~ps). Note that expression for $T_{\rm BA}(t_{\rm A})$ may be obtained from  Eq.~(\ref{eq:nt-BA+**}) by interchanging $A\leftrightarrow B$.

\subsection{Dual One-Way Range (DOWR) observables on GRAIL}
\label{sec:dowr}

To develop an analytical form for the DOWR observable, we note that Eq.~(\ref{eq:dn-lt}) could be used to express either $t_{\rm A0}$ as a function of $t_{\rm B}$ or vice versa. As observables on GRAIL are time-stamped using the time of reception (that is, $t_{\rm B}$), in the following we treat $t_{\rm A0}$ as a function of $t_{\rm B}$, i.e., $t_{\rm A0}=t_{\rm A0}(t_{\rm B})$. Furthermore, we can write $T_{\rm AB}(t_{\rm A0},t_{\rm B})=T_{\rm AB}(t_{\rm B}).$ This allow us to present Eq.~(\ref{eq:dn-ab+-+}) as
{}
\begin{equation}
d n_{\rm AB}=f_{\rm B0}\Big(\frac{d\tau}{dt}\Big)_{\rm B}dt_{\rm B}-f_{\rm A0}\Big(\frac{d\tau}{dt}\Big)_{\rm A}d\Big(t_{\rm B}-T_{\rm AB}(t_{\rm B})\Big).
\label{eq:dn-ab+-+0}
\end{equation}

To integrate Eq.~(\ref{eq:dn-ab+-+0}), we rely on Eq.~(\ref{eq:proper-A-t}) and introduce a function $u_{\rm B}(t_{\rm B})$ that allows us to write ${d\tau_{\rm B}}/{dt}$ as:
{}
\begin{eqnarray}
\frac{d\tau_{\rm B}}{dt}&=& 1-\frac{1}{c^2}u_{\rm B}(t_{\rm B})+{\cal O}({c^{-4}}), \qquad {\rm where} \qquad u_{\rm B}(t_{\rm B})=\frac{{\vec v}_{\rm A}^2}{2}+\sum_b\frac{GM_b}{r_{b{\rm A}}}
+{\cal O}({c^{-2}}).
\label{eq:Larg}
\end{eqnarray}

Using this definition of $u_{\rm B}(t_{\rm B})$ given in Eq.~(\ref{eq:Larg}) allows us to integrate the first term in Eq.~(\ref{eq:dn-ab+-+0}) as
{}
\begin{equation}
\int_{t^0_{\rm B}}^{t_{\rm B}} f_{\rm B0}\Big(\frac{d\tau}{dt}\Big)_{\rm B}dt_{\rm B}=f_{\rm B0}\Big(\frac{d\tau}{dt}\Big)_{\rm B}\big(t_{\rm B}-t_{\rm B}^0\big)+
\frac{1}{c^2}f_{\rm B0}\Big(u_{\rm B}(t_{\rm B})\big(t_{\rm B}-t_{\rm B}^0\big)-
\int_{t_{\rm B}^0}^{t_{\rm B}} u_{\rm B}(t'_{\rm B})dt'_{\rm B}\Big)+{\cal O}({c^{-4}}),
\label{eq:int-A}
\end{equation}
where $t_{\rm B}^0$ is used to denote the (for now, arbitrary) start of the integration interval.

Similarly, we have the following expression for the second term of Eq.~(\ref{eq:dn-ab+-+0}):
{}
\begin{eqnarray}
\int_{t^0_{\rm B}}^{t_{\rm B}} f_{\rm A0}\Big(\frac{d\tau}{dt}\Big)_{\rm A}d\Big(t_{\rm B}-T_{\rm AB}(t_{\rm B})\Big)&=&f_{\rm A0}\Big(\frac{d\tau}{dt}\Big)_{\rm A}\Big(t_{\rm B}-T_{\rm AB}(t_{\rm B})-\big(t_{\rm B}^0-T_{\rm AB}(t_{\rm B}^0)\big)\Big)+\nonumber\\
&&\hskip -120pt +\,\frac{1}{c^2}f_{\rm A0}\Big(u_{\rm A}(t_{\rm B})\Big(t_{\rm B}-T_{\rm AB}(t_{\rm B})-\big(t_{\rm B}^0-T_{\rm AB}(t_{\rm B}^0)\big)\Big)-
\int_{t_{\rm B}^0-T_{\rm AB}(t_{\rm B}^0)}^{t_{\rm B}-T_{\rm AB}(t_{\rm B})} u_{\rm A}(t'_{\rm B})dt'_{\rm B}\Big)+{\cal O}({c^{-4}}).
\label{eq:int-A+}
\end{eqnarray}

Transforming from proper to coordinate frequencies, as
$f_{\rm B}=f_{\rm B0}({d\tau}/{dt})_{\rm B}$ and
$f_{\rm A}=f_{\rm A0}({d\tau}/{dt})_{\rm A}$,
and using all the results developed in this section,
we can integrate Eq.~(\ref{eq:dn-ab+-+0}) and present the result in terms of the phase difference as
{}
\begin{equation}
\Delta n_{\rm AB}(t_{\rm B})=f_{\rm B}\,t_{\rm B}-f_{\rm A}\Big(t_{\rm B}-T_{\rm AB}(t_{\rm B})\Big)+
\epsilon_{\rm AB}
+\delta n_{\rm AB}
+{\cal O}(c^{-4}),
\label{eq:n-ab}
\end{equation}
where $\epsilon_{\rm AB}\equiv\epsilon_{\rm AB}(t_{\rm B}^0,t_{\rm B})$ is given by
{}
\begin{eqnarray}
\epsilon_{\rm AB}(t_{\rm B}^0,t_{\rm B})&=&\frac{1}{c^2}\Big\{f_{\rm B0}\Big(u_{\rm B}(t_{\rm B})(t_{\rm B}-t^0_{\rm B})-\int_{t^0_{\rm B}}^{t_{\rm B}} u_{\rm B}(t'_{\rm B})dt'_{\rm B}\Big)-\nonumber\\
&&\hskip 10pt -\,
f_{\rm A0}\Big(u_{\rm A}(t_{\rm B})\Big(t_{\rm B}-T_{\rm AB}(t_{\rm B})-\big(t_{\rm B}^0-T_{\rm AB}(t_{\rm B}^0)\big)\Big)-
\int_{t_{\rm B}^0-T_{\rm AB}(t_{\rm B}^0)}^{t_{\rm B}-T_{\rm AB}(t_{\rm B})} u_{\rm A}(t'_{\rm B})dt'_{\rm B}\Big)\Big\}+{\cal O}({c^{-4}}),
\label{eq:n-ab-eps}
\end{eqnarray}
and $\delta n_{\rm AB}\equiv \delta n_{\rm AB}(t_{\rm B}^0)$ is an integration constant determined by the initial conditions:
{}
\begin{eqnarray}
\delta n_{\rm AB}(t_{\rm B}^0)&=&-f_{\rm B}\,t^0_{\rm B}+f_{\rm A}\Big(t^0_{\rm B}-T_{\rm AB}(t^0_{\rm B})\Big)+{\cal O}({c^{-4}}),
\label{eq:n-ab-n}
\end{eqnarray}
The second observable $\Delta n_{\rm BA}(t_{\rm A})$ that deals with the signal propagation from $\rm B_0$ to $\rm A$ is derived in an analogous way.

To formulate the relativistic model for the dual one-way range (DOWR) observables on GRAIL, we need the expressions derived above for the instantaneous phase differences measured at both spacecraft, $n_{\rm AB}(t_{\rm B})$ and $n_{\rm BA}(t_{\rm A})$, which are given by Eqs.~(\ref{eq:n-ab}), together with the instantaneous delays measured at the points of signal reception at both spacecraft, $T_{\rm AB}(t_{\rm B})$ and $T_{\rm BA}(t_{\rm A})$, as given by Eqs.~(\ref{eq:nt-AB+**}) and (\ref{eq:nt-BA+**}). As a result, Eq.~(\ref{eq:n-ab}) becomes:
\begin{eqnarray}
\Delta n_{\rm AB}(t_{\rm B})&=&(f_{\rm B}-f_{\rm A})\,t_{\rm B}+f_{\rm A}\,T_{\rm AB}(t_{\rm B})+\epsilon_{\rm AB}
+\delta n_{\rm AB} +{\cal O}(c^{-4})\nonumber\\[3pt]
&=&(f_{\rm B}-f_{\rm A})\,t_{\rm B}+f_{\rm A}\Big(
\frac{d_{\rm AB}}{c}+ \frac{2GM_{\rm M}}{c^3}\ln\Big[\frac{
r_{\rm A} + r_{\rm B} + d_{\rm AB}}{r_{\rm A} + r_{\rm B} - d_{\rm AB}}\Big]\Big)+
\nonumber\\
&&\hskip17pt +\,
f_{\rm A}\Big(\frac{({\vec d}_{\rm AB}\cdot {\vec v}_{\rm A})}{c^2}
+\frac{d_{\rm AB}}{2c^3}\left({\vec v}_{\rm A}^2+({{\vec n}}_{\rm AB}\cdot {\vec v}_{\rm A})^2-({\vec d}_{\rm AB}\cdot {\vec a}_{\rm A})\right)\Big)+ \epsilon_{\rm AB} +\delta n_{\rm AB} +{\cal O}(c^{-4}),
\label{eq:n-ab1}
\end{eqnarray}
where $M_{\rm M}$ is the mass of the Moon.
An expression for $\Delta n_{\rm BA}(t_{\rm A})$ may be obtained from (\ref{eq:n-ab1}) by interchanging $A\leftrightarrow B$.

The authors of Ref.~\cite{TimingMemo:2010} discuss the interpolation algorithm realized on GRAIL to synchronize the LGRS clocks on both spacecraft in coordinate time. Here we just assume that synchronization is achieved, so that $t_{\rm A}=t_{\rm B}=t$ and $t_{\rm A}^0=t_{\rm B}^0=t_0$. We now can form a quantity, that is called dual one-way range (DOWR):
\begin{equation}
{R}_{\rm dowr}(t)=c\frac{\Delta n_{\rm AB}(t)+ \Delta n_{\rm BA}(t)}{f_{\rm A}+f_{\rm B}}.
\label{eq:DOWR0}
\end{equation}
Substituting Eq.~(\ref{eq:n-ab1}), we have the following result for ${\rm R}_{\rm dowr}$:
{}
\begin{eqnarray}
{R}_{\rm dowr}(t)&=&d_{\rm AB}+
\frac{\big((f_{\rm A}{\vec v}_{\rm A}-f_{\rm B}{\vec v}_{\rm B})\cdot{\vec d}_{\rm AB}\big)}{c(f_{\rm A}+f_{\rm B})}+\frac{d_{\rm AB}}{2c^2}\frac{
\big((f_{\rm B}{\vec a}_{\rm B}-f_{\rm A}{\vec a}_{\rm A})\cdot{\vec d}_{\rm AB}\big)}{f_{\rm A}+f_{\rm B}}+\nonumber\\[3pt]
&&+\,
\frac{d_{\rm AB}}{2c^2}\frac{f_{\rm A}\big({\vec v}^2_{\rm A}+({\vec n}_{\rm AB}\cdot{\vec v}_{\rm A})^2\big)+
f_{\rm B}\big({\vec v}^2_{\rm B}+({\vec n}_{\rm AB}\cdot{\vec v}_{\rm B})^2\big)}{f_{\rm A}+f_{\rm B}}+ \frac{2GM_{\rm M}}{c^2}\ln\Big[
\frac{r_{\rm A}+r_{\rm B}+d_{\rm AB}}{r_{\rm A}+r_{\rm B}-d_{\rm AB}}\Big]+\nonumber\\[3pt]
&&+\,
\frac{c\big(\epsilon_{\rm AB}+\epsilon_{\rm BA}\big)}{{f_{\rm A}+f_{\rm B}}}+\frac{c\big(\delta n_{\rm AB}+\delta n_{\rm BA}\big)}{{f_{\rm A}+f_{\rm B}}}+
{\cal O}(c^{-3}),
\label{eq:DOWR+0}
\end{eqnarray}
where $\epsilon_{\rm AB}$ and $\delta n_{\rm AB}$ depend on the choice of the start of the integration intervals, i.e., $t_{\rm A}^0$ and $t_{\rm B}^0$.

We now discuss each of the seven terms present in Eq.~(\ref{eq:DOWR+0}) and evaluate their magnitudes and relevance for GRAIL. To develop numerical estimates for the magnitude of the various terms that we consider, we use mission parameters that are provided in Table~\ref{tb:params}.

The first term in Eq.~(\ref{eq:DOWR+0}) is the instantaneous Euclidean distance $d_{\rm AB}\simeq 200$~km (see Table~\ref{tb:params}) between the two lunar orbiters.

The next three terms are the first ($\sim 1/c$) and the second (1/$c^2$) order Sagnac effects. These terms are due to the fact that representing the observables only in terms of the received times $t_{\rm A}$ and $t_{\rm B}$ on the two spacecraft is equivalent to a rotation of the reference system. To evaluate the second term in Eq.~(\ref{eq:DOWR+0}), we use the identity
\begin{equation}
(f_{\rm A}{\vec v}_{\rm A}-f_{\rm B}{\vec v}_{\rm B})=-{\textstyle\frac{1}{2}}(f_{\rm A}+f_{\rm B})({\vec v}_{\rm B}-{\vec v}_{\rm A})-{\textstyle\frac{1}{2}}(f_{\rm B}-f_{\rm A})({\vec v}_{\rm A}+{\vec v}_{\rm B}).
\label{eq:ident1}
\end{equation}
Given $\Delta f_{\rm AB}=|f_{\rm B}-f_{\rm A}|\sim 10^3$~Hz, we get:
\begin{eqnarray}
\frac{\big((f_{\rm A}{\vec v}_{\rm A}-f_{\rm B}{\vec v}_{\rm B})\cdot{\vec d}_{\rm AB}\big)}{c(f_{\rm A}+f_{\rm B})}&=&-\frac{({\vec v}_{\rm AB}\cdot{\vec d}_{\rm AB})}{2c}-\left(\frac{f_{\rm B}-f_{\rm A}}{f_{\rm A}+f_{\rm B}}\right)\frac{\big(({\vec v}_{\rm A}+{\vec v}_{\rm B})\cdot{\vec d}_{\rm AB}\big)}{2c}=\nonumber\\
&=&
(-0.061423+3\times 10^{-7})~{\rm m},
\label{eq:Sagnac1}
\end{eqnarray}
where ${\vec v}_{\rm AB}=\vec{v}_{\rm B}-\vec{v}_{\rm A}$. Therefore, the second term in Eq.~(\ref{eq:Sagnac1}) is less than 1~$\mu$m and it can be omitted.

%%%%%%%%%%%%%%%%%%%%%%%%%%%%%%
%\begin{wraptable}{R}{0.50\textwidth}
\begin{table*}[t!]
\vskip-15pt
\caption{Select parameters of the GRAIL mission (some taken from Ref.~\cite{Park-etal:2012}) and the Earth-Moon system, along with corresponding symbols and approximate formulae used in the text.\label{tb:params}}
\begin{tabular}{|l|c|c|}\hline
Parameter  &Symbol(s)  &Values used\\\hline\hline
\multicolumn{3}{|l|}{\tt GRAIL Mission}\\\hline
Inter-spacecraft range &$d_{\rm AB}$ &\phantom{0}200~km\phantom{/s}\\
Inter-spacecraft range-rate &$\dot d_{\rm AB}=(\vec{n}_{\rm AB}\cdot\vec{v}_{\rm AB})$&\phantom{000}2~m/s\phantom{k}\\
Lunar altitude &$h_{\rm G}$ &\phantom{00}55~km\phantom{/s}\\
Lunicentric velocity      & $v_{\rm A0} =|{\vec v}_{\rm A0}|\simeq |{\vec v}_{\rm B0}|$ &\phantom{.}1.65~km/s\phantom{}\\
Relative spacecraft velocity &$v_{\rm AB}\simeq v_{\rm A} d_{\rm AB}/(R_{\rm M}+h_{\rm G})$&\phantom{0.}185~m/s
\phantom{k}\\
Lunicentric acceleration & $a_{\rm A0} =|{\vec a}_{\rm A0}| \simeq |{\vec a}_{\rm B0}|$ &\phantom{0}1.53~m/s$^2$\phantom{k}\\
Relative spacecraft acceleration &$a_{\rm AB}\simeq a_{\rm A} d_{\rm AB}/(R_{\rm M}+h_{\rm G})$&\phantom{0.}0.17~m/s$^2$
\phantom{k}\\
Ka-band frequency &$f_{\rm A} \simeq f_{\rm B}$ & 32~GHz\\
Frequency difference &$\Delta f_{\rm AB}=f_{\rm B}-f_{\rm A}$                            &$\sim 10^3$~Hz\\\hline
\multicolumn{3}{|l|}{\tt Earth-Moon system}\\\hline
Moon's geocentric velocity&---&\phantom{00.}1~km/s\phantom{}\\
EMB orbital velocity      &---&\phantom{0.} 30~km/s\phantom{}\\
DSN geocentric velocity &---&\phantom{0.}465~m/s\phantom{k}\\
Earth mass parameter      &$GM_{\rm E}$&
$3.98\times 10^{14}$~m$^3$/s$^2$\\
Moon mass parameter       &$GM_{\rm M}$ &
$4.90\times 10^{12}$~m$^3$/s$^2$\\
Earth radius  &$R_{\rm E}$ & $6.371\times 10^6$~m\\
Moon radius   &$R_{\rm M}$ &$1.737\times 10^6$~m\\\hline
\end{tabular}
\end{table*}
\vskip-0pt
%\end{wraptable}
%%%%%%%%%%%%%%%%%%%%%%%%%%%%%%

The third term in Eq.~(\ref{eq:DOWR+0}) is the second order $(\sim1/c^2$) acceleration-dependent Sagnac effect. We evaluate this term in a manner similar to Eq.~(\ref{eq:Sagnac1}) and obtain the magnitude:
{}
\begin{eqnarray}
\frac{d_{\rm AB}}{2c^2}\frac{
\big((f_{\rm B}{\vec a}_{\rm B}-f_{\rm A}{\vec a}_{\rm A})\cdot{\vec d}_{\rm AB}\big)}{f_{\rm A}+f_{\rm B}}&=&d_{\rm AB}\frac{({\vec a}_{\rm AB}\cdot{\vec d}_{\rm AB})}{4c^2}-d_{\rm AB}\left(\frac{f_{\rm B}-f_{\rm A}}{f_{\rm A}+f_{\rm B}}\right)\frac{\big(({\vec a}_{\rm A}+{\vec a}_{\rm B})\cdot{\vec d}_{\rm AB}\big)}{4c^2}
\nonumber\\
&=&
(2\times 10^{-8}+1.2\times 10^{-14})~{\rm m}.
\label{eq:Sagnac2}
\end{eqnarray}
Thus, the entire third term in Eq.~(\ref{eq:DOWR+0}) may be safely omitted.

The fourth term on the right-hand side of Eq.~(\ref{eq:DOWR+0}) is the second order ($\sim 1/c^2$) Sagnac effect. As a result this term may be evaluated as
\begin{eqnarray}
&&\frac{d_{\rm AB}}{2c^2}\frac{f_{\rm A}\big({\vec v}^2_{\rm A}+({\vec n}_{\rm AB}\cdot{\vec v}_{\rm A})^2\big)+
f_{\rm B}\big({\vec v}^2_{\rm B}+({\vec n}_{\rm AB}\cdot{\vec v}_{\rm B})^2\big)}{f_{\rm A}+f_{\rm B}}=\frac{d_{\rm AB}}{4c^2}\Big({\vec v}^2_{\rm A}+({\vec n}_{\rm AB}\cdot{\vec v}_{\rm A})^2+{\vec v}^2_{\rm B}+({\vec n}_{\rm AB}\cdot{\vec v}_{\rm B})^2\Big)+\nonumber\\
&&
\hskip 30pt+\,
\frac{d_{\rm AB}}{4c^2}\left(\frac{f_{\rm B}-f_{\rm A}}{f_{\rm A}+f_{\rm B}}\right)\Big(\big(\vec{v}_{\rm AB}\cdot({\vec v}_{\rm B}+{\vec v}_{\rm A})\big)+({\vec n}_{\rm AB}\cdot{\vec v}_{\rm AB})\big({\vec n}_{\rm AB}\cdot({\vec v}_{\rm B}+\vec{v}_{\rm A})\big)\Big)=
(0.002 + 2\times 10^{-13})~{\rm m}.~~~~
\label{eq:Sagnac3}
\end{eqnarray}
Thus, the first term in (\ref{eq:Sagnac3}) must be kept in the model. One can further evaluate this term by representing the barycentirc velocities of the GRAIL twins as ${\vec v}_{\rm A}={\vec v}_{\rm M}+{\vec v}_{\rm A0}$ and ${\vec v}_{\rm B}={\vec v}_{\rm M}+{\vec v}_{\rm B0}$, where ${\vec v}_{\rm M}$ is the barycentric velocity of the Moon and ${\vec v}_{\rm A0}$ and ${\vec v}_{\rm B0}$ are the lunicentric velocities of the two orbiters. By doing this, one can see that there will be three terms, each of which is important for the GRAIL model. The term $\sim d_{\rm AB}(v_{\rm M}/c)^2$ contributes up to 2 mm to the DOWR. The term $\sim d_{\rm AB}(v_{\rm M} v_{\rm A0}/c^2)$ contributes up to 110~$\mu$m to the DOWR, and the last term $\sim d_{\rm AB}(v_{\rm A0}/c)^2$, also frequency-dependent, contributes up to 6~$\mu$m to this observable. Thus, each of these terms must be accounted for in the relativistic model of GRAIL observables.

The fifth term in Eq.~(\ref{eq:DOWR+0}) is the Shapiro gravitational time delay. Assuming a spacecraft altitude $h_{\rm G}=55$~km, this term contributes $(4GM_{\rm M}/c^2)(d_{\rm AB}/(r_{\rm A}+r_{\rm B}))=12~\mu$m to the DOWR and, thus, it may be accounted for in the range model in the following approximated form, keeping just the largest (12~$\mu$m) term:
\begin{equation}
\frac{2GM_{\rm M}}{c^2}\ln\Big[
\frac{r_{\rm A}+r_{\rm B}+d_{\rm AB}}{r_{\rm A}+r_{\rm B}-d_{\rm AB}}\Big]\approx\frac{4GM_{\rm M}}{c^2}\frac{d_{\rm AB}}{r_{\rm A}+r_{\rm B}}+\frac{4GM_{\rm M}}{3c^2}\frac{d^3_{\rm AB}}{(r_{\rm A}+r_{\rm B})^3}=12~\mu{\rm m}+1.3\times 10^{-8}~{\rm m}.
\label{eq:shapiro-r}
\end{equation}

Concerning the sixth term in Eq.~(\ref{eq:DOWR+0}), in Appendix~\ref{sec:int-error} we show that, for the times-scales of signal propagation realized on GRAIL ($d_{\rm AB}/c\simeq 1$ ms), this term is of the order of $1/c^4$ and contributes less than
$1\times10^{-15}(t-t_0)^2~{\rm m/s}^2$ to the DOWR. An acceleration error of this magnitude yields a range error of less than 1~$\mu$m over the course of 6 hours, and it is thus completely negligible.

The last term in Eq.~(\ref{eq:DOWR+0}) is of ${\cal O}(c^{-2})$. This term represents the phase ambiguity in the DOWR observable at $t_0$. A method dealing with this term was outlined in Ref.~\cite{TimingMemo:2010}. We denote this term as $\delta n_0\equiv \delta n_0(t_0)$ and keep it in the model.

As a result, Eq.~(\ref{eq:DOWR+0}) can be presented in the following simplified form:
{}
\begin{eqnarray}
R_{\rm dowr}(t)&=&d_{\rm AB}\Big\{1-\frac{({\vec v}_{\rm AB}\cdot{\vec n}_{\rm AB})}{2c}+
\frac{1}{4c^2}\Big({\vec v}^2_{\rm A}+({\vec n}_{\rm AB}\cdot{\vec v}_{\rm A})^2+{\vec v}^2_{\rm B}+({\vec n}_{\rm AB}\cdot{\vec v}_{\rm B})^2\Big)+ \frac{4GM_{\rm M}}{c^2(r_{\rm A}+r_{\rm B})}\Big\}
+{\cal O}(0.5~\mu{\rm m}).~~~~~
\label{eq:DOWR+}
\end{eqnarray}

Up to this point, we treated the start $t_0$ of the integration interval in Eq.~(\ref{eq:int-A}) as arbitrary. We now see that after negligible contributions are omitted, the start of the integration interval enters Eq.~(\ref{eq:DOWR+}) only in the form of the definition of the phase ambiguity $\delta n(t_0)$. As we indicated above, dealing with this term is discussed in Ref.~\cite{TimingMemo:2010}. Once the effects of this phase ambiguity are accounted for, our formulation of the instantaneous DOWR observable, in the form of Eq.~(\ref{eq:DOWR+}), becomes independent of the choice of the start of the integration interval, and thus $t_0$ is truly arbitrary, even as we maintain an instantaneous range accuracy better than 1~$\mu$m, as needed for the GRAIL mission.

From Fig.~\ref{fig:grail} we can see that the vectors ${\vec R}_{\rm A}$ and ${\vec R}_{\rm B}$ are given as
\begin{eqnarray}
{\vec R}_{\rm A} &=& \vec{x}_{\rm EM}+\vec{x}_{\rm M}+\vec{y}_{\rm A} \qquad {\rm and}\qquad {\vec R}_{\rm B} = \vec{x}_{\rm EM}+\vec{x}_{\rm M}+\vec{y}_{\rm B}.
\label{eq:Vect}
\end{eqnarray}
These vectors are measured simultaneously with the signal reception in TBD and are needed to compute Eq.~(\ref{eq:DOWR+}).

\subsection{Dual One-Way Range-Rate (DOWRR) observables on GRAIL}
\label{sec:dowrr}

To develop an analytical form for the DOWRR observable, we use Eq.~(\ref{eq:dn-ab+-+}) to express it as
{}
\begin{equation}
\dot n_{\rm AB}(t_{\rm B})= \frac{d n_{\rm AB}}{dt_{\rm B}}=f_{\rm B0}\Big(\frac{d\tau_{\rm B}}{dt_{\rm B}}\Big)-f_{\rm A0}\Big(\frac{d\tau_{\rm A}}{dt_{\rm A}}\Big)\frac{dt_{\rm A0}}{dt_{\rm B}}.
\label{eq:dn-AB-0}
\end{equation}
Using the notation $f_{\rm B}=f_{\rm B0}({d\tau_{\rm B}}/{dt_{\rm B}})$ and
$f_{\rm A}=f_{\rm A0}({d\tau_{\rm A}}/{dt_{\rm A}})$ for the coordinate  frequencies of the two clocks, we can present Eq.~(\ref{eq:dn-AB-0}) as
{}
\begin{equation}
\dot n_{\rm AB}(t_{\rm B})=f_{\rm B}-f_{\rm A}\frac{dt_{\rm A0}}{dt_{\rm B}}.
\label{eq:dn-AB-1}
\end{equation}

As with the DOWR, the second observable DOWRR deals with the signal propagation from $\rm B_0$ to $\rm A$ is derived in an analogous way. Similarly to Eq.~(\ref{eq:dn-AB-1}), at the time $t_{\rm A}$ on the spacecraft $\rm A$ we have
{}
\begin{equation}
\dot n_{\rm BA}(t_{\rm A})=f_{\rm A}-f_{\rm B}\frac{dt_{\rm B0}}{dt_{\rm A}},
\label{eq:dn-BA-1}
\end{equation}
where the time of signal's emission $t_{\rm B0}$ may be presented as a function of signal reception $t_{\rm A}$ as $t_{\rm B0}=t_{\rm B0}(t_{\rm A})$.

Following the procedure outlined in Ref.~\cite{TimingMemo:2010}, we assume that the LGRS clocks on both spacecraft are synchronized, such that $t_{\rm A}=t_{\rm B}=t$. We now can form a quantity that is called dual one-way range-rate (DOWRR):
\begin{equation}
{v}_{\rm dowrr}(t)=c\frac{\dot n_{\rm AB}(t)+\dot n_{\rm BA}(t)}{f_{\rm A}+f_{\rm B}}=c\Big(1-\frac{f_{\rm A}(dt_{\rm A0}/dt_{\rm B})+f_{\rm B}(dt_{\rm B0}/dt_{\rm A})}{f_{\rm A}+f_{\rm B}}\Big).
\label{eq:DOWRR0}
\end{equation}

The ratio of coordinate times ${dt_{\rm A0}}/{dt_{\rm B}}$ (and similarly ${dt_{\rm B0}}/{dt_{\rm A}}$) can be computed by differentiating the coordinate time transfer equation (\ref{eq:dn-lt}) for $t_{\rm B} - t_{\rm A0}=T_{\rm AB}(t_{\rm A0},t_{\rm B})$ (and similarly for $t_{\rm A} - t_{\rm B0}=T_{\rm BA}(t_{\rm A},t_{\rm B0})$) with respect to the reception time $t_{\rm B}$. This procedure was already performed in Appendix~\ref{sec:gr-delay}, resulting in Eq.~(\ref{eq:rat-AB0+}) for $dt_{\rm A0}/dt_{\rm B}$. From this equation, the ratio ${dt_{\rm B0}}/{dt_{\rm A}}$ is obtained by interchanging $\rm A \leftrightarrow \rm B$.

Substituting these results for ${dt_{\rm A0}}/{dt_{\rm B}}$ and ${dt_{\rm B0}}/{dt_{\rm A}}$ into Eq.~(\ref{eq:DOWRR0}) we obtain the following expression for ${v}_{\rm dowrr}$:
{}
\begin{eqnarray}
{v}_{\rm dowrr}(t)&=&({\vec n}_{\rm AB}\cdot {\vec v}_{\rm AB})-\frac{1}{c}\bigg\{\frac{\big((f_{\rm B}{\vec v}_{\rm B}-f_{\rm A}{\vec v}_{\rm A})\cdot {\vec v}_{\rm AB}\big)}{f_{\rm A}+f_{\rm B}}+\frac{\big((f_{\rm B}{\vec a}_{\rm B}-f_{\rm A}{\vec a}_{\rm A})\cdot {\vec d}_{\rm AB}\big)}{f_{\rm A}+f_{\rm B}}\bigg\}+\nonumber\\
&&\hskip 53pt+\,
\frac{1}{c^2}\bigg\{
\frac{f_{\rm A}({\vec n}_{\rm AB}\cdot {\vec v}_{\rm A})(\vec{v}_{\rm AB}\cdot\vec{v}_{\rm A})+f_{\rm B}({\vec n}_{\rm AB}\cdot {\vec v}_{\rm B})(\vec{v}_{\rm AB}\cdot\vec{v}_{\rm B})}{f_{\rm A}+f_{\rm B}}\bigg\}-
\nonumber\\
&&\hskip 53pt-\,
\frac{4GM_{\rm M}}{c^2}\frac{d^{}_{\rm AB}}{(r_{\rm A}+r_{\rm B})^2}\Big(({\vec n}_{\rm A}\cdot {\vec v}_{\rm A})+({\vec n}_{\rm B}\cdot {\vec v}_{\rm B})\Big)+{\cal O}(c^{-2}).
\label{eq:DOWRR022}
\end{eqnarray}

The first term in Eq.~(\ref{eq:DOWRR022}) is the first order
($\sim1/c$) Doppler term, which may be as high as 2~m/s (see Table~\ref{tb:params}); it clearly must be kept in the model.

The second term on the right hand side of Eq.~(\ref{eq:DOWRR022}) can be evaluated using Eq.~(\ref{eq:ident1}) as
\begin{eqnarray}
\frac{\big((f_{\rm B}{\vec v}_{\rm B}-f_{\rm A}{\vec v}_{\rm A})\cdot{\vec v}_{\rm AB}\big)}{c(f_{\rm A}+f_{\rm B})}&=&\frac{{\vec v}^2_{\rm AB}}{2c}+\left(\frac{f_{\rm B}-f_{\rm A}}{f_{\rm A}+f_{\rm B}}\right)\frac{\big(({\vec v}_{\rm A}+{\vec v}_{\rm B})\cdot{\vec v}_{\rm AB}\big)}{2c}= (5.2\times 10^{-5}+6\times 10^{-10})~{\rm m/s}.
\label{eq:dowrr22}
\end{eqnarray}
Therefore, the second term in Eq.~(\ref{eq:dowrr22}) can be dropped, but the $\vec{v}_{\rm AB}^2/2c$ term must be kept in the model.

The third term in Eq.~(\ref{eq:DOWRR022}) is the second order $(\sim1/c^2$) acceleration-dependent Sagnac effect. To evaluate this term, we note that the acceleration vectors of the spacecraft point in different directions due to the $\sim$200~km separation between the two craft. The vector difference can be calculated as $\vec{a}_{\rm AB}\simeq 0.17$~m/s$^2$. We evaluate this term in a  manner similar to Eq.~(\ref{eq:Sagnac2}) to obtain a magnitude of
\begin{equation}
\frac{
\big((f_{\rm B}{\vec a}_{\rm B}-f_{\rm A}{\vec a}_{\rm A})\cdot{\vec d}_{\rm AB}\big)}{c(f_{\rm A}+f_{\rm B})}=\frac{({\vec a}_{\rm AB}\cdot{\vec d}_{\rm AB})}{2c}+\left(\frac{f_{\rm B}-f_{\rm A}}{f_{\rm A}+f_{\rm B}}\right)\frac{\big(({\vec a}_{\rm A}+{\vec a}_{\rm B})\cdot{\vec d}_{\rm AB}\big)}{2c}=
(6\times 10^{-5}+4\times 10^{-11})~{\rm m/s}.~~~
\label{eq:dowrr33}
\end{equation}
We see that the second term in Eq.~(\ref{eq:dowrr33}) is less than the needed accuracy of 1~$\mu$m/s and it can be omitted; the $(\vec{a}_{\rm AB}\cdot\vec{d}_{\rm AB})/2c$ term, however, must be kept in the model.

The $(1/c^2)$ term on the second line of (\ref{eq:DOWRR022}) can be presented as
{}
\begin{eqnarray}
&&\hskip -14pt \frac{1}{c^2}\Big\{
\frac{f_{\rm A}({\vec n}_{\rm AB}\cdot {\vec v}_{\rm A})(\vec{v}_{\rm AB}\cdot\vec{v}_{\rm A})+f_{\rm B}({\vec n}_{\rm AB}\cdot {\vec v}_{\rm B})(\vec{v}_{\rm AB}\cdot\vec{v}_{\rm B})}{f_{\rm A}+f_{\rm B}}\Big\}=
\frac{1}{2c^2}\Big\{({\vec n}_{\rm AB}\cdot {\vec v}_{\rm A})(\vec{v}_{\rm AB}\cdot\vec{v}_{\rm A})+({\vec n}_{\rm AB}\cdot {\vec v}_{\rm B})(\vec{v}_{\rm AB}\cdot\vec{v}_{\rm B})\Big\}+
\nonumber\\
&&\hskip 25pt+\,
\frac{1}{2c^2}\left(\frac{f_{\rm B}-f_{\rm A}}{f_{\rm B}+f_{\rm A}}\right)\Big\{({\vec n}_{\rm AB}\cdot {\vec v}_{\rm B})(\vec{v}_{\rm AB}\cdot\vec{v}_{\rm B})-({\vec n}_{\rm AB}\cdot {\vec v}_{\rm A})(\vec{v}_{\rm AB}\cdot\vec{v}_{\rm A})\Big\}=
\,(2\times 10^{-6}+6\times 10^{-14}) ~{\rm m/s}.
\label{eq:DOWRR022+0}
\end{eqnarray}
Therefore, only the first of the two terms on the right-hand side of Eq.~(\ref{eq:DOWRR022+0}) must be retained.

As a result, given the strict formation configuration implemented on the GRAIL mission, the model for DOWRR on GRAIL given by Eq.~(\ref{eq:DOWRR022}) has the following form:
{}
\begin{eqnarray}
{v}_{\rm dowrr}(t)&=&({\vec n}_{\rm AB}\cdot {\vec v}_{\rm AB})
-\frac{1}{2c}\Big\{\vec{v}_{\rm AB}^2+(\vec{a}_{\rm AB}\cdot\vec{d}_{\rm AB})\Big\} +
\frac{1}{2c^2}\Big\{({\vec n}_{\rm AB}\cdot {\vec v}_{\rm A})(\vec{v}_{\rm AB}\cdot\vec{v}_{\rm A})+({\vec n}_{\rm AB}\cdot {\vec v}_{\rm B})(\vec{v}_{\rm AB}\cdot\vec{v}_{\rm B})\Big\}-
\nonumber\\
&&\hskip 54pt -\,
\frac{4GM_{\rm M}}{c^2}\frac{d^{}_{\rm AB}}{(r_{\rm A}+r_{\rm B})^2}\Big(({\vec n}_{\rm A}\cdot {\vec v}_{\rm A})+({\vec n}_{\rm B}\cdot {\vec v}_{\rm B})\Big)+{\cal O}(0.1~\mu{\rm m}/{\rm s}).~~~~~\,
\label{eq:DOWRR022+=}
\end{eqnarray}

Equation (\ref{eq:DOWRR022+=}) represents the instantaneous DOWRR observable for the GRAIL mission, developed to a level of accuracy better than 1~$\mu$m/s. One can verify that the result given in Eq.~(\ref{eq:DOWRR022+=}) may be obtained directly from Eq.~(\ref{eq:DOWR+}) by simply differentiating Eq.~(\ref{eq:DOWR+}) with respect to time and retaining terms to the appropriate order.

\section{Conclusions and recommendations}
\label{sec:sonc}

We considered the formulation of a relativistic model for the observables of the GRAIL mission. We addressed some practical aspects of implementing the relevant computations. We derived an analytic expression that characterizes the process of forming the Ka-band ranging observables of GRAIL and developed a model for the dual one-way range (DOWR) observable. We also briefly addressed the transformation of relativistic gravitational potentials. This material can be used to improve the accuracy of modeling of the GRAIL fundamental observables.

We presented a hierarchy of relativistic coordinate reference frames that are needed to GRAIL. In this respect, we introduced the barycentric (BCRS), geocentric (GCRS), topocentric (TCRS), lunicentric (LCRS) and spacecraft (SCRS) coordinate reference systems, together with the structure of the corresponding metric tensors in each of these systems and the form of the proper relativistic gravitational potentials---all presented at the accuracy required for GRAIL. We advocate a definition for the LCRS with its proper time, which we call the TCL. We presented the rules for transforming time and position measurements between the reference frames involved.

The formula given by Eq.~(\ref{eq:DOWR+}) is the main result of this paper. It is derived for the first time at this high level of accuracy including the terms of the $1/c^2$ order. The final expression (\ref{eq:DOWR+}) is relatively simple and easy to utilize in practice. The equations we provide for time and frequency transfers are accurate to the level of $1~\mu$m when used to analyze GRAIL ranging data. Modeling the DOWR observable at this level of accuracy is the most important priority for the mission and must be taken into account for the science data analysis.

Most of the relativistic computations for GRAIL are done implicitly and are based on the models and tools available within the framework of JPL's Multiple Interferometric Ranging Analysis and GPS Ensemble (MIRAGE) software \cite{Park-etal:2012}. General relativistic equations of motion form the ``back-bone'' of the entire suite of models in MIRAGE and rely on the formulation given in  Ref.~\cite{Moyer:2003}. To navigate the GRAIL spacecraft, the code transforms the proper time of each of the GRAIL spacecraft to the time based on the SSB frame  and integrates the spacecraft's barycentric equations of motion. To determine the inter-spacecraft range, the code then iteratively solves the barycentric light-time equations in terms of instantaneous distance (by recomputing the transmitter's position bearing in mind the elapsed light-time) in the presence of the Shapiro term. The analytical closed-form solution for DOWR (\ref{eq:DOWR+}) is not only more elegant, it allows for direct investigation of the observables and possible error terms under various circumstances in data analysis.

We also developed a similarly accurate formulation for the DOWRR observable. Equation~(\ref{eq:DOWRR022+=}) allows us to calculate the value of this observable with an accuracy that is significantly better than 1~$\mu$m/s. In the form presented, Eqs.~(\ref{eq:DOWR+}) and (\ref{eq:DOWRR022+=}) can be readily incorporated into computer code used to model orbits and radio-metric observables. All the quantities in these equations are directly computable once the numerical positions and velocities of the spacecraft and Solar System bodies are known. It is also relatively straightforward to compute partial derivatives of these equations with respect to the the GCRS, which facilitates their use in fast numerical integration codes and optimizing solvers. Lastly, although we presented ideal, noise-free solutions, one can add relevant noise sources, including those in Ref.~\cite{TimingMemo:2010}.

In a practical sense, the small relativistic terms that we calculated are easily absorbed into constant and periodic ad-hoc biases that are introduced during data analysis, with no impact whatsoever on mission objectives or the quality of the mission's results. Yet the existence of these terms, and the fact that they are observable at the level of sensitivity of the GRAIL mission demonstrate that GRAIL is already a practical instrument for relativistic geodesy, especially after the mission is complete and all data sets for the primary and extended mission phases are assembled \cite{Zuber2012,Asmar2012}. For future spacecraft that operate at even greater accuracy, accounting for these relativistic terms will be essential.

Although this paper was aimed specifically at discussing the range and range-rate observables of the GRAIL mission, we note that the solutions presented here are also applicable to other, similar missions. Foremost comes to mind the GRACE with a mission design very similar to that of GRAIL. Indeed, the calculations presented here may help shed light on the origin of small residual terms that were seen in the GRACE range and range rate observables \cite{Bertiger2002}. We will further investigate this possibility with results to be reported elsewhere. Clearly, the model to be developed for the GRACE Follow-on mission \cite{Loomis:2009,Watkins-etal:2011} must include similar higher-order terms to reach the anticipated DOWR and DOWRR at the level of few nm and nm/s correspondingly. We will address these issues in a subsequent publications.

\begin{acknowledgments}
We thank Sami W. Asmar, William M. Folkner, Nathaniel E. Harvey, Alexander S. Konopliv, Gerhard L. Kruizinga, Ryan S. Park, Michael M. Watkins, James G. Williams, Dah-Ning Yuan of JPL and Maria T. Zuber of MIT for their interest and support during the work and preparation of this manuscript. We also thank Sergey M. Kopeikin and Sergey A. Klioner for their insightful comments and suggestions. We also thank the anonymous referee for valuable comments on this manuscript. This work was performed at the Jet Propulsion Laboratory, California Institute of Technology, under a contract with the National Aeronautics and Space Administration.
\end{acknowledgments}

%\bibliography{grail-gr}

\begin{thebibliography}{35}
\expandafter\ifx\csname natexlab\endcsname\relax\def\natexlab#1{#1}\fi
\expandafter\ifx\csname bibnamefont\endcsname\relax
  \def\bibnamefont#1{#1}\fi
\expandafter\ifx\csname bibfnamefont\endcsname\relax
  \def\bibfnamefont#1{#1}\fi
\expandafter\ifx\csname citenamefont\endcsname\relax
  \def\citenamefont#1{#1}\fi
\expandafter\ifx\csname url\endcsname\relax
  \def\url#1{\texttt{#1}}\fi
\expandafter\ifx\csname urlprefix\endcsname\relax\def\urlprefix{URL }\fi
\providecommand{\bibinfo}[2]{#2}
\providecommand{\eprint}[2][]{\url{#2}}

\bibitem[{\citenamefont{Kim}(2000)}]{Kim:2000}
\bibinfo{author}{\bibfnamefont{J.~R.} \bibnamefont{Kim}}, Ph.D. thesis,
  \bibinfo{school}{University of Texas at Austin} (\bibinfo{year}{2000}).

\bibitem[{\citenamefont{Roncoli and Fujii}(2010)}]{Roncoli-Fujii:2010}
\bibinfo{author}{\bibfnamefont{R.}~\bibnamefont{Roncoli}} \bibnamefont{and}
  \bibinfo{author}{\bibfnamefont{K.}~\bibnamefont{Fujii}},
  \bibinfo{journal}{{AIAA Guidance, Navigation, and Control Conference,
  Toronto, AIAA Paper 2010-8383}}  (\bibinfo{year}{2010}).

\bibitem[{\citenamefont{Park et~al.}(2012)\citenamefont{Park, Asmar,
  Fahnestock, Konopliv, Lu, and Watkins}}]{Park-etal:2012}
\bibinfo{author}{\bibfnamefont{R.~S.} \bibnamefont{Park}},
  \bibinfo{author}{\bibfnamefont{S.~W.} \bibnamefont{Asmar}},
  \bibinfo{author}{\bibfnamefont{E.~G.} \bibnamefont{Fahnestock}},
  \bibinfo{author}{\bibfnamefont{A.~S.} \bibnamefont{Konopliv}},
  \bibinfo{author}{\bibfnamefont{W.}~\bibnamefont{Lu}}, \bibnamefont{and}
  \bibinfo{author}{\bibfnamefont{M.~M.} \bibnamefont{Watkins}},
  \bibinfo{journal}{J. of Spacecraft and Rockets}
  \textbf{\bibinfo{volume}{49}}, \bibinfo{pages}{390} (\bibinfo{year}{2012}).

\bibitem[{\citenamefont{{Zuber} et~al.}(2012)\citenamefont{{Zuber}, {Smith},
  {Lehman}, {Hoffman}, {Asmar}, and {Watkins}}}]{Zuber2012}
\bibinfo{author}{\bibfnamefont{M.~T.} \bibnamefont{{Zuber}}},
  \bibinfo{author}{\bibfnamefont{D.~E.} \bibnamefont{{Smith}}},
  \bibinfo{author}{\bibfnamefont{D.~H.} \bibnamefont{{Lehman}}},
  \bibinfo{author}{\bibfnamefont{T.~L.} \bibnamefont{{Hoffman}}},
  \bibinfo{author}{\bibfnamefont{S.~W.} \bibnamefont{{Asmar}}},
  \bibnamefont{and} \bibinfo{author}{\bibfnamefont{M.~M.}
  \bibnamefont{{Watkins}}}, \bibinfo{journal}{Space Sci. Rev., submitted}
  (\bibinfo{year}{2012}).

\bibitem[{\citenamefont{{Asmar} et~al.}(2012)\citenamefont{{Asmar}, {Konopliv},
  {Park}, {Watkins}, {Kruizinga}, {Zuber}, {Smith}, {Williams}, {Paik}, {Yuan}
  et~al.}}]{Asmar2012}
\bibinfo{author}{\bibfnamefont{S.~W.} \bibnamefont{{Asmar}}},
  \bibinfo{author}{\bibfnamefont{A.~S.} \bibnamefont{{Konopliv}}},
  \bibinfo{author}{\bibfnamefont{S.~R.} \bibnamefont{{Park}}},
  \bibinfo{author}{\bibfnamefont{M.~M.} \bibnamefont{{Watkins}}},
  \bibinfo{author}{\bibfnamefont{G.}~\bibnamefont{{Kruizinga}}},
  \bibinfo{author}{\bibfnamefont{M.~T.} \bibnamefont{{Zuber}}},
  \bibinfo{author}{\bibfnamefont{D.~E.} \bibnamefont{{Smith}}},
  \bibinfo{author}{\bibfnamefont{J.~G.} \bibnamefont{{Williams}}},
  \bibinfo{author}{\bibfnamefont{M.}~\bibnamefont{{Paik}}},
  \bibinfo{author}{\bibfnamefont{D.-N.} \bibnamefont{{Yuan}}},
  \bibnamefont{et~al.}, \bibinfo{journal}{Space Sci. Rev., submitted}
  (\bibinfo{year}{2012}).

\bibitem[{\citenamefont{Kruizinga and Bertiger}(2009)}]{TimingMemo:2010}
\bibinfo{author}{\bibfnamefont{G.~L.~H.} \bibnamefont{Kruizinga}}
  \bibnamefont{and} \bibinfo{author}{\bibfnamefont{W.~I.}
  \bibnamefont{Bertiger}}, \bibinfo{journal}{{GRAIL Project Memorandum}}
  (\bibinfo{year}{2009}).

\bibitem[{\citenamefont{Soffel et~al.}(2003)\citenamefont{Soffel, Klioner,
  Petit, Kopeikin, Bretagnon, Brumberg, Capitaine, Damour, Fukushima, Guinot
  et~al.}}]{Soffel-etal:2003}
\bibinfo{author}{\bibfnamefont{M.}~\bibnamefont{Soffel}},
  \bibinfo{author}{\bibfnamefont{S.~A.} \bibnamefont{Klioner}},
  \bibinfo{author}{\bibfnamefont{G.}~\bibnamefont{Petit}},
  \bibinfo{author}{\bibfnamefont{S.~M.} \bibnamefont{Kopeikin}},
  \bibinfo{author}{\bibfnamefont{P.}~\bibnamefont{Bretagnon}},
  \bibinfo{author}{\bibfnamefont{V.~A.} \bibnamefont{Brumberg}},
  \bibinfo{author}{\bibfnamefont{N.}~\bibnamefont{Capitaine}},
  \bibinfo{author}{\bibfnamefont{T.}~\bibnamefont{Damour}},
  \bibinfo{author}{\bibfnamefont{T.}~\bibnamefont{Fukushima}},
  \bibinfo{author}{\bibfnamefont{B.}~\bibnamefont{Guinot}},
  \bibnamefont{et~al.}, \bibinfo{journal}{Astron. J.}
  \textbf{\bibinfo{volume}{126}}, \bibinfo{pages}{2687} (\bibinfo{year}{2003}).

\bibitem[{\citenamefont{{McCarthy}}(2010)}]{IERS2010}
\bibinfo{author}{\bibfnamefont{D.~D.} \bibnamefont{{McCarthy}}},
  \bibinfo{type}{Tech. Rep.}, \bibinfo{institution}{{U.S. Naval Observatory}}
  (\bibinfo{year}{2010}), \bibinfo{note}{{IERS Conventions}},
  \urlprefix\url{http://www.iers.org/IERS/EN/Publications/TechnicalNotes/tn36.%
html}.

\bibitem[{\citenamefont{{Kopeikin} and {Xie}}(2010)}]{Kopeikin2010}
\bibinfo{author}{\bibfnamefont{S.}~\bibnamefont{{Kopeikin}}} \bibnamefont{and}
  \bibinfo{author}{\bibfnamefont{Y.}~\bibnamefont{{Xie}}},
  \bibinfo{journal}{Celestial Mechanics and Dynamical Astronomy}
  \textbf{\bibinfo{volume}{108}}, \bibinfo{pages}{245} (\bibinfo{year}{2010}).

\bibitem[{\citenamefont{{Klioner}}(2004)}]{Klioner2004}
\bibinfo{author}{\bibfnamefont{S.~A.} \bibnamefont{{Klioner}}},
  \bibinfo{journal}{\prd} \textbf{\bibinfo{volume}{69}}, \bibinfo{eid}{124001}
  (\bibinfo{year}{2004}), \eprint{arXiv:0311540 [astro-ph]}.

\bibitem[{\citenamefont{Klioner}(2008)}]{Klioner:2008}
\bibinfo{author}{\bibfnamefont{S.~A.} \bibnamefont{Klioner}},
  \bibinfo{journal}{Astron. Astrophys.} \textbf{\bibinfo{volume}{478}},
  \bibinfo{pages}{951} (\bibinfo{year}{2008}).

\bibitem[{\citenamefont{Klioner et~al.}(2010)\citenamefont{Klioner, Capitaine,
  Folkner, Guinot, Huang, Kopeikin, Petit, Pitjeva, Seidelmann, and
  Soffel}}]{Klioner-etal:2010}
\bibinfo{author}{\bibfnamefont{S.~A.} \bibnamefont{Klioner}},
  \bibinfo{author}{\bibfnamefont{N.}~\bibnamefont{Capitaine}},
  \bibinfo{author}{\bibfnamefont{W.}~\bibnamefont{Folkner}},
  \bibinfo{author}{\bibfnamefont{B.}~\bibnamefont{Guinot}},
  \bibinfo{author}{\bibfnamefont{T.~Y.} \bibnamefont{Huang}},
  \bibinfo{author}{\bibfnamefont{S.}~\bibnamefont{Kopeikin}},
  \bibinfo{author}{\bibfnamefont{G.}~\bibnamefont{Petit}},
  \bibinfo{author}{\bibfnamefont{E.}~\bibnamefont{Pitjeva}},
  \bibinfo{author}{\bibfnamefont{P.~K.} \bibnamefont{Seidelmann}},
  \bibnamefont{and} \bibinfo{author}{\bibfnamefont{M.}~\bibnamefont{Soffel}},
  in \emph{\bibinfo{booktitle}{{Relativity in Fundamental Astronomy: Dynamics,
  Reference Frames, and Data Analysis}}}, edited by
  \bibinfo{editor}{\bibfnamefont{S.}~\bibnamefont{Klioner}},
  \bibinfo{editor}{\bibfnamefont{P.~K.} \bibnamefont{Seidelmann}},
  \bibnamefont{and} \bibinfo{editor}{\bibfnamefont{M.}~\bibnamefont{Soffel}}
  (\bibinfo{publisher}{Cambridge University Press}, \bibinfo{year}{2010}).

\bibitem[{\citenamefont{Kopeikin et~al.}(2011)\citenamefont{Kopeikin,
  Efroimsky, and Kaplan}}]{Kopeikin-book}
\bibinfo{author}{\bibfnamefont{S.~M.} \bibnamefont{Kopeikin}},
  \bibinfo{author}{\bibfnamefont{M.}~\bibnamefont{Efroimsky}},
  \bibnamefont{and} \bibinfo{author}{\bibfnamefont{G.}~\bibnamefont{Kaplan}},
  \emph{\bibinfo{title}{{Relativistic Celestial Mechanics of the Solar
  System}}} (\bibinfo{publisher}{Wiley-VCH}, \bibinfo{year}{2011}).

\bibitem[{\citenamefont{Folkner et~al.}(2009)\citenamefont{Folkner, Williams,
  and Boggs}}]{FolknerWilliamsBoggs:2009}
\bibinfo{author}{\bibfnamefont{W.~M.} \bibnamefont{Folkner}},
  \bibinfo{author}{\bibfnamefont{J.~G.} \bibnamefont{Williams}},
  \bibnamefont{and} \bibinfo{author}{\bibfnamefont{D.~H.} \bibnamefont{Boggs}},
  \bibinfo{journal}{IPN Progress Report} \textbf{\bibinfo{volume}{42-178}},
  \bibinfo{pages}{1} (\bibinfo{year}{2009}), \eprint{{\tt
  http://ipnpr.jpl.nasa.gov/}\hspace{0pt}{\tt
  progress\_report/42-178/178C.pdf}}.

\bibitem[{\citenamefont{Turyshev and Toth}(2012)}]{Turyshev:2012}
\bibinfo{author}{\bibfnamefont{S.~G.} \bibnamefont{Turyshev}} \bibnamefont{and}
  \bibinfo{author}{\bibfnamefont{V.~T.} \bibnamefont{Toth}},
  \bibinfo{journal}{in preparation}  (\bibinfo{year}{2012}).

\bibitem[{\citenamefont{Turyshev et~al.}(2012)\citenamefont{Turyshev,
  Minazzoli, and Toth}}]{Turyshev-etal:2012}
\bibinfo{author}{\bibfnamefont{S.~G.} \bibnamefont{Turyshev}},
  \bibinfo{author}{\bibfnamefont{O.~L.} \bibnamefont{Minazzoli}},
  \bibnamefont{and} \bibinfo{author}{\bibfnamefont{V.~T.} \bibnamefont{Toth}},
  \bibinfo{journal}{J. Math. Phys.} \textbf{\bibinfo{volume}{53}},
  \bibinfo{pages}{032501} (\bibinfo{year}{2012}), \eprint{arXiv:1109.1796
  [gr-qc]}.

\bibitem[{\citenamefont{Einstein et~al.}(1938)\citenamefont{Einstein, Infeld,
  and Hoffmann}}]{EIH1938}
\bibinfo{author}{\bibfnamefont{A.}~\bibnamefont{Einstein}},
  \bibinfo{author}{\bibfnamefont{L.}~\bibnamefont{Infeld}}, \bibnamefont{and}
  \bibinfo{author}{\bibfnamefont{B.}~\bibnamefont{Hoffmann}},
  \bibinfo{journal}{The Annals of Mathematics} \textbf{\bibinfo{volume}{39}},
  \bibinfo{pages}{65} (\bibinfo{year}{1938}).

\bibitem[{\citenamefont{Moyer}(2003)}]{Moyer:2003}
\bibinfo{author}{\bibfnamefont{T.~D.} \bibnamefont{Moyer}},
  \emph{\bibinfo{title}{{Formulation for Observed and Computed Values of Deep
  Space Network Data Types for Navigation}}}, JPL Deep-Space Communications and
  Navigation Series (\bibinfo{publisher}{Wiley-Interscience},
  \bibinfo{year}{2003}).

\bibitem[{\citenamefont{Montenbruck and Gill}(2005)}]{MG2005}
\bibinfo{author}{\bibfnamefont{O.}~\bibnamefont{Montenbruck}} \bibnamefont{and}
  \bibinfo{author}{\bibfnamefont{B.}~\bibnamefont{Gill}},
  \emph{\bibinfo{title}{Satellite Orbits}} (\bibinfo{publisher}{Springer},
  \bibinfo{year}{2005}), \bibinfo{edition}{3rd} ed.

\bibitem[{\citenamefont{Turyshev}(2008)}]{Turyshev:2008}
\bibinfo{author}{\bibfnamefont{S.~G.} \bibnamefont{Turyshev}},
  \bibinfo{journal}{Ann. Rev. Nucl. Part. Sci.} \textbf{\bibinfo{volume}{58}},
  \bibinfo{pages}{207} (\bibinfo{year}{2008}), \eprint{arXiv:0806.1731
  [gr-qc]}.

\bibitem[{\citenamefont{Standish and Williams}(in press,
  2012)}]{Standish-Williams:2012}
\bibinfo{author}{\bibfnamefont{E.~M.} \bibnamefont{Standish}} \bibnamefont{and}
  \bibinfo{author}{\bibfnamefont{J.~G.} \bibnamefont{Williams}}
  (\bibinfo{publisher}{Mill Valley: University Science Books},
  \bibinfo{year}{in press, 2012}), Explanatory Supplement to the American
  Ephemeris and Nautical Almanac, P.~K. Seidelmann, ed.,
  chap.~\bibinfo{chapter}{8}.

\bibitem[{\citenamefont{{Teyssandier} and {Le
  Poncin-Lafitte}}(2008)}]{Teyssandier2008}
\bibinfo{author}{\bibfnamefont{P.}~\bibnamefont{{Teyssandier}}}
  \bibnamefont{and} \bibinfo{author}{\bibfnamefont{C.}~\bibnamefont{{Le
  Poncin-Lafitte}}}, \bibinfo{journal}{Classical and Quantum Gravity}
  \textbf{\bibinfo{volume}{25}}, \bibinfo{pages}{145020}
  (\bibinfo{year}{2008}), \eprint{arXiv:0803.0277 [gr-qc]}.

\bibitem[{\citenamefont{{Ashby} and {Bertotti}}(2010)}]{Ashby2010}
\bibinfo{author}{\bibfnamefont{N.}~\bibnamefont{{Ashby}}} \bibnamefont{and}
  \bibinfo{author}{\bibfnamefont{B.}~\bibnamefont{{Bertotti}}},
  \bibinfo{journal}{Classical and Quantum Gravity}
  \textbf{\bibinfo{volume}{27}}, \bibinfo{pages}{145013}
  (\bibinfo{year}{2010}), \eprint{arXiv:0912.2705 [gr-qc]}.

\bibitem[{\citenamefont{Will}(2000)}]{Will}
\bibinfo{author}{\bibfnamefont{C.~M.} \bibnamefont{Will}},
  \emph{\bibinfo{title}{{Theory and experiment in gravitational physics}}}
  (\bibinfo{publisher}{Cambridge University Press}, \bibinfo{year}{2000}),
  \bibinfo{edition}{2nd} ed.

\bibitem[{\citenamefont{Williams et~al.}(2001)\citenamefont{Williams, Boggs,
  Yoder, Ratcliff, and Dickey}}]{Williams-etal:2001}
\bibinfo{author}{\bibfnamefont{J.~G.} \bibnamefont{Williams}},
  \bibinfo{author}{\bibfnamefont{D.~H.} \bibnamefont{Boggs}},
  \bibinfo{author}{\bibfnamefont{C.~F.} \bibnamefont{Yoder}},
  \bibinfo{author}{\bibfnamefont{J.~T.} \bibnamefont{Ratcliff}},
  \bibnamefont{and} \bibinfo{author}{\bibfnamefont{J.~O.}
  \bibnamefont{Dickey}}, \bibinfo{journal}{J. Geophys. Res.}
  \textbf{\bibinfo{volume}{106}}, \bibinfo{pages}{27933}
  (\bibinfo{year}{2001}).

\bibitem[{\citenamefont{Goossens and
  Matsumoto}(2008)}]{Goossens-Matsumoto:2008}
\bibinfo{author}{\bibfnamefont{S.}~\bibnamefont{Goossens}} \bibnamefont{and}
  \bibinfo{author}{\bibfnamefont{K.}~\bibnamefont{Matsumoto}},
  \bibinfo{journal}{Geophys. Res. Lett.} \textbf{\bibinfo{volume}{35}},
  \bibinfo{pages}{L02204} (\bibinfo{year}{2008}).

\bibitem[{\citenamefont{Tommei et~al.}(2010)\citenamefont{Tommei, Milani, and
  Vokrouhlicky}}]{Tommei-etal:2010}
\bibinfo{author}{\bibfnamefont{G.}~\bibnamefont{Tommei}},
  \bibinfo{author}{\bibfnamefont{A.}~\bibnamefont{Milani}}, \bibnamefont{and}
  \bibinfo{author}{\bibfnamefont{D.}~\bibnamefont{Vokrouhlicky}},
  \bibinfo{journal}{Celest. Mech. Dynam. Astron.}
  \textbf{\bibinfo{volume}{107}}, \bibinfo{pages}{285} (\bibinfo{year}{2010}).

\bibitem[{\citenamefont{Landau and Lifshitz}(1988)}]{Landau-Lifshitz:1988}
\bibinfo{author}{\bibfnamefont{L.~D.} \bibnamefont{Landau}} \bibnamefont{and}
  \bibinfo{author}{\bibfnamefont{E.~M.} \bibnamefont{Lifshitz}},
  \emph{\bibinfo{title}{{The Classical Theory of Fields {\rm (in Russian)}}}}
  (\bibinfo{publisher}{Nauka, Moscow}, \bibinfo{year}{1988}),
  \bibinfo{edition}{7th} ed.

\bibitem[{\citenamefont{Misner et~al.}(1973)\citenamefont{Misner, Thorne, and
  Wheeler}}]{MTW:1973}
\bibinfo{author}{\bibfnamefont{C.~W.} \bibnamefont{Misner}},
  \bibinfo{author}{\bibfnamefont{K.~S.} \bibnamefont{Thorne}},
  \bibnamefont{and} \bibinfo{author}{\bibfnamefont{J.~A.}
  \bibnamefont{Wheeler}}, \emph{\bibinfo{title}{{Gravitation}}}
  (\bibinfo{publisher}{W. H. Freeman \& Co. (San Francisco)},
  \bibinfo{year}{1973}).

\bibitem[{\citenamefont{Blanchet et~al.}(2001)\citenamefont{Blanchet, Salomon,
  Teyssandier, and Wolf}}]{Blanchet-etal:2001}
\bibinfo{author}{\bibfnamefont{L.}~\bibnamefont{Blanchet}},
  \bibinfo{author}{\bibfnamefont{C.}~\bibnamefont{Salomon}},
  \bibinfo{author}{\bibfnamefont{P.}~\bibnamefont{Teyssandier}},
  \bibnamefont{and} \bibinfo{author}{\bibfnamefont{P.}~\bibnamefont{Wolf}},
  \bibinfo{journal}{Astron. Astrophys.} \textbf{\bibinfo{volume}{370}},
  \bibinfo{pages}{320} (\bibinfo{year}{2001}).

\bibitem[{\citenamefont{Bertiger et~al.}(2002)\citenamefont{Bertiger,
  Bar-Sever, Bettadpur, Desai, Dunn, Haines, Kruizinga, Kuang, Nandi, Romans
  et~al.}}]{Bertiger2002}
\bibinfo{author}{\bibfnamefont{W.}~\bibnamefont{Bertiger}},
  \bibinfo{author}{\bibfnamefont{Y.}~\bibnamefont{Bar-Sever}},
  \bibinfo{author}{\bibfnamefont{S.}~\bibnamefont{Bettadpur}},
  \bibinfo{author}{\bibfnamefont{S.}~\bibnamefont{Desai}},
  \bibinfo{author}{\bibfnamefont{C.}~\bibnamefont{Dunn}},
  \bibinfo{author}{\bibfnamefont{B.}~\bibnamefont{Haines}},
  \bibinfo{author}{\bibfnamefont{G.}~\bibnamefont{Kruizinga}},
  \bibinfo{author}{\bibfnamefont{D.}~\bibnamefont{Kuang}},
  \bibinfo{author}{\bibfnamefont{S.}~\bibnamefont{Nandi}},
  \bibinfo{author}{\bibfnamefont{L.}~\bibnamefont{Romans}},
  \bibnamefont{et~al.}, \bibinfo{journal}{{Proceedings of ION GPS 2002,
  Portland OR}}  (\bibinfo{year}{2002}).

\bibitem[{\citenamefont{Loomis}(2009)}]{Loomis:2009}
\bibinfo{author}{\bibfnamefont{B.}~\bibnamefont{Loomis}}, Ph.D. thesis,
  \bibinfo{school}{University of Texas at Austin} (\bibinfo{year}{2009}).

\bibitem[{\citenamefont{{Watkins} et~al.}(2011)\citenamefont{{Watkins},
  {Flechtner}, {Morton}, {Gross}, and {Bettadpur}}}]{Watkins-etal:2011}
\bibinfo{author}{\bibfnamefont{M.~M.} \bibnamefont{{Watkins}}},
  \bibinfo{author}{\bibfnamefont{F.}~\bibnamefont{{Flechtner}}},
  \bibinfo{author}{\bibfnamefont{P.}~\bibnamefont{{Morton}}},
  \bibinfo{author}{\bibfnamefont{M.~A.} \bibnamefont{{Gross}}},
  \bibnamefont{and} \bibinfo{author}{\bibfnamefont{S.~V.}
  \bibnamefont{{Bettadpur}}}, \bibinfo{journal}{AGU Fall Meeting Abstracts}
  p.~\bibinfo{pages}{A7} (\bibinfo{year}{2011}).

\bibitem[{\citenamefont{Kopeikin}(2009)}]{Kopeikin:2009}
\bibinfo{author}{\bibfnamefont{S.~M.} \bibnamefont{Kopeikin}},
  \bibinfo{journal}{Mon. Not. R. Astron. Soc.} \textbf{\bibinfo{volume}{399}},
  \bibinfo{pages}{1539} (\bibinfo{year}{2009}).

\bibitem[{\citenamefont{Sazhin et~al.}(2010)\citenamefont{Sazhin, Vlasov,
  Sazhina, and Turyshev}}]{Sazhin-etal:2010}
\bibinfo{author}{\bibfnamefont{M.~V.} \bibnamefont{Sazhin}},
  \bibinfo{author}{\bibfnamefont{I.~Y.} \bibnamefont{Vlasov}},
  \bibinfo{author}{\bibfnamefont{O.~S.} \bibnamefont{Sazhina}},
  \bibnamefont{and} \bibinfo{author}{\bibfnamefont{V.~G.}
  \bibnamefont{Turyshev}}, \bibinfo{journal}{Astronomy Reports}
  \textbf{\bibinfo{volume}{54}}, \bibinfo{pages}{959} (\bibinfo{year}{2010}).

\end{thebibliography}

\appendix

\section{The phase of an electromagnetic signal in gravitational field}
\label{sec:phase-delay}

In this Appendix, we present derivations of the formulas use for time and frequency transfer in the GRAIL experiment. The derivation presented in this appendix is based on material that can be found in standard textbooks such as Refs.~\cite{MTW:1973} and \cite{Will}. A general solution is presented to the problem of light propagation in a gravitational field in the linearized approximation.

\subsection{General-relativistic post-Minkowskian space-time}
\label{sec:post-Mink}

To develop the solution to the equations of the general theory of relativity in the post-Minkowskian approximation, we introduce the post-Minkowskian decomposition of the metric tensor $g_{mn}$ as
\begin{equation}
g_{mn} = \gamma_{mn} + h_{mn}+{\cal O}(G^2),
\label{eq:g-linear}
\end{equation}
where $h_{mn}$ denotes the post-Minkowskian perturbation of the Minkowski metric tensor $\gamma_{mn}$. Following \cite{MTW:1973}, we impose the harmonic gauge condition on the metric tensor $g_{mn}$, given in the form
\begin{equation}
\partial_m\left(\sqrt{-g}g^{mn}\right)=0, \qquad\qquad {\rm or} \qquad\qquad
\partial_mh^{mn}-{\textstyle\frac{1}{2}}\partial^n h={\cal O}(G^2),
\label{eq:harmonic}
\end{equation}
where $h=\gamma_{kl}h^{kl}+{\cal O}(G^2)$. In the first post-Minkowskian approximation of general relativity \cite{MTW:1973,Turyshev:2008}, Einstein's equations $R^{mn}=16\pi G/c^4\big(T^{mn} -{\textstyle\frac{1}{2}}g^{mn}T\big)$ take the following form in arbitrary harmonic coordinates $\{x^m\} = (ct, {\vec x})$:
\begin{equation}
\left(\frac{1}{c^2}\frac{\partial^2}{\partial t^2}-\nabla^2\right)h^{mn}= \frac{16\pi G}{c^4}\left(
T^{mn} -{\textstyle\frac{1}{2}}\gamma^{mn}T\right)+{\cal O}(G^2),
\label{eq:GR-linear}
\end{equation}
where $T^{mn}$ is the stress-energy tensor describing a body that deflects a light ray and $T=\gamma_{kl}T^{kl}+{\cal O}(G)$. In the linearized approximation and neglecting the higher multipole moments, this tensor is given as \cite{Landau-Lifshitz:1988}:
\begin{equation}
T^{mn}(t,{\vec x})=Mu^m u^n\sqrt{1-\beta^2}\delta^3(\vec{x}-\vec{z}(t))+{\cal O}(G),
\label{eq:TEM}
\end{equation}
where $M$ is the rest mass of the body, $\vec z (t )$ its time-dependent spatial coordinate, ${\bf \beta} = c^{-1}d{\vec z}/dt$ and $\delta^3(\vec{x})$ is the three-dimensional Dirac delta function. The body's normalized four-velocity $u^m$, such that $u_mu^m=1$, is given by
\begin{equation}
u^m=(u^0,u^\alpha)=u^0\left(1,\frac{v^\alpha}{c}\right), \qquad u^0 =\frac{dt}{ds}= \frac{1}{\sqrt{1 - \beta^2}}+{\cal O}(G), \qquad u^\alpha = \frac{v^\alpha/c}{\sqrt{1 - \beta^2}}+{\cal O}(G).
\label{eq:4-vel}
\end{equation}
Note that in Eq.~(\ref{eq:TEM}) we have neglected the factor of $\sqrt{-g}$. This is done because in the linearized approximation $\sqrt{-g}=1 + {\cal O}(G)$ and the quadratic terms $\propto G^2$ are irrelevant in $T^{mn}$ since the corresponding time-dependent terms of the second post-Minkowskian order are currently unobservable in measurements made in the Solar System.

We can now write down the Green's function solution to Eqs.~(\ref{eq:GR-linear}):
\begin{equation}
h^{mn}(t,{\vec x})=\frac{4GM}{c^2}\int \sqrt{1-\beta'^2}\left({u'^mu'^n-{\textstyle\frac{1}{2}\gamma^{mn}}}\right)\delta^3(\vec{x'}-\vec{z}(t'))G(t,{\vec x}; t',{\vec x}')d^3{\vec x}'dt' +{\cal O}(G^2),
\label{eq:LW-pot-Green}
\end{equation}
where $G(t,{\vec x}; t',{\vec x}')$ is the Green's function
\begin{equation}
G(t,{\vec x}; t',{\vec x}')=G(t-t';{\vec x}-{\vec x}')=\frac{1}{4\pi}\frac{1}{|{\vec x}-{\vec x}'|}\delta\Big(t-t' - \frac{1}{c}|{\vec x}-{\vec x}'|\Big).
\label{eq:LW-Green-funct}
\end{equation}
Integrating Eq.~(\ref{eq:LW-pot-Green}), one obtains the post-Minkowskian metric tensor perturbation in terms of retarded Li\'enard-Wiechert tensor potentials \cite{MTW:1973,Kopeikin:2009}:
\begin{equation}
h^{mn}(t,{\vec x})=\frac{4GM}{c^2}\frac{u^mu^n-{\textstyle\frac{1}{2}\gamma^{mn}}}{u_mr^m}+{\cal O}(G),
\label{eq:LW-pot}
\end{equation}
where $r^m = x^m - z^m(t_{\rm ret})=[c(t-t_{\rm ret}), ({\vec x}-{\vec z}(t_{\rm ret})].$ In Eq.~(\ref{eq:LW-pot}), all time-dependent quantities are taken at a retarded time $t_{\rm ret}$ (defined by Eq.~(\ref{eq:t-ret}) below): $u^m \equiv u^m(t_{\rm ret}) = c^{-1}dz^m(t_{\rm ret})/dt_{\rm ret}$ is the body's four-velocity, ${\bf \beta}(t_{\rm ret}) = c^{-1}dz(t_{\rm ret})/dt_{\rm ret}$ is the body's coordinate velocity normalized to the speed of light $c$.

In the solution to Eq.~(\ref{eq:GR-linear}), given in terms of the retarded Li\'enard-Wiechert potentials, all quantities involved including the distance $r^m = x^m - z^m(t_{\rm ret})$, the body's world-line $z^m(t_{\rm ret}) = [ct_{\rm ret}, {\vec z}(t_{\rm ret})]$, and the four-velocity $u^m(t_{\rm ret})$ are functions of the retarded time $t_{\rm ret}$. It is known (see, for instance, \cite{Kopeikin:2009}) that the retarded time in the first post-Newtonian approximation may be found from the null-cone equation
\begin{equation}
\gamma_{mn} r^mr^n\equiv \gamma_{mn}[x^m - z^m(t_{\rm ret})][x^n - z^n(t_{\rm ret})]=0,
\label{eq:r-ret}
\end{equation}
suggesting that the retarded time $t_{\rm ret} = t_{\rm ret}(t , {\vec x})$ is established as a solution to the equation
\begin{equation}
t_{\rm ret}= t - \frac{1}{c}|x - z(t_{\rm ret})|.
\label{eq:t-ret}
\end{equation}
Note that Eq.~(\ref{eq:t-ret}) has an analytic solution only in the case of uniform motion of the gravitating body along a straight line.

\subsection{Phase of the electromagnetic wave}
\label{sec:em-phase}

The phase of an electromagnetic wave is a scalar function that is invariant under a set of general coordinate transformations. In the geometric optics approximation, the phase is found as a solution to the eikonal equation \cite{Landau-Lifshitz:1988,MTW:1973,Kopeikin:2009}:
\begin{equation}
g^{mn}\partial_m\varphi\partial_n\varphi=0,
\label{eq:eq_eik}
\end{equation}
with $g^{mn} = \gamma^{mn}-h^{mn}+{\cal O}(G^2)$. Equation~(\ref{eq:eq_eik}) is a direct consequence of Maxwell's equations. Its solution describes the front of an electromagnetic wave propagating in curved space-time. The solution's geometric properties are defined by the metric tensor (\ref{eq:g-linear}), where $h_{mn}$ (\ref{eq:LW-pot}) is the solution of the linearized Einstein equations (\ref{eq:GR-linear}) with stress-energy tensor (\ref{eq:TEM}).

To solve Eq.~(\ref{eq:eq_eik}), we introduce a covector of the electromagnetic wavefront in curved space-time, $K_m = \partial_m\varphi$. We use $\lambda$ to denote an affine parameter along the trajectory of a light ray being orthogonal to the wavefront $\varphi$. Vector $K^m = dx^m/d\lambda = g^{mn}\partial_n\varphi$ is tangent to the light ray. Equation~(\ref{eq:eq_eik}) states that the vector $K^m$ simply is null or $g_{mn}K^mK^n = 0$. Therefore, the light rays are null geodesics \cite{Landau-Lifshitz:1988} described by
\begin{equation}
\frac{dK_m}{d\lambda} = \frac{1}{2}\partial_m g_{kl}K^kK^l.
\label{eq:eq_eik-K}
\end{equation}
Since eikonal and light-ray equations, given by Eqs.~(\ref{eq:eq_eik}) and (\ref{eq:eq_eik-K}) respectively, have equivalent physical content in the general theory of relativity, one can use either of them to study the properties of an electromagnetic wave. However, the eikonal equation offers a more straightforward way to study the propagation of a wave. To find a solution of Eq.~(\ref{eq:eq_eik}), we expand the eikonal $\varphi$ with respect to the gravitational constant $G$ assuming that the unperturbed solution of Eq.~(\ref{eq:eq_eik}) is a plane wave. The expansion may be given as
\begin{equation}
\varphi(t,{\vec x}) = \varphi_0+\int k_m dx^m+\varphi_G (t,{\vec x})+{\cal O}(G^2),
\label{eq:eq_eik-phi}
\end{equation}
where $\varphi_0$ is an integration constant and $k^m = k^0(1, {\vec k})$ is a constant null vector (i.e., $\gamma_{mn}k^mk^n=0$) along the direction of propagation of the unperturbed electromagnetic wavefront. Furthermore, $k^0=\omega/c$ where $\omega$ is the constant frequency of the unperturbed wave, and $\varphi_G$ is the perturbation of the eikonal to ${\cal O}(G)$, which is yet to be determined.
Substituting Eqs.~(\ref{eq:LW-pot}) and (\ref{eq:eq_eik-K}) into (\ref{eq:eq_eik}) and keeping only first order terms in $G$, we obtain an ordinary differential equation to determine $\varphi_G$:
\begin{equation}
\frac{d\varphi_G}{d\lambda}= \frac{1}{2}h^{mn}k_mk_n= \frac{2GM}{c^2}\frac{(k_mu^m)^2}{u_mr^m} +{\cal O}(G^2),
\label{eq:eq_eik-phi-lamb}
\end{equation}
which alternatively can be obtained as a first integral of the null geodesic equation (\ref{eq:eq_eik-K}). We can now integrate (\ref{eq:eq_eik-phi-lamb}) while keeping in mind that $dk^m=0$ and employing the exact relationship \cite{Kopeikin:2009}:
\begin{equation}
\frac{d\lambda}{u_mr^m}= \frac{ds}{k_mr^m}= \frac{1}{k_mu^m}d\ln\left[k_mr^m\right].
\label{eq:eq_eik-relat}
\end{equation}
Neglecting the body's acceleration (or $du^m=0$), a plane-wave solution of Eq.~(\ref{eq:eq_eik-phi-lamb}) has the form
\begin{equation}
\varphi_G (t,{\vec x})= \frac{2GM}{c^2}(k_mu^m)\ln\left[k_mr^m\right],
\label{eq:eik-sol}
\end{equation}
where all quantities on the right-hand side are taken at the retarded instant of time $t_{\rm ret}$ in agreement with (\ref{eq:t-ret}). Therefore, we can now write the post-Minkowskian expansion for the phase of the electromagnetic wave as:
\begin{equation}
\varphi(t,{\vec x}) = \varphi_0+\int k_m dx^m+\frac{2GM}{c^2}(k_mu^m)\ln\left[k_mr^m\right]+{\cal O}(G^2),
\label{eq:eq_eik-phi+}
\end{equation}
which can be presented in the following form
\begin{equation}
\varphi(t,{\vec x}) = \varphi_0+k_0\Big(ct-{\vec k}\cdot\vec{x}+\frac{2GM}{c^2}\frac{1-({\vec v}\cdot{\vec k})/c}{\sqrt{1-v^2/c^2}}\ln\left[k^0(r-{\vec k}\cdot{\vec r})\right]\Big)+{\cal O}(G^2),
\label{eq:eq_eik-phi+v}
\end{equation}
where all the quantities in the last term are taken at the retarded time $t_{\rm ret}$ defined by Eq.~(\ref{eq:t-ret}).

Let us now consider signal propagation from a point $(ct_{\rm A},{\vec x}_{\rm A})$ to a point $(ct,{\vec x})$. Then along the signal's path the phase (\ref{eq:eq_eik-phi+v}) will change according to
\begin{equation}
\varphi(t,{\vec x}) = k_0\Big(ct-R_{\rm A}+\frac{2GM}{c^2}\frac{1-({\vec v}\cdot{\vec k})/c}{\sqrt{1-v^2/c^2}}\ln\Big[\frac{r-{\vec k}\cdot{\vec r}}{r_{\rm A}-{\vec k}\cdot{\vec r}_{\rm A}}\Big]\Big)+{\cal O}(G^2),
\label{eq:eq_eik-phi+v-rat}
\end{equation}
where we used the following notations:
$
{\vec k}={\vec R}_{\rm A}/{R_{\rm A}}, {\vec R}_{\rm A}={\vec x}-{\vec x}_{\rm A}, R_{\rm A}=|{\vec R}_{\rm A}|$ and ${\vec r}={\vec x}-{\vec z}(t_{\rm ret}), r=|{\vec r}|.
$

One can further simplify the argument of the logarithmic term in Eq.~(\ref{eq:eq_eik-phi+v-rat}) as
\begin{equation}
\frac{r-{\vec k}\cdot{\vec r}}{r_{\rm A}-{\vec k}\cdot{\vec r}_{\rm A}}=\frac{r_{\rm A}+r-R_{\rm A}}{r_{\rm A}+r+R_{\rm A}}.
\label{eq:eq_eik-phi+v++}
\end{equation}

Thus, Eq.~(\ref{eq:eq_eik-phi+v-rat}) takes the form:
\begin{equation}
\varphi(t,{\vec x}) = k_0\Big(ct-R_{\rm A}-\frac{2GM}{c^2}\frac{1-({\vec v}\cdot{\vec k})/c}{\sqrt{1-v^2/c^2}}\ln\Big[\frac{r_{\rm A}+r+R_{\rm A}}{r_{\rm A}+r-R_{\rm A}}\Big]\Big)+{\cal O}(G^2).
\label{eq:eq_eik-phi+v-rat++}
\end{equation}

Effects of order of $vG/c^3$ are very small and one can neglect them in Eq.~(\ref{eq:eq_eik-phi+v-rat++}). As a result, the post-Minkowskian phase of the electromagnetic wave, with an accuracy appropriate for modern-day Solar System experiments \cite{Blanchet-etal:2001,Kopeikin:2009}, can be presented as:
\begin{equation}
\varphi(t,{\vec x}) = k_0\Big(ct-R_{\rm A}-\frac{2GM}{c^2}\ln\Big[\frac{r_{\rm A}+r+R_{\rm A}}{r_{\rm A}+r-R_{\rm A}}\Big]\Big)+{\cal O}(G^2, c^{-3}).
\label{eq:eq_eik-phi0}
\end{equation}

Along the signal's path the phase stays constant and equal to $\varphi(t_{\rm A},{\vec x}_{\rm A})=k_0 ct_{\rm A}$.

\subsection{Coordinate gravitational time delay}
\label{sec:gr-delay}

Consider the case of one-way signal transmission. Let $\rm A$ be the emitting station, with BCRS position ${\vec x}_{\rm A}(t)$, and $\rm B$ the receiving station, with position ${\vec x}_{\rm B}(t)$. Also, ${\vec z}(t)$ is the vector connecting the SSB to a gravitating body. We denote by $t_{\rm A}$ the coordinate time at the instant of emission of a radio signal, and by $t_{\rm B}$ the coordinate time at the instant of reception. We put ${\vec r}_{\rm A} = {\vec x}_{\rm A}(t_{\rm A})-{\vec z}(t_{\rm A})$, ${\vec r}_{\rm B} = {\vec x}_{\rm B}(t_{\rm B})-{\vec z}(t_{\rm B})$, also ${\vec R}_{\rm AB} = {\vec x}_{\rm B}(t_{\rm B}) - {\vec x}_{\rm A}(t_{\rm A})$, ${\vec N}_{\rm AB} = {\vec R}_{\rm AB}/R_{\rm AB}$ and $R_{\rm AB} = |{\vec R}_{\rm AB}|$, $r_{\rm A}=|{\vec r}_{\rm A}|, r_{\rm B}=|{\vec r}_{\rm B}|$ are the Euclidean norms of these vectors.

We know that along the signal's path the phase stays constant. Thus, equating the eikonal of the wave given by Eq.~(\ref{eq:eq_eik-phi0}) at the two points $\rm A$ and $\rm B$ as $\varphi(t_{\rm A},{\vec x}_{\rm A})=\varphi(t_{\rm B},{\vec x_{\rm B}})$, we determine the gravitational delay of the signal moving through a stationary space-time. Indeed, up to ${\cal O}(c^{-3})$, the coordinate time transfer $T_{\rm AB} = t_{\rm B} - t_{\rm A}$ is given by \cite{Blanchet-etal:2001,Turyshev:2012}:
\begin{equation}
T_{\rm AB} =t_{\rm B} - t_{\rm A}= \frac{R_{\rm AB}}{c}+ \frac{2GM}{c^3}\ln\Big[\frac{
r_{\rm A} + r_{\rm B} + R_{\rm AB}}{r_{\rm A} + r_{\rm B} - R_{\rm AB}}\Big],
\label{eq:t-AB}
\end{equation}
where the logarithmic term represents the Shapiro time delay.

The ratio of coordinate times ${dt_{\rm A0}}/{dt_{\rm B}}$ (and similarly ${dt_{\rm B0}}/{dt_{\rm A}}$) can be determined directly by differentiating the coordinate time transfer equation (\ref{eq:dn-lt}) for $t_{\rm B} - t_{\rm A0}=T_{\rm AB}(t_{\rm A0},t_{\rm B})$ (and similarly for $t_{\rm A} - t_{\rm B0}=T_{\rm BA}(t_{\rm A},t_{\rm B0})$) with respect to the reception time $t_{\rm B}$.  In other words, we must evaluate
\begin{equation}
\frac{d}{dt_{\rm B}}(t_{\rm B}-t_{\rm A0}) =\frac{d}{dt_{\rm B}}\Big\{\frac{1}{c}|{\vec r}_{\rm B}(t_{\rm B}) - {\vec r}_{\rm A}(t_{\rm A0})|
+\frac{2GM}{c^3}\ln\Big(\frac{r_{\rm A} + r_{\rm B} + R_{\rm AB}}{r_{\rm A} + r_{\rm B} - R_{\rm AB}}\Big)\Big\}.
\label{eq:qa-qb}
\end{equation}
While performing the differentiation, we must account for the fact that the coordinate distance between $\rm A_0$ and $\rm B$ depends on both the times of emission and reception, {\it i.e.} $R_{\rm AB} = |{\vec r}_{\rm B}(t_{\rm B}) - {\vec r}_{\rm A}(t_{\rm A0})|$. For instance, we have
\begin{equation}
\frac{dR_{\rm AB}}{dt_{\rm B}}
= {\vec N}_{\rm AB}\cdot \Big({\vec v}_{\rm B} - {\vec v}_{\rm A}\frac{dt_{\rm A0}}{dt_{\rm B}}\Big).
\label{eq:qa-qb+}
\end{equation}
For notational convenience, we henceforth denote the vectors by ${\vec r}_{\rm A} = {\vec x}_{\rm A}(t_{\rm A})-{\vec z}(t_{\rm A})$ and ${\vec r}_{\rm B} = {\vec x}_{\rm B}(t_{\rm B})-{\vec z}(t_{\rm B})$. As before, we have $r_{\rm A} = |{\vec r}_{\rm A}|$ and $r_{\rm B} =|{\vec r}_{\rm B}|$, as well as the coordinate velocities ${\vec v}_{\rm A} = \dot{\vec x}_{\rm A}(t_{\rm A})-\dot{\vec z}(t_{\rm A})$ and ${\vec v}_{\rm B} = \dot{\vec x}_{\rm B}(t_{\rm B})-\dot{\vec z}(t_{\rm B})$. Performing the differentiation in Eq.~(\ref{eq:qa-qb}) to order $1/c^3$, the last factor in Eq.~(\ref{eq:fA-fB}) ${dt_{\rm A0}}/{dt_{\rm B}}$ can be given by the ratio (in agreement with the recent work on the ACES \cite{Blanchet-etal:2001} and RadioAstron \cite{Sazhin-etal:2010} missions; for convenience, we adopt notations similar to those introduced in Ref.~\cite{Blanchet-etal:2001}):
{}
\begin{equation}
\frac{dt_{\rm A0}}{dt_{\rm B}}=\frac{q_{\rm B}}{q_{\rm A0}},
\label{eq:rat-a0b}
\end{equation}
where $q_{\rm A0}$ and $q_{\rm B}$ are derived from
{}
\begin{eqnarray}
q_{\rm A0} &=& 1 -\frac{1}{c}({\vec N}_{\rm AB}\cdot {\vec v}_{\rm A})- \frac{4GM}{c^3}
\frac{(r_{\rm A} + r_{\rm B})({\vec N}_{\rm AB}\cdot {\vec v}_{\rm A}) + R_{\rm AB}({\vec r}_{\rm A}\cdot {\vec v}_{\rm A})/r_{\rm A}}{(r_{\rm A} + r_{\rm B})^2 - R^2_{\rm AB}},
\label{eq:qA5}\\
q_{\rm B} &=& 1 -\frac{1}{c}({\vec N}_{\rm AB}\cdot {\vec v}_{\rm B})
-\frac{4GM}{c^3}\frac{(r_{\rm A} + r_{\rm B})({\vec N}_{\rm AB}\cdot {\vec v}_{\rm B}) - R_{\rm AB}({\vec r}_{\rm B}\cdot {\vec v}_{\rm B})/r_{\rm B}}{(r_{\rm A} + r_{\rm B})^2 - R^2_{\rm AB}}.
\label{eq:qB5}
\end{eqnarray}

In an experiment, the position of the transmitter $\rm A_0$ may be recorded at the time of reception $t_{\rm B}$ rather than at the time of emission $t_{\rm A0}$, i.e. we may have more direct access to ${\vec x}_{\rm A}(t_{\rm B})$ rather than ${\vec x}_{\rm A}(t_{\rm A0})$, and the formulae (\ref{eq:qA5})--(\ref{eq:qB5}) get modified by Sagnac correction terms consistently to the order $1/c^3$:
\begin{eqnarray}
q_{\rm A0} &=& 1 -\frac{1}{c}({\vec n}_{\rm AB}\cdot {\vec v}_{\rm A})- \frac{1}{c^2}\Big({\vec v}^2_{\rm A}-({\vec n}_{\rm AB}\cdot {\vec v}_{\rm A})^2-({\vec a}_{\rm A}\cdot {\vec d}_{\rm AB})\Big)+\nonumber\\[3pt]
&&~~+\frac{1}{2c^3}\Big\{({\vec v}_{\rm A}^2-({\vec n}_{\rm AB}\cdot{\vec v}_{\rm A})^2)({\vec n}_{\rm AB}\cdot{\vec v}_{\rm A})+d_{\rm AB}\Big(3({\vec a}_{\rm A}\cdot{\vec v}_{\rm A})-({{\vec n}}_{\rm AB}\cdot{\vec a}_{\rm A})({\vec n}_{\rm AB}\cdot{\vec v}_{\rm A})-(\dot{\vec a}_{\rm A}\cdot{\vec d}_{\rm AB})\Big)\Big\}-\nonumber\\[3pt]
&&~~-\frac{4GM}{c^3}
\frac{(r_{\rm A} + r_{\rm B})({\vec n}_{\rm AB}\cdot {\vec v}_{\rm A}) + d_{\rm AB}({\vec r}_{\rm A}\cdot {\vec v}_{\rm A})/r_{\rm A}}{(r_{\rm A} + r_{\rm B})^2 - d^2_{\rm AB}}+{\cal O}(c^{-4}),
\label{eq:qA-tB+}\\[3pt]
q_{\rm B} &=& 1 -\frac{1}{c}({\vec n}_{\rm AB}\cdot {\vec v}_{\rm B})- \frac{1}{c^2}\Big(({\vec v}_{\rm A}\cdot{\vec v}_{\rm B})-({\vec n}_{\rm AB}\cdot{\vec v}_{\rm A})({\vec n}_{\rm AB}\cdot{\vec v}_{\rm B})\Big)+\nonumber\\[3pt]
&&~~+\frac{1}{2c^3}\Big\{({\vec v}_{\rm A}^2-({\vec n}_{\rm AB}\cdot{\vec v}_{\rm A})^2)({\vec n}_{\rm AB}\cdot{\vec v}_{\rm B})+d_{\rm AB}\Big(({\vec a}_{\rm A}\cdot{\vec v}_{\rm B})-({\vec n}_{\rm AB}\cdot{\vec a}_{\rm A})({\vec n}_{\rm AB}\cdot{\vec v}_{\rm B})\Big)\Big\}-\nonumber\\[3pt]
&&~~
-\frac{4GM}{c^3}\frac{(r_{\rm A} + r_{\rm B})({\vec n}_{\rm AB}\cdot {\vec v}_{\rm B}) - d_{\rm AB}({\vec r}_{\rm B}\cdot {\vec v}_{\rm B})/r_{\rm B}}{(r_{\rm A} + r_{\rm B})^2 - d^2_{\rm AB}}+{\cal O}(c^{-4}).
\label{eq:qB-tB+}
\end{eqnarray}

For the GRAIL spacecraft in lunar orbit, in the lunicentric reference frame the first, $1/c$ term in Eqs.~(\ref{eq:qA-tB+})--(\ref{eq:qB-tB+}) is $\sim 10^{-4}$ in the barycentric frame and $\sim 5.5 \times 10^{-6}$ in the lunicentric frame. The second term is the $1/c^2$ Sagnac term, which is $\sim 10^{-8}$ for the BCRS and $\sim 10^{-11}$ for the LCRS. The $1/c^3$ terms are $\sim 8\times 10^{-17}$ for the Sagnac term and $\sim 1.4\times 10^{-16}$ for the Shapiro term. Note that the result given in Eqs.~(\ref{eq:qA-tB+})--(\ref{eq:qB-tB+}) has been obtained assuming that the field of the Earth is spherically symmetric. Indeed, the $J_2$-terms in the factor $(q_{\rm A}/q_{\rm B})$ do not exceed $10^{-15}$.

As a result, the expression ${dt_{\rm A0}}/{dt_{\rm B}}$ from Eq.~(\ref{eq:rat-a0b}) has the form
{}
\begin{eqnarray}
\frac{dt_{\rm A0}}{dt_{\rm B}}=\frac{q_{\rm B}}{q_{\rm A0}}&=&1 -\frac{1}{c}({\vec n}_{\rm AB}\cdot {\vec v}_{\rm AB})- \frac{1}{c^2}\Big(({\vec v}_{\rm A}\cdot {\vec v}_{\rm AB})+({\vec a}_{\rm A}\cdot {\vec d}_{\rm AB})\Big)-\nonumber\\
&&\hskip 8pt -\,
\frac{1}{2c^3}\Big\{2({\vec v}_{\rm A}\cdot {\vec v}_{\rm AB})({\vec n}_{\rm AB}\cdot {\vec v}_{\rm A})+
({\vec v}^2_{\rm A}-({\vec n}_{\rm AB}\cdot {\vec v}_{\rm A})^2)({\vec n}_{\rm AB}\cdot {\vec v}_{\rm AB})+\nonumber\\
&&\hskip -10pt +\,
d_{\rm AB}\Big(2({\vec a}_{\rm A}\cdot {\vec v}_{\rm A})+2({\vec n}_{\rm AB}\cdot{\vec a}_{\rm A})({\vec n}_{\rm AB}\cdot {\vec v}_{\rm A})-({\vec a}_{\rm A}\cdot {\vec v}_{\rm AB})-({\vec n}_{\rm AB}\cdot {\vec a}_{\rm A})({\vec n}_{\rm AB}\cdot {\vec v}_{\rm AB})-(\dot{\vec a}_{\rm A}\cdot {\vec d}_{\rm AB})\Big)\Big\}-\nonumber\\
&&\hskip 8pt -\,
\frac{4GM}{c^3}\frac{(r_{\rm A}+r_{\rm B})({\vec n}_{\rm AB}\cdot {\vec v}_{\rm AB})-d^{}_{\rm AB}\big(({\vec r}_{\rm B}\cdot {\vec v}_{\rm B})/r_{\rm B})+({\vec r}_{\rm A}\cdot {\vec v}_{\rm A})/r_{\rm A})\big)}{(r_{\rm A}+r_{\rm B})^2-d^2_{\rm AB}}
+{\cal O}(c^{-3}).
\label{eq:rat-AB0}
\end{eqnarray}

We can evaluate the magnitude of each of the seven terms in Eq.~(\ref{eq:rat-AB0}) using the values from Table~\ref{tb:params}. We will use these basic formation parameters  to evaluate the terms in (\ref{eq:DOWRR022}), keeping only those that contribute to the range-rate model more than $\Delta v_{\rm dowrr}=1~\mu$m/s. The error terms in these expressions must be less than  $\Delta v_{\rm dowrr}/c=3\times 10^{-15}$.

Keeping these numbers in mind we see that the ($1/c$) term in (\ref{eq:rat-AB0}) is of the order $7\times10^{-9}$ and must be kept in the model. The two $1/c^2$ terms were evaluated to be of the order of
$\big(({\vec v}_{\rm A}\cdot {\vec v}_{\rm AB})+({\vec a}_{\rm A}\cdot {\vec d}_{\rm AB})\big)/c^2\approx 6\times 10^{-11}+3\times 10^{-12}$. Thus, both of these terms must be in the model.

Among the $1/c^3$ terms, the first one is $\sim 6\times 10^{-15}$ and must be kept. Next is the term that contains $(\vec{n}_{\rm AB}\cdot{\vec v}_{\rm AB})$, which was evaluated to be at most $3\times 10^{-17}$ and is too small to be in the model. The remaining $1/c^3$ term, which is $\propto d_{\rm AB}$, is at most $\sim 5\times 10^{-16}$ and thus, this entire term may be neglected.

Lastly, the Shapiro term in Eq.~(\ref{eq:rat-AB0})  was evaluated to be
\begin{eqnarray}
-\,\frac{4GM}{c^3}\frac{({\vec n}_{\rm AB}\cdot {\vec v}_{\rm AB})}{(r_{\rm A}+r_{\rm B})}+
\frac{4GM}{c^3}\frac{d^{}_{\rm AB}\big(({\vec n}_{\rm A}\cdot {\vec v}_{\rm A})+({\vec n}_{\rm B}\cdot {\vec v}_{\rm B})\big)}{(r_{\rm A}+r_{\rm B})^2}\approx 8\times 10^{-19}+3\times 10^{-15}.
\label{eq:shap}
\end{eqnarray}
Therefore, we will keep only the second part of the Shapiro term in the model.

As a result, the expression (\ref{eq:rat-AB0}) may be presented in the following simplified form:
{}
\begin{eqnarray}
\frac{dt_{\rm A0}}{dt_{\rm B}} &=&1 -\frac{1}{c}({\vec n}_{\rm AB}\cdot {\vec v}_{\rm AB})- \frac{1}{c^2}\Big(({\vec v}_{\rm A}\cdot {\vec v}_{\rm AB})+({\vec a}_{\rm A}\cdot {\vec d}_{\rm AB})\Big)-
\nonumber\\
&&\hskip 8pt-\,
\frac{1}{c^3}\Big(({\vec n}_{\rm AB}\cdot {\vec v}_{\rm A})(\vec{v}_{\rm AB}\cdot\vec{v}_{\rm A})\Big)+
\frac{4GM}{c^3}\frac{d^{}_{\rm AB}}{(r_{\rm A}+r_{\rm B})^2}\Big(({\vec n}_{\rm A}\cdot {\vec v}_{\rm A})+({\vec n}_{\rm B}\cdot {\vec v}_{\rm B})\Big)+{\cal O}(5\times10^{-16}),
\label{eq:rat-AB0+}
\end{eqnarray}
where ${\vec n}_{\rm A}={\vec r}_{\rm A}/{r}_{\rm A}$ and
${\vec n}_{\rm B}={\vec r}_{\rm B}/{r}_{\rm B}$. This expression accounts for all the terms that may have magnitudes larger than $5\times10^{-16}$ and are thus relevant to the GRAIL mission configuration.

\section{Evaluating the DOWR integral expression}
\label{sec:int-error}

In Sec.~\ref{sec:dowr} we defined the quantities $\epsilon_{\rm AB}(t_{\rm B})$ and  $\epsilon_{\rm BA}(t_{\rm A})$, which were given in Eq.~(\ref{eq:n-ab-eps}).
% and (\ref{eq:n-ba-eps}).
We now construct the quantity $\epsilon_{\rm AB}(t_{\rm B})+\epsilon_{\rm BA}(t_{\rm A})$:
{}
\begin{eqnarray}
\epsilon_{\rm AB}(t_{\rm B})+\epsilon_{\rm BA}(t_{\rm A})&=&\frac{1}{c^2}\Big\{f_{\rm B0}\Big(u_{\rm B}(t_{\rm B})(t_{\rm B}-t^0_{\rm B})-u_{\rm B}(t_{\rm A})\Big(t_{\rm A}-T_{\rm BA}(t_{\rm A})-\big(t_{\rm A}^0-T_{\rm BA}(t^0_{\rm A})\big)\Big)-
\nonumber\\
&&\hskip 112pt-\,
\int_{t^0_{\rm B}-t_{\rm A}^0+T_{\rm BA}(t^0_{\rm A})}^{t_{\rm B}-t_{\rm A}+T_{\rm BA}(t_{\rm A})} u_{\rm B}(t')dt'\Big)+\nonumber\\
&&\hskip 8pt+\,f_{\rm A0}\Big(u_{\rm A}(t_{\rm A})(t_{\rm A}-t^0_{\rm A})-u_{\rm A}(t_{\rm B})\Big(t_{\rm B}-T_{\rm AB}(t_{\rm B})-\big(t_{\rm B}^0-T_{\rm AB}(t_{\rm B}^0)\big)\Big)-\nonumber\\
&&\hskip 112pt-\,\int_{t^0_{\rm A}-t_{\rm B}^0+T_{\rm AB}(t_{\rm B}^0)}^{t_{\rm A}-t_{\rm B}+T_{\rm AB}(t_{\rm B})} u_{\rm A}(t')dt'\Big)\Big\}+{\cal O}({c^{-4}}).~~~~~
\label{eq:n-abba-eps-a}
\end{eqnarray}

To estimate the magnitude of this quantity, we can assume that the clocks $A$ and $B$ are perfectly synchronized. In reality, nothing is perfect and the GRAIL mission's design relies on a synchronization procedure (see Ref.~\cite{TimingMemo:2010} for details) that, of course, leaves synchronization errors  that percolate in the data analysis. One can study the impact of synchronization errors on the measurement accuracy, using the approach outlined here. However, presently we concerned with the evaluation of the magnitude of the error that may arise in the ideal case, if we were to drop the $\epsilon_{\rm AB}$ and $\epsilon_{\rm BA}$ terms. Thus, we assume that $t_{\rm A}=t_{\rm B}=t$ and $t^0_{\rm A}=t^0_{\rm B}=t^0$, so that the quantity given by Eq.~(\ref{eq:n-abba-eps-a}) is reduced to
{}
\begin{eqnarray}
\epsilon_{\rm AB}(t)+\epsilon_{\rm BA}(t)&=&\frac{1}{c^2}\Big\{f_{\rm A0}\Big(u_{\rm A}(t)\big(T_{\rm AB}(t)-T_{\rm AB}(t^0)\big)-
\int_{T_{\rm AB}(t^0)}^{T_{\rm AB}(t)} u_{\rm A}(t')dt'\Big)+\nonumber\\
&&\hskip 8pt+\,
f_{\rm B0}\Big(u_{\rm B}(t)\big(T_{\rm BA}(t)-T_{\rm BA}(t^0)\big)-
\int_{T_{\rm BA}(t^0)}^{T_{\rm BA}(t)} u_{\rm B}(t')dt'\Big)\Big\}+{\cal O}({c^{-4}}).~~~~~
\label{eq:n-abba-eps-a2}
\end{eqnarray}

We series expand the integrals in Eq.~(\ref{eq:n-abba-eps-a2}) around the start $t^0$ of the integration interal. For the integral term multiplied by $f_{\rm A0}$ inside the second set of square brackets in Eq.~(\ref{eq:n-abba-eps-a2}), we have  $u_{\rm A}(t')=u_{\rm A}(t^0)+\dot u_{\rm A}(t^0)(t'-t^0)+{\cal O}(\Delta t^2)$ (where $\Delta t=t-t^0$), and obtain:
\begin{eqnarray}
&&u_{\rm A}(t)\big(T_{\rm AB}(t)-T_{\rm AB}(t_0)\big)-
\int_{T_{\rm AB}(t_0)}^{T_{\rm AB}(t)} u_{\rm A}(t')dt'={\textstyle\frac{1}{2}}\dot u_{\rm A}(t_0)\big(T_{\rm AB}(t)-T_{\rm AB}(t_0)\big)^2+{\cal O}(\Delta t^3).~~~~
\end{eqnarray}
The integral term in the first set of square brackets in Eq.~(\ref{eq:n-abba-eps-a2}), which is multiplied by $f_{\rm B0}$, can be evaluated in a similar way. Keeping just the leading terms ($\sim d_{\rm AB}/c$) in $T_{\rm AB}$ and $T_{\rm BA}$, we present, for instance, $T_{\rm AB}(t)-T_{\rm AB}(t_0)=c^{-1}\dot d_{\rm AB}(t-t_0)+{\cal O}({c^{-2}})$, so that Eq.~(\ref{eq:n-abba-eps-a2}) becomes:
{}
\begin{eqnarray}
\epsilon_{\rm AB}(t)+\epsilon_{\rm BA}(t)&=&\frac{1}{2c^4}\Big(f_{\rm A0}\dot u_{\rm A}(t_0)+f_{\rm B0}\dot u_{\rm B}(t_0)\Big)\dot d_{\rm AB}^2(t-t_0)^2+{\cal O}({c^{-4}})={\cal O}({c^{-4}}).
\label{eq:n-abba-eps-a3}
\end{eqnarray}
Remembering the form of $u_{\rm B}(t)$ in Eq.~(\ref{eq:Larg}), and denoting by $f_0$ the typical GRAIL radio frequency ($f_{\rm A0}\simeq f_{\rm B0}\simeq f_0$, see Table~\ref{tb:params}), we see   that the term above is of the order of $\sim(v_{\rm A} a_{\rm A} \dot d_{\rm AB}^2/2c^4)(t-t_0)^2f_0\ll 7.7\times 10^{-24}\,(t-t_0)^2f_0$ at most. This term is multiplied by $\sim c/2f_0$ as its contribution to the DOWR measurement is calculated in Eq.~(\ref{eq:DOWR+0}): the magnitude of this contribution is therefore less than $1\times 10^{-15}(t-t_0)^2~{\rm m/s}^2$, which is negligible for GRAIL.

\section{Relativistic frequency transfer between spacecraft and a DSN antenna}
\label{sec:frqxfr}

The DOWR observable is defined in terms of the coordinate frequencies $f_{\rm A}$ and $f_{\rm B}$ of both signals generated on-board the two GRAIL spacecraft. However, these frequencies are measured by a ground-based DSN station. Specifically, in the case of the GRAIL spacecraft, while the spacecraft-to-ground transmission takes place using frequency bands that are different from the frequencies used for inter-spacecraft communication, the two frequencies are synthesized using the same timing source (USO) on board. Thus, measuring the frequency of the spacecraft-to-ground transmission is, in effect, a measurement of the inter-spacecraft frequency as well.

Below, we simply assume that the spacecraft-to-ground transmission does serve as a means to measure $f_{\rm A}$ and $f_{\rm B}$ without going into further detail.  We discuss how one can use these measurements, taken using the TT time coordinate, and transform them to the TDB time coordinate.
We focus only on relativistic frequency transformations between the frames involved and will neglect frequency-dependent media effects, which are easy to reinstate when needed. Although communication between the spacecraft is conducted using Ka-band microwave signals and spacecraft-to-DSN is done relying on X-band signals, these frequencies are related by simple numeric factors. Therefore, in a slight abuse of notation, we use the same symbol for the inter-spacecraft and spacecraft-to-ground frequencies.

\subsection{One-way frequency transfer}
\label{sec:1-way-freq}

We consider the situation when a signal with frequency $f$ is measured by an electronic counter whose register is incremented by 1 each time the magnitude of the signal changes from minus to plus.  The number of cycles $dn$ measured by this counter in the interval of proper time  $d\tau$ is then $dn=fd\tau$. Thus, the frequency transfer between transmitter (at point {\rm A}) and receiver (at point C)  requires the determination of the ratio $f_{\rm A}^{\rm C}/f_{\rm A0}$ between the proper frequencies $f_{\rm A0}$ transmitted by satellite ({\rm A}) and $f_{\rm A}^{\rm C}$ on the ground ({\rm C}). The infinitesimal proper time intervals $d\tau_{\rm A}$ and $d\tau_{\rm C}$ correspond to the infinitesimal number of cycles, $dn$, at the transmission and reception points {\rm A} and {\rm C}, so that $dn_{\rm A}^{\rm C}=dn_{\rm A}$. Therefore, the one-way frequency shift during the transfer from {\rm A} to {\rm C} is
{}
\begin{equation}
\frac{f_{\rm A}^{\rm C}}{f_{\rm A0}}=\frac{dn_{\rm A}^{\rm C}}{d\tau_{\rm C}}\frac{d\tau_{\rm A}}{dn_{\rm A}}=\frac{d\tau_{\rm A}}{d\tau_{\rm C}}=\frac{\left(d\tau/dt\right)_{\rm A}}{\left(d\tau/dt\right)_{\rm C}}\frac{dt_{\rm A}}{dt_{\rm C}}.
\label{eq:fA-fB}
\end{equation}

For $(d\tau/dt)_{\rm C}/(d\tau/dt)_{\rm A}$, we get
\begin{equation}
\frac{(d\tau/dt)_{\rm A}}{(d\tau/dt)_{\rm C}}=\frac{1 - c^{-2}\left[U({\vec r}_{\rm A}) + {\textstyle\frac{1}{2}}{\vec v}^2_{\rm A}\right]}{1 - c^{-2}\left[U({\vec r}_{\rm C}) + {\textstyle\frac{1}{2}}{\vec v}^2_{\rm C}\right]}+{\cal O}(c^{-4}),
\label{eq:nu_AB+}
\end{equation}
where $U({\vec r}_{\rm A})=\sum_b U_b({\vec r}_{b{\rm M}}+{\vec y}_{\rm A})$
and $U({\vec r}_{\rm C})=\sum_b U_b({\vec r}_{b{\rm E}}+{\vec y}_{\rm C})$
are the Newtonian potentials at the points $\rm A$ and $\rm C$ and vectors ${\vec v}_{\rm A}$ and ${\vec v}_{\rm C}$ are the barycentric velocities of the spacecraft and the DSN station correspondingly. The factor given by Eq.~(\ref{eq:nu_AB+}) consists of the Einstein gravitational red-shift and second-order Doppler effects, both of order $1/c^2$. Therefore, Eq.~(\ref{eq:fA-fB}) becomes
{}
\begin{equation}
\frac{f_{\rm A}^{\rm C}}{f_{\rm A0}}=\frac{d\tau_{\rm A}}{d\tau_{\rm C}}=\frac{1 - c^{-2}\left[U({\vec r}_{\rm A}) + {\textstyle\frac{1}{2}}{\vec v}^2_{\rm A}\right]}{1 - c^{-2}\left[U({\vec r}_{\rm C}) + {\textstyle\frac{1}{2}}{\vec v}^2_{\rm C}\right]}~\frac{dt_{\rm A}}{dt_{\rm C}}+{\cal O}(c^{-4}).
\label{eq:fA-fB*+}
\end{equation}

The $dt_{\rm A}/dt_{\rm C}$ term on the right-hand-side of Eq.~(\ref{eq:fA-fB*+}) is the ratio of coordinate periods of the same signal at $\rm A$ and $\rm C$. It contains both the $(\sim 1/c)$ Doppler effect and the $(\sim 1/c^3)$ terms that we seek. To compute this factor at the accuracy of $3 \times 10^{-15}$, it is sufficient to treat the Earth and Moon gravitational potentials as monopole potentials, i.e., the approximation $U=GM/r$ and consequently, the use of Eq.~(\ref{eq:proper-A-t}) is sufficient.

Similarly to the discussion in Sec.\ref{sec:gr-delay}, the ratio $dt_{\rm A}/dt_{\rm C}$ can be computed by a direct differentiation of the coordinate time transfer $T_{\rm AC} = t_{\rm C} - t_{\rm A}$ with respect to the emission time $t_{\rm A}$. In Appendix~\ref{sec:gr-delay} we computed all the relevant terms,  accurate to the $\sim1/c^3$ order. Using the time of reception $t_{\rm C}$ (and expressing the time of emission as a function of the time of reception, $t_{\rm A}=t_{\rm A}(t_{\rm C})$), the factor ${dt_{\rm A}}/{dt_{\rm C}}$ in Eq.~(\ref{eq:fA-fB}) can be given (similar to Eq.~(\ref{eq:rat-a0b})) by
\begin{equation}
\frac{dt_{\rm A}}{dt_{\rm C}}=\frac{q_{\rm C}}{q_{\rm A}}
\label{eq:t_a/t_B}
\end{equation}
where $q_{\rm A}$ and $q_{\rm C}$ are then given to ${\cal O}(c^{-3})$ from Eqs.~(\ref{eq:qA-tB+})--(\ref{eq:qB-tB+}) as
{}
\begin{eqnarray}
q_{\rm A} &=& 1 -\frac{1}{c}({\vec  n}_{\rm AC}\cdot {\vec v}_{\rm A})- \frac{1}{c^2}\Big({\vec v}^2_{\rm A}-({\vec v}_{\rm A}\cdot{\vec  n}_{\rm AC})^2-({\vec a}_{\rm A}\cdot {\vec d}_{\rm AC})\Big)+{\cal O}(c^{-3}),
\label{eq:qA-tB*}\\
q_{\rm C} &=& 1 -\frac{1}{c}({\vec n}_{\rm AC}\cdot {\vec v}_{\rm C})- \frac{1}{c^2}\Big(({\vec v}_{\rm A}\cdot{\vec v}_{\rm C})-({\vec v}_{\rm A}\cdot{\vec  n}_{\rm AC})({\vec v}_{\rm C}\cdot{\vec  n}_{\rm AC})\Big)+{\cal O}(c^{-3}),
\label{eq:qB-tB*}
\end{eqnarray}
where ${\vec d}_{\rm AC} = {\vec x}_{\rm C}(t_{\rm C})-{\vec x}_{\rm A}(t_{\rm C})$ is the coordinate distance between $A$ and $C$ at the moment of reception at $C$ (we have $d_{\rm AC} = |{\vec d}_{\rm AC}|$ and ${\vec n}_{\rm AC}={\vec d}_{\rm AC}/d_{\rm AC}$), where ${\vec v}_{\rm A}(t_{\rm C})$ denotes the coordinate velocity of the station $A$ at that instant, and where ${\vec a}_{\rm C}$ is the acceleration of $A$. Therefore, the ratio ${q_{\rm C}}/{q_{\rm A}}$ can be given as
{}
\begin{equation}
\frac{q_{\rm C}}{q_{\rm A}}=1-\frac{1}{c}({\vec n}_{\rm AC}\cdot {\vec v}_{\rm AC})+\frac{1}{c^2}\Big({\textstyle\frac{1}{2}}{\vec v}_{\rm A}^2-{\textstyle\frac{1}{2}}{\vec v}_{\rm C}^2+{\textstyle\frac{1}{2}}{\vec v}^2_{\rm AC}-({\vec a}_{\rm A}\cdot{\vec d}_{\rm AC})\Big)+{\cal O}(c^{-3}),
\label{eq:qB/qA}
\end{equation}
where all the quantities involved are given at the time of reception $t_{\rm C}$.

Therefore, for the one-way frequency transfer, we have
\begin{eqnarray}
\frac{f_{\rm A}^{\rm C}}{f_{\rm A0}}=\frac{1 - c^{-2}\left[U({\vec r}_{\rm A}) + \frac{1}{2}{\vec v}^2_{\rm A}\right]}{1 - c^{-2}\left[U({\vec r}_{\rm C}) + \frac{1}{2}{\vec v}^2_{\rm C}\right]}\cdot\frac{q_{\rm C}}{q_{\rm A}},
\label{eq:nu_AB}
\end{eqnarray}
with the ratio $q_{\rm A}/q_{\rm C}$ given to sufficient accuracy by Eq.~(\ref{eq:qB/qA}). Up to ${\cal O}(c^{-3})$, this expression becomes:
{}
\begin{equation}
\frac{f_{\rm A}^{\rm C}}{f_{\rm A0}}=1-\frac{1}{c}({\vec n}_{\rm AC}\cdot {\vec v}_{\rm AC})+\frac{1}{c^2}\Big(
{\textstyle\frac{1}{2}}{\vec v}^2_{\rm AC}+\sum_b \big[U_b({\vec r}_{b{\rm E}}+{\vec y}_{\rm C})-U_b({\vec r}_{b{\rm M}}+{\vec y}_{\rm A})\big]-({\vec a}_{\rm A}\cdot{\vec d}_{\rm AC})\Big)+{\cal O}(5\times 10^{-14}).
\label{eq:f_AB}
\end{equation}

Note that all the quantities in Eq.~(\ref{eq:f_AB}) are taken at the reception time. This relation determines the frequency displacement due to the combined motion of the emitter and the receiver (intrinsic Doppler effect) and the difference in the gravitational potentials at the points of emission and reception of the signal (gravitational displacement of the frequency).

In the case of GRAIL, the one-way frequency transfer given by Eqs.~(\ref{eq:qA-tB*})--(\ref{eq:qB-tB*}) can be evaluated numerically as follows. The first-order Doppler effect is $|({\vec n}_{\rm AC}\cdot {\vec v}_{\rm A})/c| = 5.5 \times 10^{-6}$; for the ground $|({\vec n}_{\rm AC}\cdot {\vec v}_{\rm C})/c| = 1.6 \times 10^{-6}$. The second-order Doppler effect is ${\vec v}^2_{\rm A}/2c^2 = 3.4 \times 10^{-10}$ for the satellite; ${\vec v}^2_{\rm C}/2c^2 = 1.3 \times 10^{-12}$ for the ground. The gravitational red-shift (Einstein) effect is given by $U_{\rm A}/c^2 = U_{\rm E}({\vec r}_{\rm A})/c^2 = 6.5 \times 10^{-10}$; $U_{\rm C}/c^2 = 6.9 \times 10^{-10}$. The ${\cal O}(c^{-3})$ terms are less than $3.6 \times 10^{-14}$ for the spacecraft and $2.2 \times 10^{-15}$ for the Earth station, thus, they are omitted.

\subsection{Integrated Doppler Effect}
\label{sec:inegrated-Doppler}

To measure the frequency of the received signal, DSN receivers count the number of cycles received from the spacecraft over intervals of time measured by high precision clocks located at the receiver station.

We consider the integrated (one-way) Doppler effect that is at the basis of forming the Doppler observable for spacecraft such as GRAIL that are equipped with a precision on-board frequency source. Consider a clock (Fig.~\ref{fig:grail-timing}) with proper frequency $f_{\rm A0}$, located at point {\rm A} (GRAIL-A), which emits light signals during the proper time interval ($\tau_{\rm A1},\tau_{\rm A2})$. These signals are received with frequency $f_{\rm A}^{\rm C}$ by a ground-based DSN station, located at {\rm C} (see Fig.~\ref{fig:grail}), over a proper time interval $(\tau_{\rm C1},\tau_{\rm C2})$. The measured quantity over the interval $(\tau_{\rm C1},\tau_{\rm C2})$ is the number of cycles received at point {\rm C} from the transmitter at point {\rm A}. The number of cycles received at point {\rm C} must equal the number of cycles emitted at point {\rm A}. Hence,
{}
\begin{equation}
\int_{\tau_{\rm C1}}^{\tau_{\rm C2}}f_{\rm A}^{\rm C}d\tau_{\rm C}=\int_{\tau_{\rm A1}}^{\tau_{\rm A2}}f_{\rm A0}d\tau_{\rm A},
\label{eq:N1}
\end{equation}
where the frequency $f_{\rm A0}$ is the constant (with respect to $\tau_{\rm A}$) precision oscillator frequency of the transmitter, and may be moved outside the integral sign. The remaining integral can be transformed to the proper time $\tau_{\rm C}$ with the use of Eq.~(\ref{eq:fA-fB*+}):
{}
\begin{equation}
\int_{\tau_{\rm A1}}^{\tau_{\rm A2}}d\tau_{\rm A}=
\int_{\tau_{\rm C1}}^{\tau_{\rm C2}}\frac{d\tau_{\rm A}}{d\tau_{\rm C}}\,d\tau_{\rm C}=
\int_{\tau_{\rm C1}}^{\tau_{\rm C2}}\frac{f_{\rm A}^{\rm C}}{f_{\rm A0}}\,d\tau_{\rm C}=\int_{t_1}^{t_2}\frac{f_{\rm A}^{\rm C}}{f_{\rm A0}}\frac{d\tau_{\rm C}}{dt}\,dt,
\label{eq:N2}
\end{equation}
with ratio $d\tau_{\rm C}/dt$ given by Eq.~(\ref{eq:proper-t-C+}) as
{}
\begin{equation}
\frac{d\tau_{\rm C}}{dt}=
1 -\frac{1}{c^2}\Big[{\textstyle\frac{1}{2}}{\vec v}_{\rm C}^2+\sum_{b}U_b({\vec r}^{}_{b{\rm E}}+\vec{y}^{}_{\rm C})
\Big]+{\cal O}(10^{-17}).
\label{eq:proper-t-C+app}
\end{equation}
Also, in Eq.~(\ref{eq:f_AB}) we expressed the ratio ${f_{\rm A}^{\rm C}}/{f_{\rm A0}}$ at the adopted accuracy using terms that are functions of $t$. Therefore,
{}
\begin{equation}
\int_{\tau_{\rm A1}}^{\tau_{\rm A2}}d\tau_{\rm A}=t_2-t_1-\frac{1}{c}\Big(d_{\rm AC}(t_2)-d_{\rm AC}(t_1)\Big)+ \frac{1}{c^2}\int_{t_1}^{t_2}\Big[{\textstyle\frac{1}{2}}{\vec v}^2_{\rm AC}-{\textstyle\frac{1}{2}}{\vec v}^2_{\rm C}-
\sum_b U_b({\vec r}_{b{\rm M}}+{\vec y}_{\rm A})-({\vec a}_{\rm A}\cdot{\vec d}_{\rm AC})\Big]dt+{\cal O}(c^{-3}),
\label{eq:N3}
\end{equation}
where we relied on the identity $({\vec n}_{\rm AC}\cdot{\vec v}_{\rm AC})=\dot d_{\rm AC}$. Using these intermediate results, we find that the total number of cycles $N^{\rm C}_{\rm A}$, received during the count interval $\Delta t=t_2-t_1$, is given by:
{}
\begin{eqnarray}
N^{\rm C}_{\rm A}=\int_{\tau_{\rm C1}}^{\tau_{\rm C2}} f_{\rm A}^{\rm C}d\tau_{\rm C}
&=&f_{\rm A0}(t_2-t_1)-\frac{1}{c}f_{\rm A0}\Big(d_{\rm AC}(t_2)-d_{\rm AC}(t_1)\Big)+\nonumber\\
&&\hskip 30pt+\,
\frac{1}{c^2}f_{\rm A0}\int_{t_1}^{t_2}\Big[{\textstyle\frac{1}{2}}{\vec v}^2_{\rm AC}-{\textstyle\frac{1}{2}}{\vec v}^2_{\rm C}-
\sum_b U_b({\vec r}_{b{\rm M}}+{\vec y}_{\rm A})-({\vec a}_{\rm A}\cdot{\vec d}_{\rm AC})\Big]dt+{\cal O}(c^{-3}).~~
\label{eq:N4}
\end{eqnarray}
Dividing the number of cycles $N^{\rm C}_{\rm A}$ by the count interval $\Delta t$ yields the Doppler observable $\hat f_{\rm A}^{\rm C}$:
{}
\begin{equation}
\hat f_{\rm A}^{\rm C}=\frac{N^{\rm C}_{\rm A}}{\Delta t}.
\end{equation}
Therefore, the spacecraft frequency $\hat f_{\rm A}$ that is measured at a DSN receiver during the count interval of proper time $\Delta t$ can be modeled as
{}
\begin{eqnarray}
\frac{\hat f_{\rm A}^{\rm C}}{f_{\rm A0}}
&=&1-\frac{d_{\rm AC}(t_2)-d_{\rm AC}(t_1)}{c\Delta t}+
\frac{1}{c^2\Delta t}\int_{t_1}^{t_2}\Big[{\textstyle\frac{1}{2}}{\vec v}^2_{\rm AC}-{\textstyle\frac{1}{2}}{\vec v}^2_{\rm C}-
\sum_b U_b({\vec r}_{b{\rm M}}+{\vec y}_{\rm A})-({\vec a}_{\rm A}\cdot{\vec d}_{\rm AC})\Big]dt+{\cal O}(c^{-3}),~~
\label{eq:N4DSN}
\end{eqnarray}
where the range $d_{\rm AC}$ is defined as $d_{\rm AC}=|{\vec x}_{\rm C}-{\vec x}_{\rm A}|$.

\subsection{Transmitted spacecraft frequency, as seen from the BCRS}

In a 1-way Doppler transmission between a GRAIL spacecraft and a DSN station, the measured frequency received at the DSN station, $\hat f^{\rm C}_{\rm A}$, is related to the transmitted frequency $f_{\rm A0}$ as given in Eq.~(\ref{eq:N4DSN}). Our objective is to express the frequency received at the DSN using the TDB time coordinate of the BCRS. To do this, we imagine a hypothetical receiver that is at rest with respect to the BCRS, and co-located with the DSN receiver at the moment of receiving the same signal. Using Eq.~(\ref{eq:f_AB}) we can relate the instantaneous frequency $f_{\rm A}$ that is received by this hypothetical receiver to the transmitted frequency $f_{\rm A0}$:
{}
\begin{equation}
\frac{f_{\rm A}}{f_{\rm A0}}=1-\frac{1}{c}({\vec n}_{\rm A}\cdot {\vec v}_{\rm A})+\frac{1}{c^2}\Big(
{\textstyle\frac{1}{2}}{\vec v}^2_{\rm A}+
\sum_b \big[U_b({\vec z}_{b})-U_b({\vec r}_{b{\rm M}}+{\vec y}_{\rm A})\big]+({\vec a}_{\rm A}\cdot{\vec d}_{\rm A})\Big)+{\cal O}(c^{-3}),
\label{eq:f_BA}
\end{equation}
where $\sum_b U_b({\vec z}_{b})$ is the gravitational potential at the origin of the SSB (see discussion after Eqs.~(\ref{eq:w_0-BCRS})--(\ref{eq:w_a-BCRS})) defined as
\begin{equation}
\sum_b\mu^*_b{\vec z}_b=0 \qquad {\rm and}\qquad
\mu^*_b=GM_{b}\Big\{1+\frac{1}{2c^2}\Big(v^2_b-\sum_{c\not=b}\frac{GM_c}{r_{bc}}\Big)\Big\}+{\cal O}(c^{-4}),~~
\end{equation}
also, ${\vec z}_{b}$ is the barycentic position vector of the body $b$, $v_{b}=|\dot{\vec z}_{b}|$ is its barycentic velocity, and all times are measured in the TDB time.

Analogously to the discussion in the Section~\ref{sec:inegrated-Doppler}, the Doppler observable $\hat f_{\rm A}$ that corresponds to the instantaneous frequency $f_{\rm A}$  is established during the count interval of TDB time of $\Delta t=t_2-t_1$ and is described as
{}
\begin{eqnarray}
\frac{\hat f_{\rm A}}{f_{\rm A0}}
&=&1-\frac{d_{\rm A}(t_{\rm 2})-d_{\rm A}(t_{\rm 1})}{c\Delta t}+
\frac{1}{c^2\Delta t}\int_{t_{\rm 1}}^{t_{\rm 2}}\Big[{\textstyle\frac{1}{2}}{\vec v}^2_{\rm A}+
\sum_b \big[U_b({\vec z}_{b})-U_b({\vec r}_{b{\rm M}}+{\vec y}_{\rm A})\big]+({\vec a}_{\rm A}\cdot{\vec d}_{\rm A})\Big]dt+{\cal O}(c^{-3}),~~
\label{eq:freq-BCRS}
\end{eqnarray}
where we used the fact that $d_{\rm A}=|x_{\rm A}|$ and $\dot d_{\rm A}=({\vec n}_{\rm A}\cdot{\vec v}_{\rm A})$.

Finally, with the help of Eqs.~(\ref{eq:N4DSN}) and (\ref{eq:freq-BCRS}), we obtain the following relation between two Doppler observables---the spacecraft's frequency $\hat f_{\rm A}$ reported at the BCRS  and the same frequency $\hat f^{\rm C}_{\rm A}$ as measured at the DSN:
{}
\begin{eqnarray}
\hat f_{\rm A}
&=&\hat f_{\rm A}^{\rm C}
\Big[\frac{\hat f_{\rm A}}{f_{\rm A0}}\Big]
\Big[\frac{\hat f_{\rm A}^{\rm C}}{f_{\rm A0}}
\Big]^{-1}+{\cal O}(c^{-3})\nonumber\\
&=&
\hat f_{\rm A}^{\rm C}
\Big\{1+\frac{1}{c\Delta t}
\Big([d_{\rm AC}(t_2)-d_{\rm AC}(t_1)]-[d_{\rm A}(t_2)-d_{\rm A}(t_1)]\Big)+
\nonumber\\
&& \hskip 65pt+\,
\frac{1}{c^2\Delta t}\Big(({\vec v}_{\rm A}\cdot{\vec d}_{\rm C})(t_2)-({\vec v}_{\rm A}\cdot{\vec d}_{\rm C})(t_1)+
\sum_b\int_{t_{\rm 1}}^{t_{\rm 2}}
 U_b({\vec z}_{b})dt\Big)
\Big\}+{\cal O}(\Delta t^{-2},c^{-3}).~~~
\label{eq:freq-freq+}
\end{eqnarray}

Note that the vectors $\vec{R}_{\rm AC}$ (given in Eq.~(\ref{eq(1a)})), ${\vec R}_{\rm A}$ and ${\vec R}_{\rm B}$, (given in Eq.~(\ref{eq:Vect})) are measured simultaneously at the time of signal reception. The result of Eq.~(\ref{eq:freq-freq+}) is what we need in order to derive the DOWR and DWORR observables of the GRAIL mission given by Eqs.~(\ref{eq:DOWR+}) and (\ref{eq:dowrr33}). With the X-band navigation on GRAIL with frequency $\sim8.4$~MHz, the result is accurate to $\sim$0.5 mHz, which is sufficient for the purposes of navigating the GRAIL twins around the Moon.

The expression for frequency transformation, Eq.~(\ref{eq:freq-freq+}), relates the frequency of the signal transmitted by the GRAIL spacecraft as reported at the BCRS to that measured by the DSN station. This result is complete up to the $1/c^2$ order, sufficient for GRAIL.

\end{document}